\documentclass[preprint2]{aastex62}

\usepackage{txfonts}
\usepackage{natbib}
\usepackage{graphicx}
\usepackage{times}
\bibpunct{(}{)}{;}{a}{}{,}
\newcommand{\bc}{\hbox{$\beta$~CMi }}
\newcommand{\be}{\hbox{$\beta$~CMi}}

\newcommand{\fotel}{{\tt FOTEL} }

\newcommand{\korel}{{\tt KOREL} }
\newcommand{\korele}{{\tt KOREL}}
\newcommand{\spefo}{{\tt SPEFO} }
\newcommand{\spefoe}{{\tt SPEFO}}
\newcommand{\pyt}{{\tt PYTERPOL} }
\newcommand{\pyte}{{\tt PYTERPOL}}
\newcommand{\breger}{{\tt PERIOD04} }

\newcommand{\ANG}{\accent'27A}
\newcommand{\Am}{\ANG~mm$^{-1}$ }
\newcommand{\Ame}{\ANG~mm$^{-1}$}

\newcommand{\ubv}{\hbox{$U\!B{}V$}}

\newcommand{\m}{$^{\rm m}\!\!.$}

\newcommand{\kms}{km~s$^{-1}$ }
\newcommand{\ks}{km~s$^{-1}$}
\newcommand{\vsin}{$v$~sin~$i$ }

\newcommand{\tef}{$T_{\rm eff}$ }
\newcommand{\lgg}{{\rm log}~$g$ }

\newcommand{\rs}{R$_{\odot}$}
\newcommand{\cd}{c$\,$d$^{-1}$}
\newcommand{\ha}{H$\alpha$ }

\begin{document}


\title{A new look into putative duplicity and pulsations
       of the Be star $\beta$~CMi
\thanks{Based on spectral observations obtained at the Dominion Astrophysical
Observatory, NRC Herzberg, Programs in Astronomy and Astrophysics,
National Research Council of Canada, Ond\v{r}ejov Observatory and
Universit\"atssternwarte Bochum, and on photometry from the Canadian
MOST satellite and \ubv\ observations from the Hvar Observatory.}
}


\author{P. Harmanec}
\affiliation{Astronomical Institute of the Charles University, Faculty of
Mathematics and Physics,\\ V~Hole\v sovi\v ck\'ach~2, CZ-180 00 Praha~8,
Czech Republic}

\author{M. \v{S}vanda}
\affiliation{Astronomical Institute of the Charles University, Faculty of
Mathematics and Physics,\\ V~Hole\v sovi\v ck\'ach~2, CZ-180 00 Praha~8,
Czech Republic}
\affiliation{Astronomical Institute of the Czech Academy of Sciences,
Fri\v{c}ova~298, CZ-251 65 Ond\v{r}ejov, Czech Republic}

\author{D. Kor\v{c}\'akov\'a}
\affiliation{Astronomical Institute of the Charles University, Faculty of
Mathematics and Physics,\\ V~Hole\v sovi\v ck\'ach~2, CZ-180 00 Praha~8,
Czech Republic}

\author{R. Chini}
\affiliation{Astronomisches Institut, Ruhr--Universit\"at Bochum,
   Universit\"atsstr. 150, 44801 Bochum, Germany}
\affiliation{Instituto de Astronom\'\i a, Universidad Cat\'olica del Norte,
Avenida Angamos~0610, Antofagasta, Chile}

\author{A. Nasseri}
\affiliation{Astronomisches Institut, Ruhr--Universit\"at Bochum,
   Universit\"atsstr. 150, 44801 Bochum, Germany}

\author{S. Yang}
\affiliation{Physics \& Astronomy Department, University of Victoria,
   PO Box 3055 STN CSC, Victoria, BC, V8W 3P6, Canada}

\author{H. Bo\v zi\'c}
\affiliation{Hvar Observatory, Faculty of Geodesy, Zagreb University,
Ka\v ci\'ceva~26, HR-10000 Zagreb, Croatia}

\author{M. \v{S}lechta}
\affiliation{Astronomical Institute of the Czech Academy of Sciences,
Fri\v{c}ova~298, CZ-251 65 Ond\v{r}ejov, Czech Republic}

\author{L. Vanzi}
\affiliation{Department of Electrical Engineering and Centre of Astro-Engineering,
Pontificia Universidad Catolica de Chile, Av. Vicu\~na Mackenna 4860,
Santiago, Chile}

\shorttitle{On the duplicity and pulsations of \be}
\shortauthors{Harmanec et al.}


\begin{abstract}
Bright Be star \bc has been identified as a non-radial pulsator on the basis
of space photometry with the MOST satellite and also as a single-line
spectroscopic binary with a period of 170\fd4. The purpose of this study is
to re-examine both these findings, using numerous electronic
spectra from the Dominion Astrophysical Observatory, Ond\v{r}ejov Observatory,
Universit\"atssterwarte Bochum, archival electronic spectra from
several observatories, and also the original MOST satellite photometry.
We measured the radial velocity of the outer wings of the double
\ha emission in all spectra at our disposal and were not able to confirm
significant radial-velocity changes. We also discuss the problems related
to the detection of very small radial-velocity changes and conclude
that while it is still possible that the star is a spectroscopic binary,
there is currently no convincing proof of it from the radial-velocity
measurements.  Wavelet analysis of the MOST photometry shows
that there is only one persistent (and perhaps slightly variable)
periodicity of 0\fd617 of the light variations, with a double-wave light
curve, all other short periods having only transient character.
Our suggestion that this dominant period is the star's rotational period
agrees with the estimated stellar radius, projected rotational velocity
and with the orbital inclination derived by two teams of investigators.
New spectral observations obtained in the whole-night series would be
needed to find out whether some possibly real, very small
radial-velocity changes cannot in fact be due to rapid line-profile changes.
\end{abstract}





\section{Introduction}

\medskip

The very nature of the Be phenomenon, i.e. the temporal presence and
absence of Balmer and other emission lines in the spectra of otherwise
seemingly normal B stars, remains a puzzle in spite of a huge effort
of several generations of astronomers. The most widely considered
hypotheses are the modern variants of the original ideas of \citet{struve31},
who suggested an outflow of material due to rotational instability at
the stellar equator, and of \citet{ger34}, who advocated an outflow due to
stellar wind.

 There were also suggestions that the duplicity of Be stars can play some
role in the whole phenomenon \citep[cf., e.g.][]{bebin75, bebin82, gies2001,
bebin2016}. \citet{bebin2001} argued that the duplicity of Be stars can
play two possible roles: (1) to be somehow responsible for the formation and
dispersal of the Be envelopes, and (2) to be responsible for various
observed types of the Be-star time variations. He also published a catalog
of known hot emission-line binaries. \citet{pano2016} suggested
yet another possible role of the duplicity: truncation
of their disks by the Roche lobes around of the Be components.

A good summary of the current views can be found in \citet{rivi2013}.

\section{Current knowledge about \be}
The bright B8Ve star \bc (3~CMi, HD~58715, HR~2845, MWC~178, HIP~36188;
$V=2$\m9) is known for the long-term stability of its envelope. Probably due
to this circumstance, its observational history is not as rich as one would
expect for such a bright star, accessible to observers from both hemispheres.
The double \ha emission-line profiles, with violet ($V$) and red ($R$)
peak intensities of some 80~\%  of the continuum level,
look similar on the high-resolution spectra
from $1989-1992$ published by \citet{hanu96} (their Fig.~35),
on the $2001-2007$ spectra in Fig.~1 of \citet{kl2015}, and on the
$2013-2014$ spectra in Fig.~1 of \citet{folsom2016}. However, a photographic
\ha spectrum with a good dispersion of 6.8~\Ame\ from March 26, 1953 published
by \citet{under53} seems to indicate a larger asymmetry of the emission
peaks. From her Fig.~1 we estimate $V=1.90$, $R=1.62$, therefore $V/R=1.17$\,.

\citet{saio2007} analysed series of \bc photometry secured with the MOST
satellite and reported that the star is a multiperiodic non-radial pulsator.

\citet{struve25} published a preliminary report on twelve new spectroscopic
binaries discovered at Yerkes and reported a radial-velocity (RV) range
from $-1$ to $+53$~\kms for \be. \citet{frost26} published those 7 Yerkes RVs
from 1904 to 1911 individually and classified the star as a binary with unknown
orbit. They remarked that the H$\beta$ line appeared as a distinct double emission
on two of their spectra. \citet{jarad89} obtained RVs from digitized 30~\Am\
photographic spectra for 18 emission-line stars including \be.
From 30 RVs for \bc they concluded that the star is a single-line
spectroscopic binary with an orbital period of 218\fd498, eccentric orbit
and a semiamplitude  of 22.4~\ks. \citet{krugov92} obtained 63 photographic
spectra of \bc with a moderate dispersion of 44~\Ame. They were secured
in night series from April 1987 to March 1988 in an effort to detect
rapid variability of the \ha line. He measured and tabulated the equivalent
widths (EW), peak intensities and the $V/R$ ratio and claimed that rapid
irregular variations were detected. In another study, \citet{krugov98}
re-analysed 45 of the same spectra using a~digitizing microdensitometer
and provided some plots of the EW, peak intensities and RVs. It is not
clear to what parts of the \ha profile these RVs refer. They also published
a selection of the eight \ha profiles.  The peak intensity is between
1.6 and 2.0 of the continuum level.

\citet{kl2015} \citep[see also][]{kl2017c} and \citet{kl2017} carried out
a~multitechnique testing of the viscous decretion disk model and studied
the structure of the outer parts of the disks around \bc and a few other
Be stars. They derived various physical parameters of the disks and central
stars and tentatively suggested that the disks could be truncated due to
presence of binary companions. Inspired by this suggestion, \citet{folsom2016}
measured the $V/R$ ratio of the double \ha emission in several hundred spectra
and found two possible periods, 368\fd23, and 182\fd83. They preferred the
shorther period, which they identified with the star binary orbital
period. Subsequently, \citet{dul2017} analysed RVs from
124 Ritter Observatory echelle spectra, measured on the steep wings of
the \ha emission by an~automatic procedure and concluded that the star is
a~single-line binary with a low semi-amplitude of ($2.25\pm0.4$)~\kms and
an~orbital period of 170\fd4$\pm$4\fd3. \citet{wang2018} carried out a~search
for hot compact companions to Be stars in the archival IUE spectra. They
were unable to detect such a companion for \be.

Since the orbital RV curve obtained by \citet{dul2017} shows a~large scatter,
and since we have access to several independent sets of electronic
spectra of \be, we decided to re-investigate the duplicity of this star,
its time variability on several time scales and its physical properties.
The results are reported in this study.

\section{Spectroscopic observations}
Our new spectroscopic material consists of
32~CCD spectra from the Dominion Astrophysical Observatory, 5~Reticon and
25~CCD spectra from Ond\v{r}ejov, and 10 echelle spectra from the
Universit\"atssternwarte Bochum near Cerro Armazones.
We also used spectra from public archives of several other observatories.
A~journal of all electronic spectra available to us is in Tab.~\ref{jousp}.
Everywhere in this paper, we shall use the abbreviation

\begin{equation}
{\rm RJD=HJD-2400000.0}
\end{equation}

\noindent for the reduced heliocentric Julian dates.

\begin{deluxetable*}{ccrcclllrrr}
\tablecaption{Journal of available electronic spectra.\label{jousp}}
\tablewidth{0pt}
\tablehead{
\colhead{Instrument} & \colhead{RJD range} & \colhead{No.}&
\colhead{Wavelength range (\ANG)}&\colhead{Spectral resolution}
}
\startdata
DAO\tablenotemark{ 1}&49745.9--54519.6&32&6160--6750&21700\\
OND R\tablenotemark{ 2}&51567.5--51661.3& 5&6260--6730&11700\\
OND C\tablenotemark{ 3}&52229.5--58198.3&25&6260--6730&11700\\
ELODIE\tablenotemark{ 4}&    52362.3     & 1&3906--6811&42000\\
FEROS\tablenotemark{ 5}&53011.7--53012.6& 2&3740--9200&48000\\
UVES\tablenotemark{ 6}&53334.8--53334.9& 2&4790--6800&40000\\
SOPHIE\tablenotemark{ 7}&    54423.7     & 2&3870--6940&75000\\
BESO\tablenotemark{ 8}&55320.5--56962.9&10&3700--8500&50000\\
CFHT\tablenotemark{ 9}&55871.1--55877.1& 2&3700--10480&68000\\
PUCHEROS\tablenotemark{10}&56350.7--57080.6& 3&4250--7300&20000\\
\enddata
\tablenotetext{1}{Dominion Astrophysical Observatory 1.22 m reflector,
    CCD detector}
\tablenotetext{2}{Ond\v{r}ejov 2.0 m reflector, coud\'e spg., Reticon 1872
                 detector.}
\tablenotetext{3}{Ond\v{r}ejov 2.0 m reflector, coud\'e spg., CCD detector.}
\tablenotetext{4}{Haute Provence Observatory 1.93 m refelector,
                 Elodie echelle spg.; \citet{baran96}}
\tablenotetext{5}{ESO 1.52 m reflector, FEROS echelle spg.;
              \citet{kaufer97}}
\tablenotetext{6}{ESO UT2 Kueyen 8 m reflector, UVES echelle spg.;
        \citet{dekker2000}}
\tablenotetext{7}{Haute Provence Observatory 1.93 m refelector,
      Sophie echelle spg.; \citet{perru2008}}
\tablenotetext{8}{Bochum Hexapod 1.5 m Telescope, BESO echelle spg.;
              \citet{fuhr2011}}
\tablenotetext{9}{Canada France Hawaii Telescope 3.6 m reflector,
      ESPaDOnS echelle spg.; \citet{grund90}}
\tablenotetext{10}{Pontificia Universidad Cat\'olica 0.5 m reflector,
      Pucheros echelle spg.; \citet{vanzi2012, arcos2018}}
\end{deluxetable*}

We also derived RJDs for the early Yerkes photographic spectra and publish
them together with the RVs published by \citet{frost26} in Tab.~\ref{oldrv}
in the Appendix.

Initial reductions of the DAO spectra and creation 1-D spectra was carried out
by SY in IRAF. The same plus the wavelength calibration was done by
M.~\v{S}lechta for the Ond\v{r}ejov CCD spectra. The initial reductions of
the BESO spectra was carried by AN with the help of an automatic pipeline.

The final reductions of all spectra, including the archival ones
(i.e. the wavelength calibration for the Ond\v{r}ejov Reticon and DAO
spectra, rectification, removal of cosmics when necessary, and the RV and line
intensity measurements of the \ha line) was carried out by PH in
\spefo \citep{spefo1,spefo3}. All these measurements are also provided
as Tab.~\ref{ourrv} in the Appendix.

\section{Analysis of electronic spectra}
\subsection{Radial velocities of \be}
Determination of precise RVs of Be stars, especially those related
to orbital motion in a binary system, has been a very challenging task.
The problem lies in the following obstacles.
\begin{enumerate}
\item The supposedly photospheric lines of \ion{He}{1} or \ion{H}{1}
are usually broadened by a high rotation and distorted by travelling
subfeatures moving across the line profiles and/or by often
unrecognized emission from their circumstellar envelopes.
\item Virtually all known binaries among Be stars have low-mass
companions so the orbital RV curves of the Be stars have low semi-amplitudes,
typically less than 10~\ks.
\item In epochs of activity, the envelopes of Be stars become elongated,
not axially symmetric, and their elongation slowly revolves
on a time scale of several years. In such situation, the self-absorptions
in the envelopes exhibit RV variations, which can have semi-amplitudes as
high as some $60-70$~\ks, thus masking much smaller and usually more
rapid orbital RV changes
\citep[e.g.][]{mclaughlin61, cop63, delplace70}.
\item In case of some asymmetries present in the line profiles, the result can
also depend on the spectral resolution used.
\end{enumerate}

\begin{figure}
\plotone{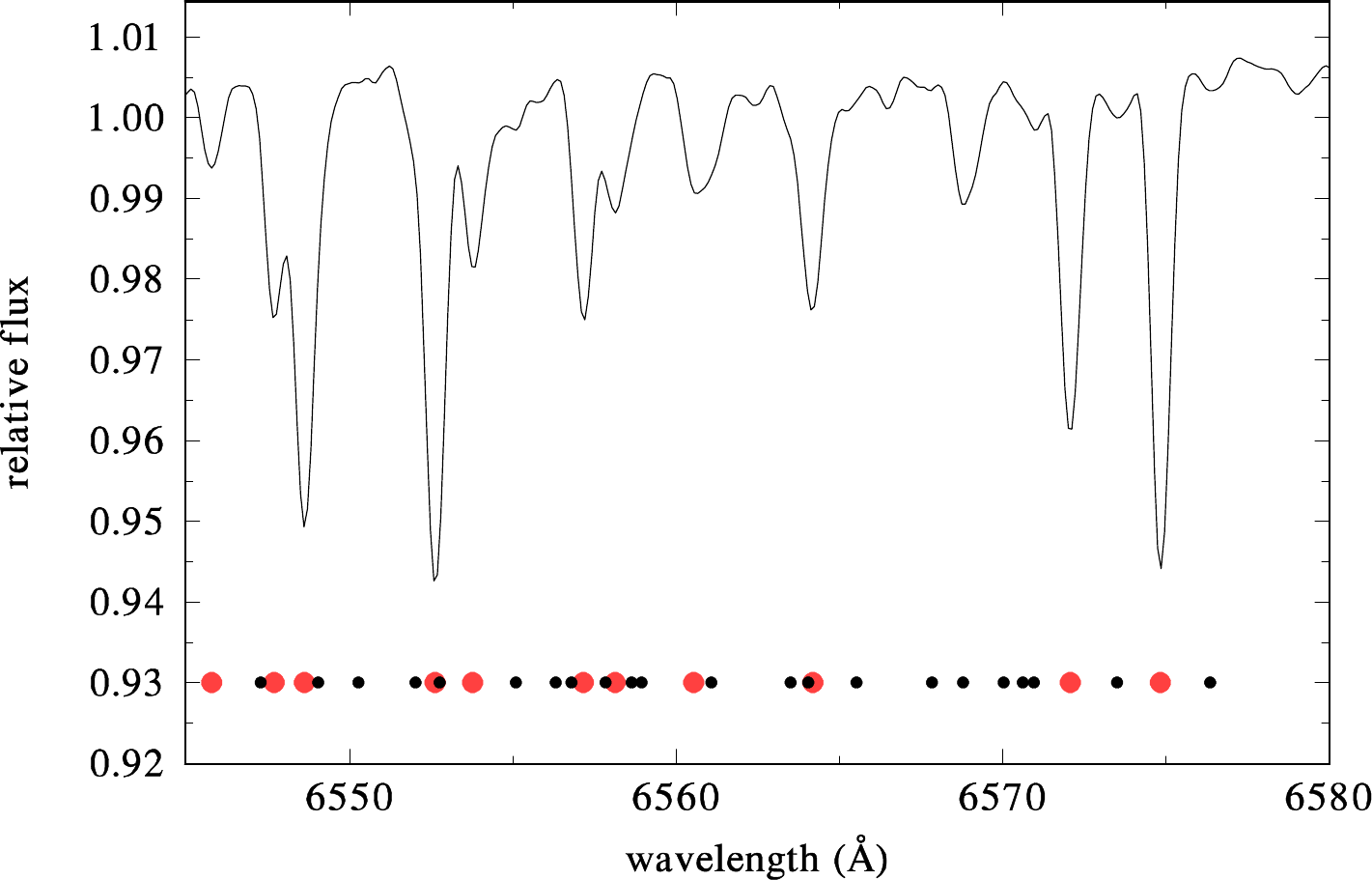}
\plotone{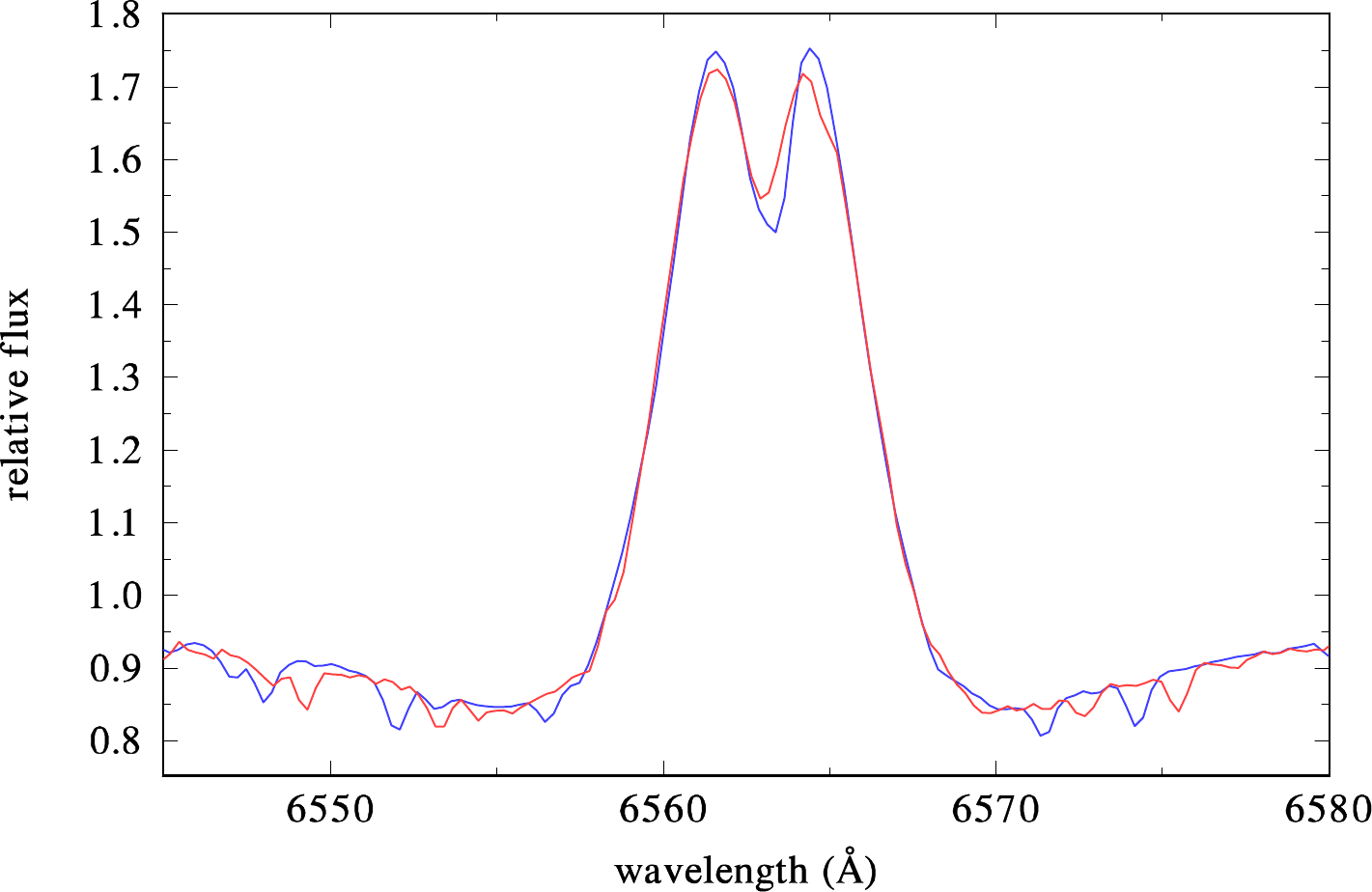}
\plotone{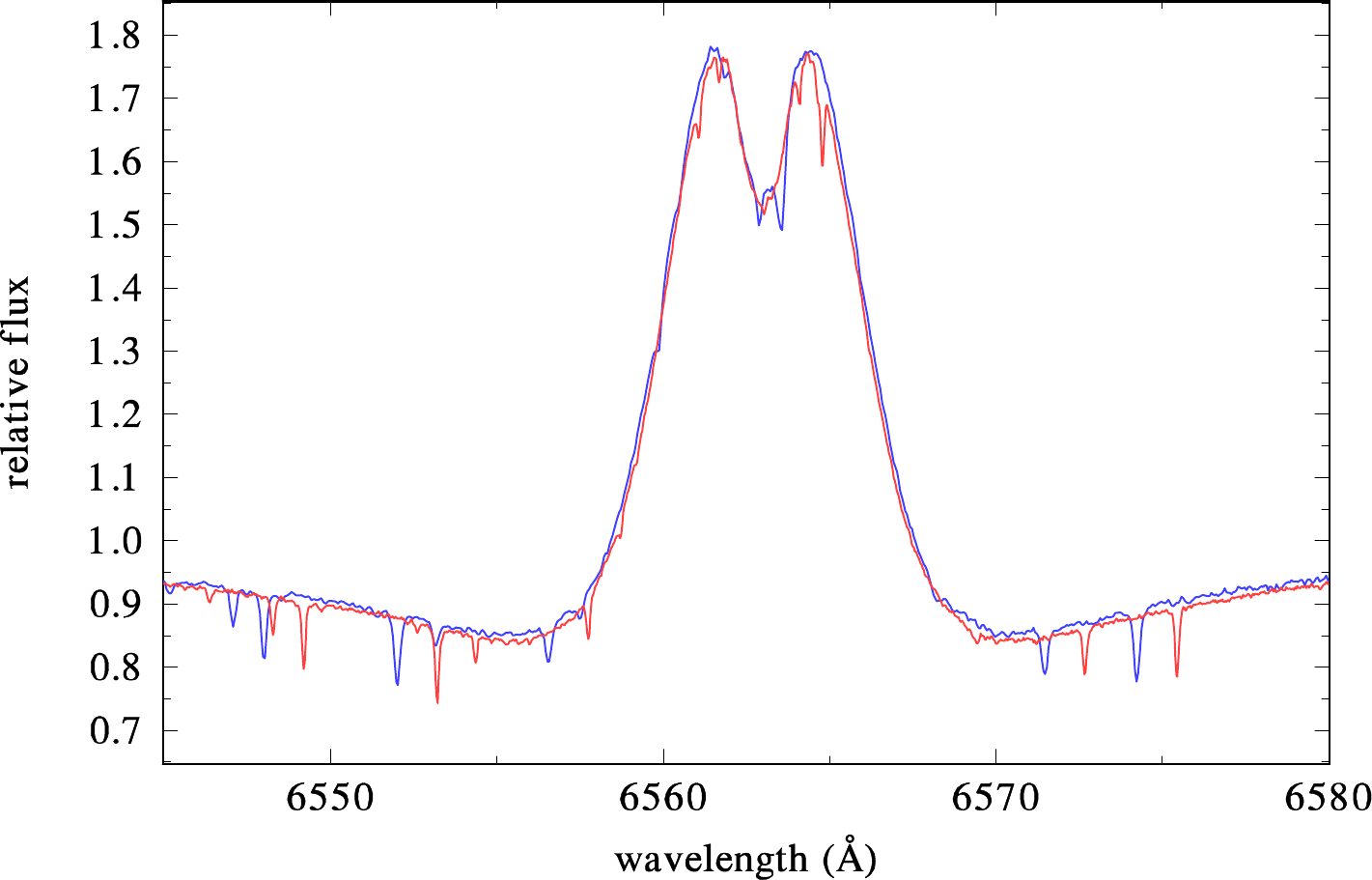}
\caption{Comparison of medium- and high-dispersion \ha spectra
of \bc from the two extrema of the heliocentric RV corrections. All
spectra are on the heliocentric wavelength scale.
Top: Telluric lines from the Ond\v{r}ejov CCD spectra disentangled with
the \korel program. Circles at the bottom of the panel show the position
of all telluric lines from the solar spectrum of \citet{moore66}, the red
ones denoting the lines with equivalent widths greater than 5~m\ANG.
Middle: Ond\v{r}ejov CCD spectra: blue is from RJD~56395.3447 with the heliocentric
RV correction $V=-29.1$~\ks, red is from RJD~55470.6614 with $V=+28.3$~\ks.
Bottom: blue is a Canada France Hawaii Telescope spectrum from RJD~52871.1482
with $V=-28.1$~\ks, red is the Elodie spectrum from RJD~55871.1482 with $V=+27.1$~\ks.}
\label{haspec}
\end{figure}

 The situation is well illustrated by the fact that in spite of great effort
after the first discoveries of X-ray+Be binaries, for virtually none of
the main-sequence Be stars the orbital period was found from the RVs of
the Be star, see, e.g. the negative search in the large collection of
photographic RVs of $\gamma$~Cas by \citet{cowley76}.

\citet{hrvoje95} were the first to demonstrate that for Be stars with a~strong
\ha emission the best estimate of the true orbital motion of the Be star
is the RV measured on the steep outer wings of the \ha emission, which
originates from the central, more or less axially symmetric parts of the
disk around the central star. This way and using the technique of the
comparison of the direct and flipped line profiles on the computer screen
as it is possible in the program \spefo \citep{spefo1,spefo2,spefo3}, they
obtained a sinusoidal RV curve for the Be component of $\varphi$~Per.

  Using the same technique, \citet{zarf20} succeeded to discover the 203\fd6
orbital period of $\gamma$~Cas and to demostrate its presence even in the
photographic RVs by \citet{cowley76}. Their finding was then confirmed
from an independent series of \ha spectra by \citet{miro2002}. These
authors, however, measured the RV of the \ha outer wings by an automatic
procedure. Both approaches were compared in the so far richest collection
of \ha spectra of $\gamma$~Cas by \citet{zarf29}, with the conclusion that
the manual measurement in \spefo provides more accurate RVs than the automatic
procedure. This is probably due the fact that the measurer can avoid
line-profile distortions due to numerous telluric lines and/or small flaws
in the region of the \ha line.

 \citet{zarf21} derived the RV curve of the Be component of V832~Cyg = 59~Cyg
also from the \ha emission wings. They noted that due to a partial filling
of the broad \ion{He}{1} lines by emission, exhibiting orbital $V/R$ changes
in phase with the orbital RV changes, RVs from these, seemingly
photospheric lines, give an RV curve with a larger semi-amplitude
than those of the \ha emission.
Their conclusion was nicely confirmed by \citet{peters2013}, who derived
the RV curves of both binary components of V832~Cyg from the UV spectra.
Other \ha RV curves were also derived for $\zeta$~Tau by \citet{zarf26}
(these authors summarized all reasons, why the \ha RV measured on the
steep wings of the \ha emission best reflects the true orbital motion of
the Be component) and by \citet{zarf28} for o~Cas.

The same approach has also been used by \citet{dul2017} to derive the RV curve
of \be. They once more used the automatic determination of the RV from the
\ha emission wings. There is a trap, however. \citet{desmet2010}
and \citet{zarf30} measured the RVs of the \ha emission wings for two Be stars
in semi-detached systems with cool, Roche-lobe filling secondaries, AU~Mon,
and BR~CMi, respectively. They found that in both cases the RVs of the
\ha emission wings led to RV curves, which were not exactly in anti-phase
with respect to the well-defined RV curves of the cool secondaries but
showed some phase shift. This could be due to presence of gas flow
deflected by the Coriolis force in these mass-exchaning systems and also due to
the fact that the \ha emission in both objects is only moderate, with peak
intensities of less than 2.0 of the continuum level. In such situations,
the RV measurements are prone to the blending effects of numerous telluric
lines much more than strong emission profiles.

   This is, unfortunately, also the case of \be, since its \ha emission reaches
only some 60 to 80 \% above the continuum level. For this particular star,
there are two additonal problems. It is located near the ecliptic, which causes
a RV shift of the telluric lines for full 60~\kms as the Earth revolves around
the Sun and it can only be observed for about half a~year. In Fig.~\ref{haspec}
we show pairs of the \ha line profiles of \bc from two extremes of the telluric-line
shifts for high-dispersion and high-resolution, and medium-dispersion and
medium-resolution spectra at our disposal. Also shown there is the mean telluric
spectrum from the numerous Ond\v{r}ejov spectra disentangled with the program
\korel \citep{korel1, korel2, korel3}. One can see how the telluric lines can
fool both, the precise RV measurements, and measurements of the $V/R$ ratio.

\begin{figure}
\includegraphics[width=0.5\textwidth]{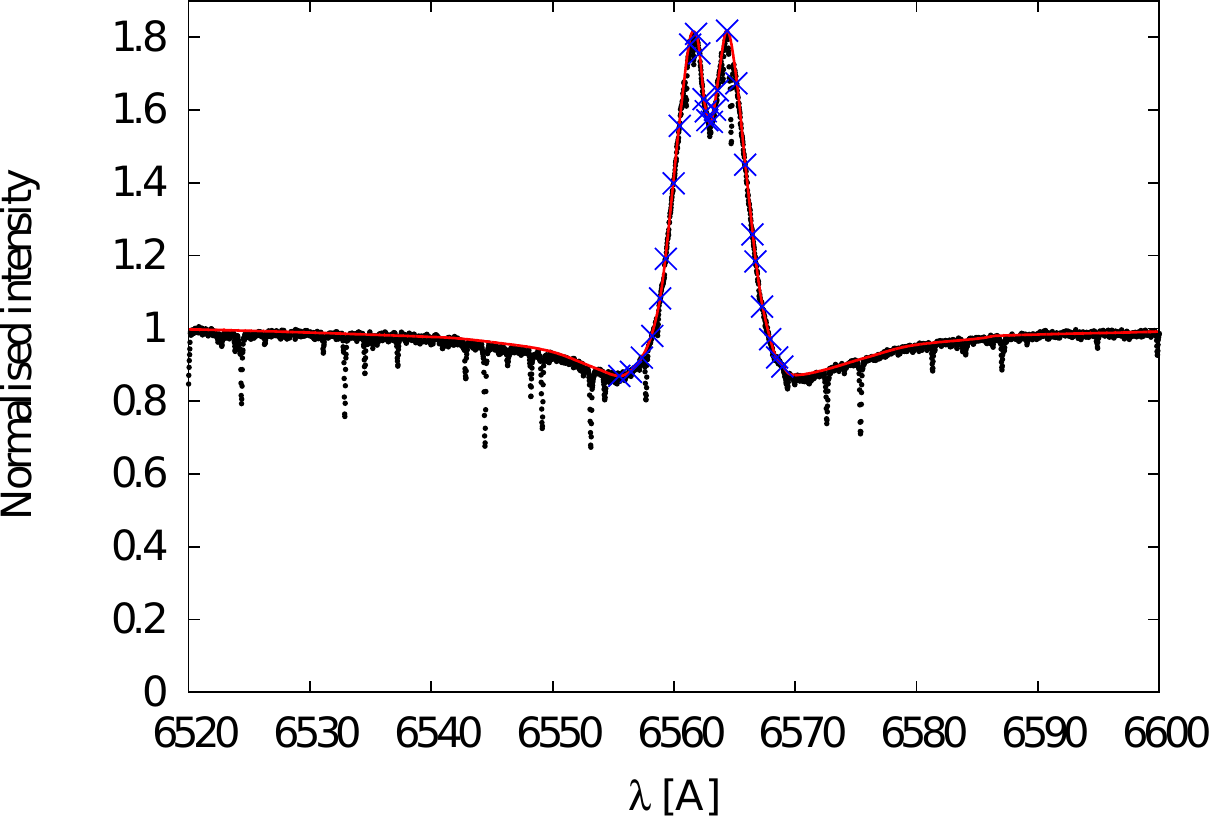}
\includegraphics[width=0.5\textwidth]{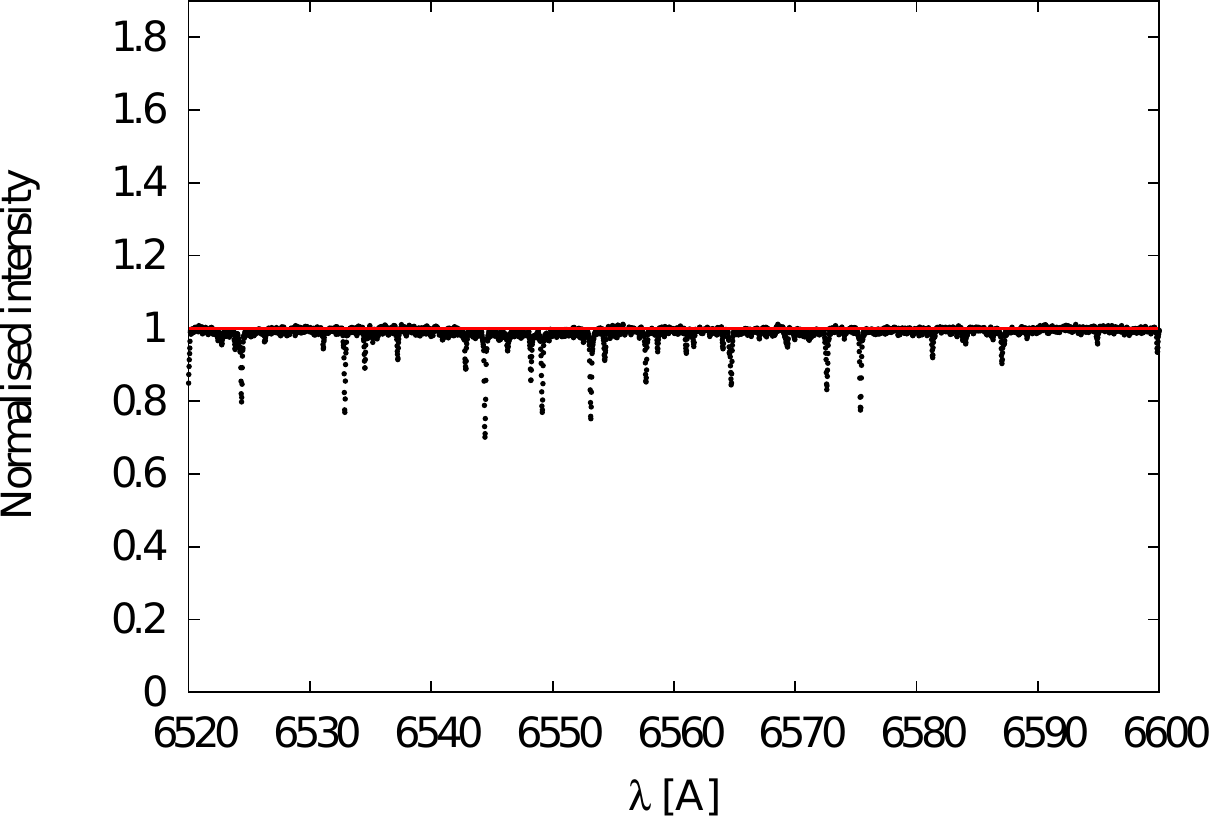}
\includegraphics[width=0.5\textwidth]{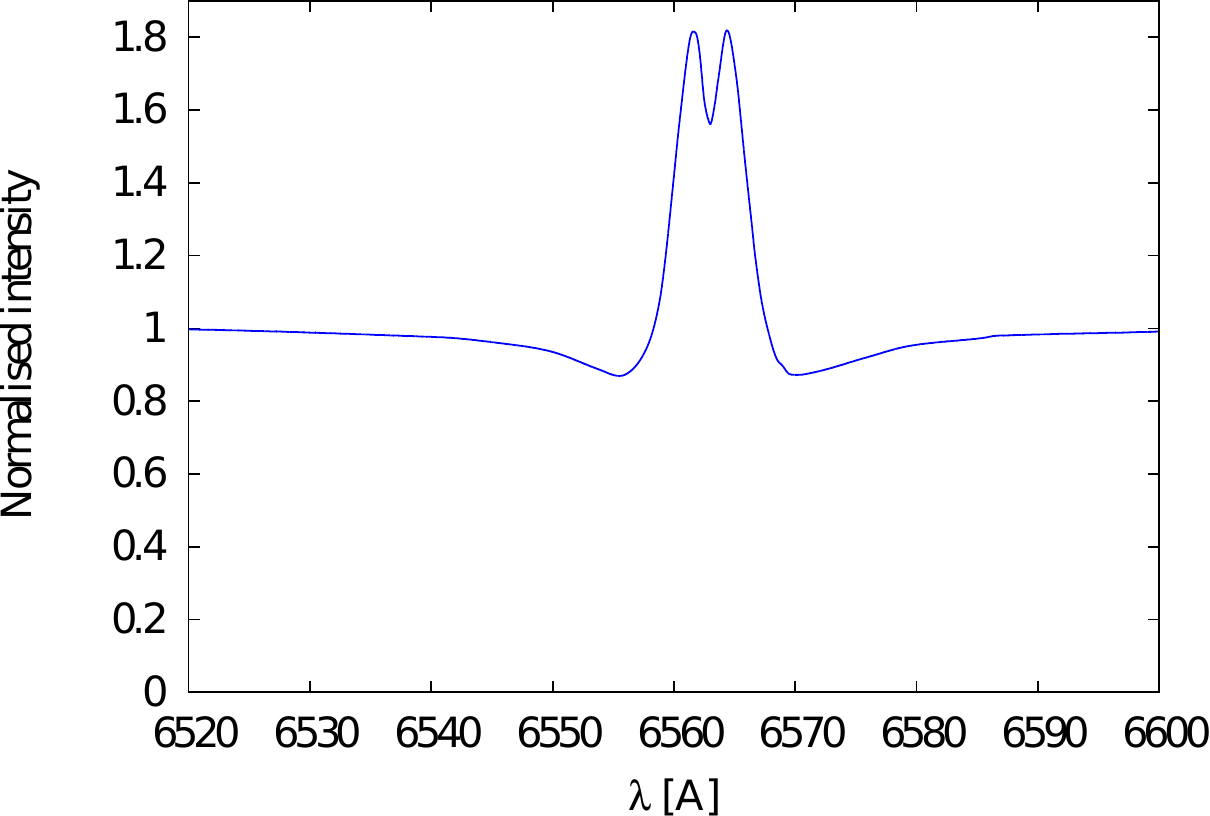}
\caption{Cleaning one SOPHIE spectrum from telluric lines. Top:  original
spectrum with an artificial rectification (red) to remove the \ha profile.
Crosses denote the points where we set pseudocontinuum within the \ha line.
Middle: Clean telluric spectrum after the above artificial rectification.
Bottom: Clean \ha profile resulting from the division of the original
spectrum by the clean telluric spectrum.}
\label{telluric}
\end{figure}

   Being aware of this problem, \citet{dul2017} applied the IRAF task
{\tt telluric} to get rid of them. Regrettably, they do not provide any details
other than that they used spectra of rapidly rotating hot stars obtained with
the same spectrograph as their spectra of \be. It is not mentioned whether
the spectra of rapidly rotating stars were obtained each night, when \bc
was observed and how exactly they re-normalized them. This technique of
the telluric-line removal is a delicate task. Given the fact that the reported
full amplitude of RV changes of $\sim 4$~\kms represent a sub-pixel resolution,
it is conceivable that the removal was not complete. This suspicion seems to be
supported by the fact that the residual absorption features seen at about
400, 530, and 1070 \kms in Fig.~1 of \citet{dul2017} clearly correspond to
the strong water vapour lines at 6572.1, and 6574.8~\ANG, and to the blend of
two such lines at 6586.6~\ANG. We do admit, however, that \citet{dul2017} do not
state explicitely whether the \ha profile in their Fig.~1 corresponds to the
spectrum with the telluric lines already removed or to the original one.

\begin{figure*}
\gridline{\fig{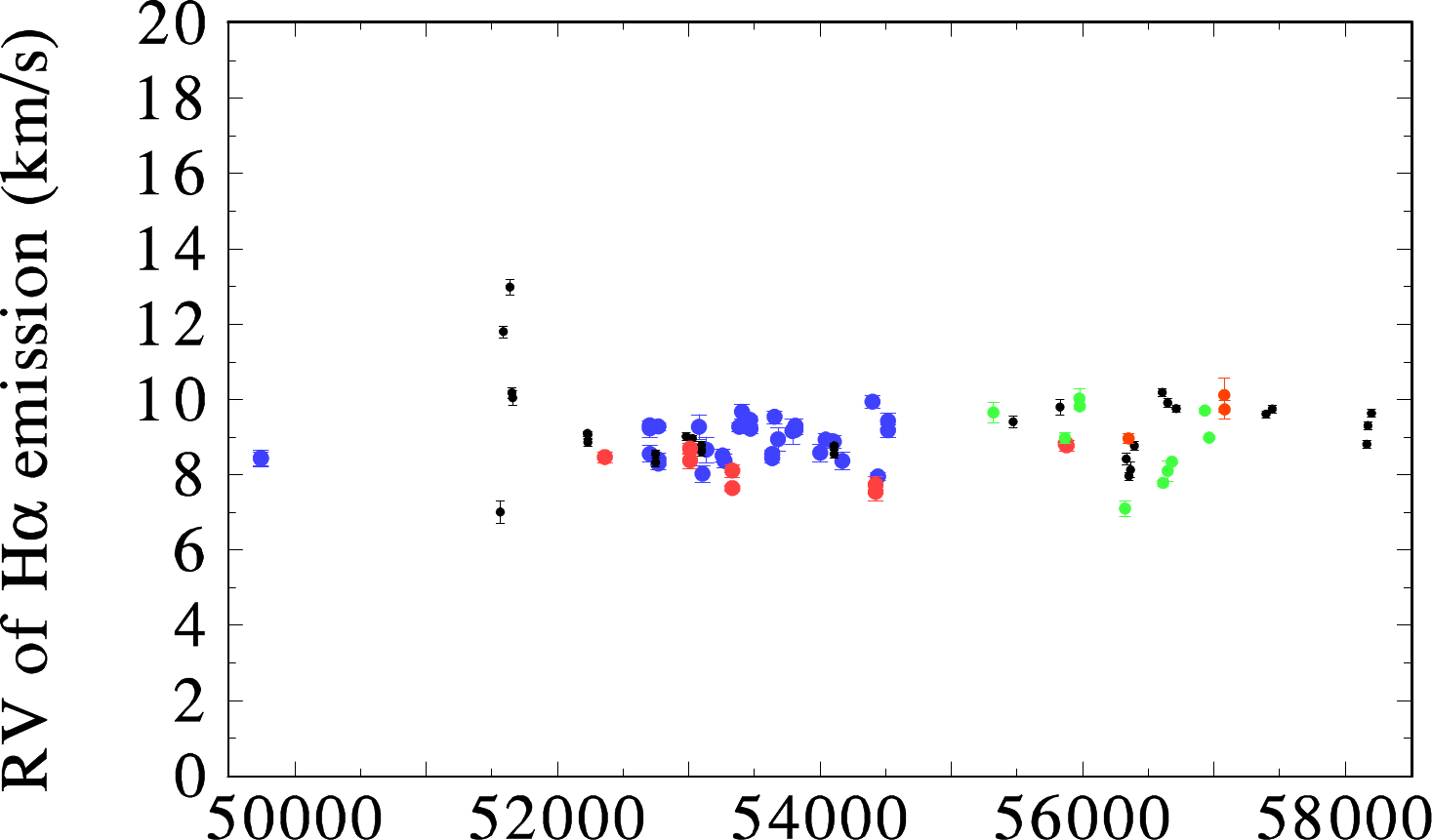}{0.46\textwidth}{(a)}
          \fig{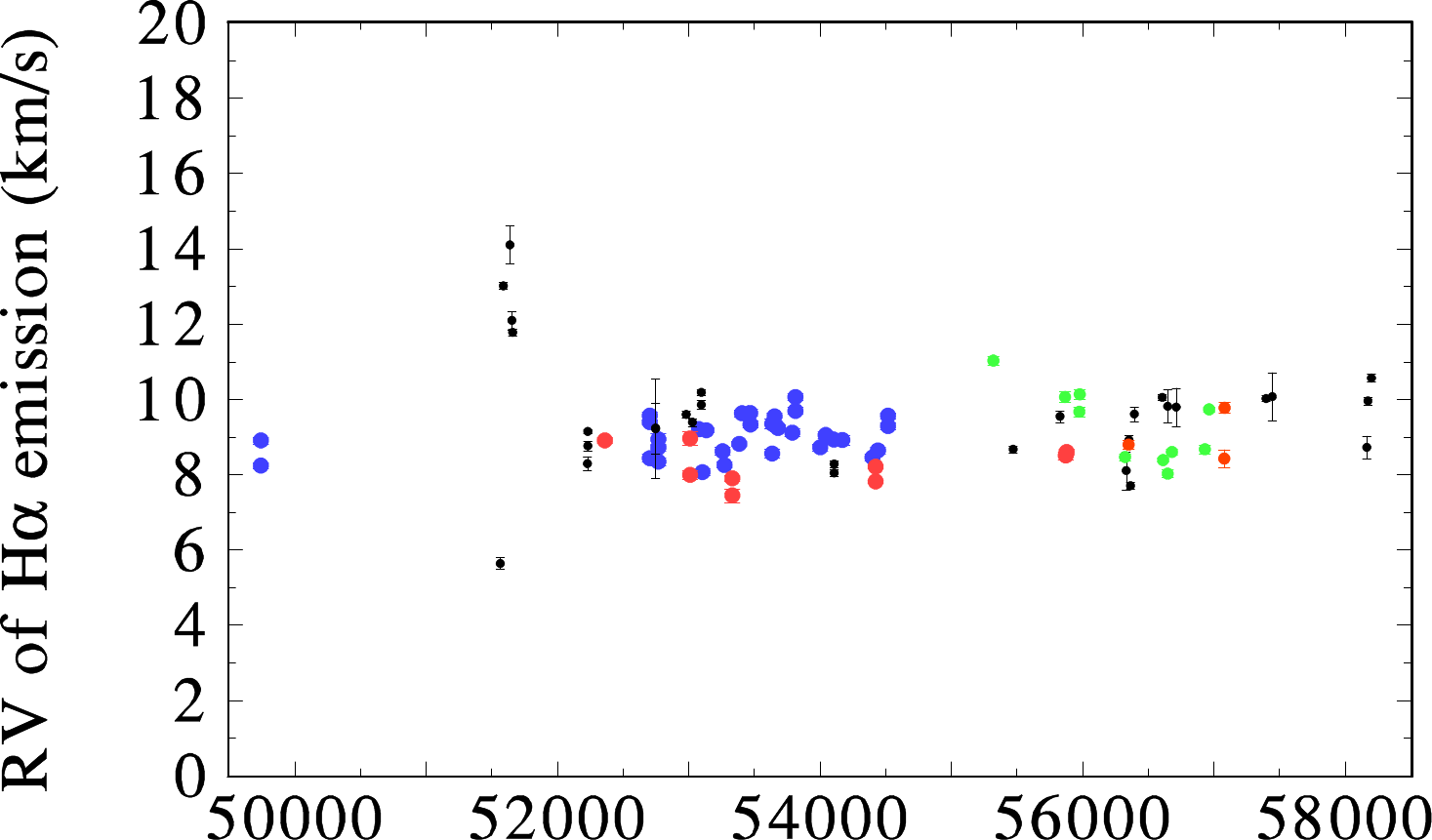}{0.46\textwidth}{(b)}
}
\gridline{\fig{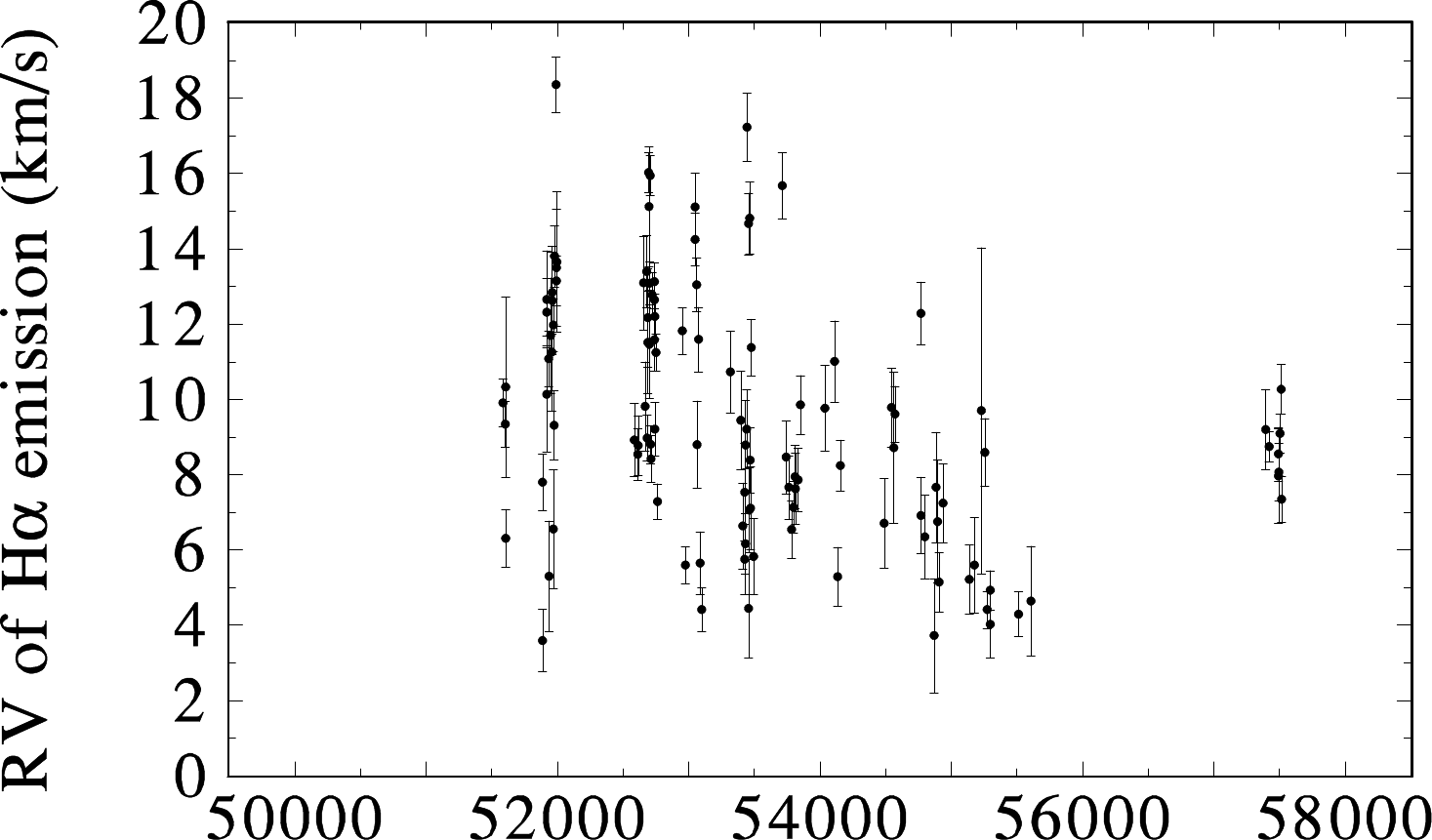}{0.46\textwidth}{(c)}
          \fig{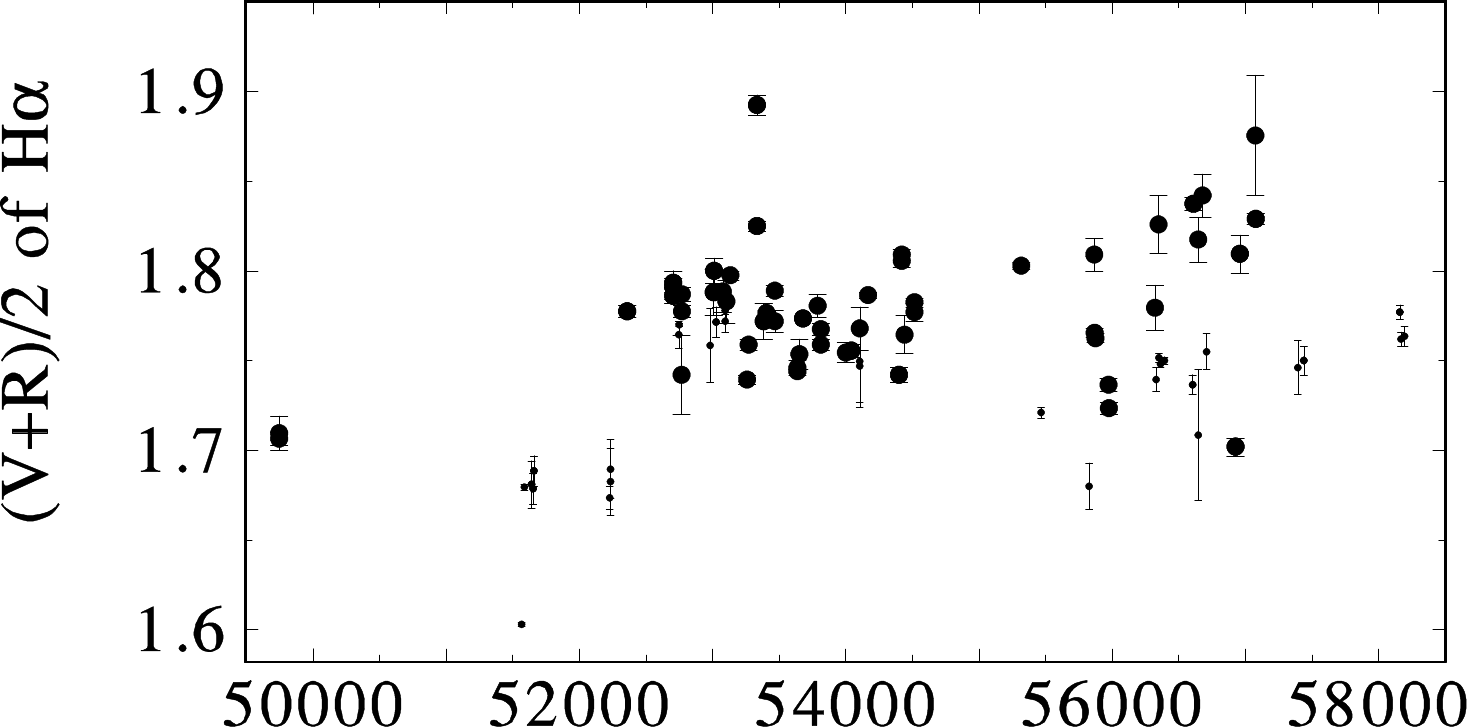}{0.46\textwidth}{(d)}
}
\caption{Individual quantities measured in the \ha profiles of all spectra
of Table~\ref{jousp} plotted vs. time (in RJD):
(a) Our emission-wing RVs measured in the original spectra in \spefoe.
Individual data sets are distinguished as follows: blue... DAO,
black... OND~R and OND~C, red... best high-S/N
ELODIE, FEROS, UVES, SOPHIE, and CFHT spectra,
green... BESO, and brown... PUCHEROS.
(b) Our emission-wing RVs measured in \spefo in the spectra with telluric lines
removed. The same notation as above.
(c) Emission-wing RVs of \citet{dul2017} measured by an automatic procedure
by them. Since their RVs were derived via cross-correlation and cluster thus
around zero, we added +9~\kms to them to make the direct comparison with our RVs
easier.
(d) Total strength of the \ha emission measured in our spectra as a mean of the
V and R peak intensities expressed in the units of local continuum.
The medium-resolution Ond\v{r}ejov spectra are shown by smaller black circles
than all other data.}
\label{timeplots}
\end{figure*}

\subsubsection{RVs of the \ha emission wings}
We measured the RVs of the \ha emission wings using two different approaches:
\begin{itemize}
\item Using our standard technique of the comparison of the direct and flipped line-profile
images on the computer screen in the \spefo program, we measured RVs of the steep emission-line
wings of \ha profiles on all spectra listed in Table~\ref{jousp}, trying to avoid the parts
of the profiles affected by telluric lines. All measurements were carried repeatably,
at different times, not to remember the previous settings to obtain some idea about
their accuracy. Moreover, we also measured a selection
of telluric lines to carry out an additional correction of the zero point
of the RV scale \citep{spefo1}. The RJDs, and RVs of \ha emission with their
rms errors are provided in the first two columns of Table~\ref{ourrv}
in Appendix, while the additional RV zero point corrections and the rms errors
of the mean measured RV of telluric lines are provided in the fourth and
fifth column.
\item Realizing that the Hermite polynomials used in \spefo for spectra normalization
can also be used for the removal of the observed \ha profiles like it is done for the
spectra of hot, rapidly rotating stars in the IRAF task {\tt telluric}, we renormalized
all our spectra to obtain clean spectra of telluric lines and divided the original spectra
by them. In our opinion, this is a suitable approach in the given case since it uses
the real strength of telluric lines and no adjustment in either the strength or velocity
of the lines is needed. It also takes out the noise, so that it has a very similar
effect to a low-pass filtering. We caution, however, that it can only be carried out
for stars with spectral lines significantly broader than the telluric lines. The procedure
worked remarkably well as it is illustrated for one spectrum
in Fig.~\ref{telluric}. We then measured the steep \ha emission wings in the clean
spectra in the same way as before. These measurements and their rms errors, based again
on several repeated measurements, are given in the third column of
Table~\ref{ourrv}. Finally, we also derived the strength of the \ha emission as the
mean of the strength of $V$ and $R$ peaks of the double emission profile. It is
given in the sixth column of Table~\ref{ourrv}.
\end{itemize}

  We started our analyses plotting the RVs of the emission wings of
\ha profiles measured by both methods versus time in Fig.~\ref{timeplots}a,b.
In Fig.~\ref{timeplots}c we also plot the \ha emission RVs obtained by
\citet{dul2017}, and Fig.~\ref{timeplots}d we display the strength of
the \ha emission. This shows a~few things.
\begin{enumerate}
\item There is an indication of a temporal increase of the RV to some 12~\kms
after RJD~51500 but it is solely based on several early Reticon spectra. It might be
real since Be stars are known to show intervals of increased activity with secular
RV changes. Also, as already mentioned, early photographic RVs are indicative of
larger RV changes. However, without independent evidence we do not assign too much
weight to it, noting that the Reticon digital convertor used only 4096 discrete
levels in recording the flux.
\item It is obvious that our RV measurements are much more precise than the automatically
measured RVs of \citet{dul2017} and show a~much lower scatter, perhaps a constant RV
within the accuracy of the measurements. The real uncertainty of RV values is hard
to estimate well. The rms errors of repeated setting on the emission-line wings are quite
low for both methods we used. However, there can be larger external errors related, e.g.,
to small local inaccuracies in the dispersion relation. From the comparison of RV derived
for the same spectra by two alternative approaches we conclude that the realistic error
of individual RV can be something like $\pm 1$~\ks. Robust mean value of RVs measured
in the original spectra is $8.93\pm0.11$~\ks, while that for the spectra with the telluric
lines removed is $9.05\pm0.13$~\ks.
\item There is a very uncertain indication of a mild secular increase of the \ha emission strength
over the interval of our observations. One UVES spectrum from
RJD~53334.8232, and a~BESO spectrum from RJD~56929.9217 exhibit rather
anomalous emission strengths. We checked the whole reduction procedure and
found no problem, so we reproduce these values as they are.
\end{enumerate}

\begin{figure*}
\gridline{\fig{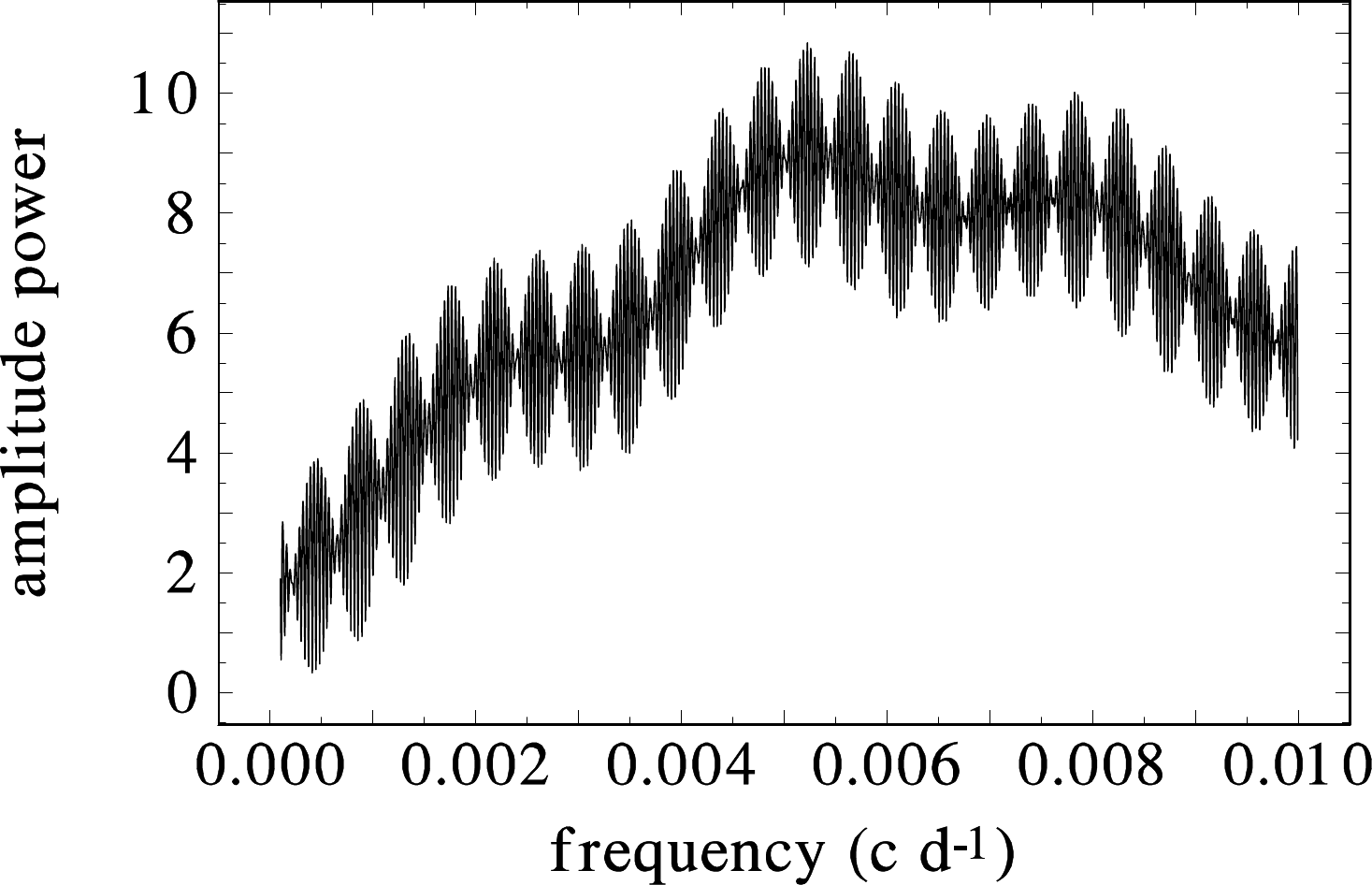}{0.4\textwidth}
         {Frost et al. (1926) \& Jarad et al.(1989)}
          \fig{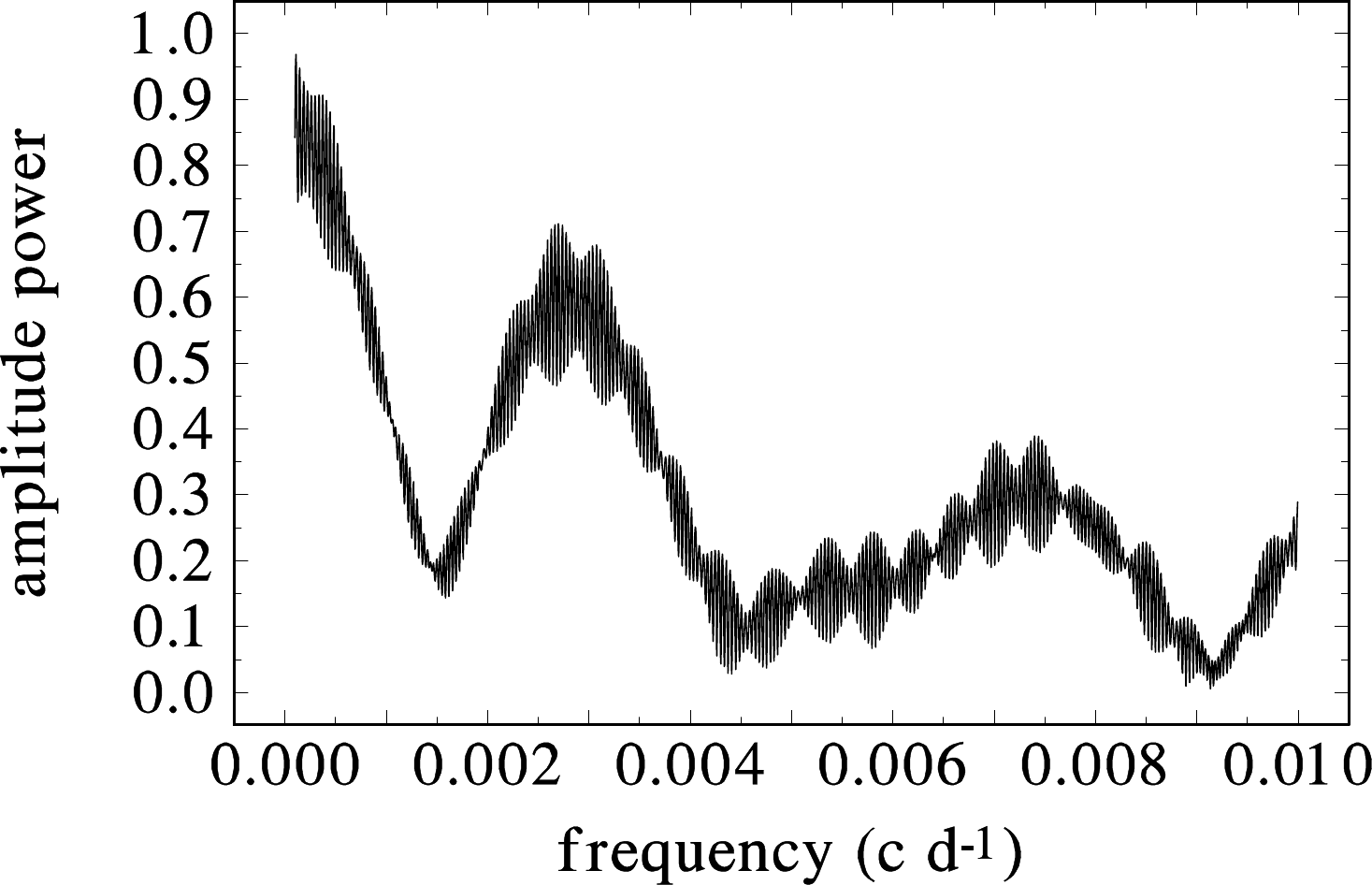}{0.4\textwidth}{spectral window}
}
\gridline{\fig{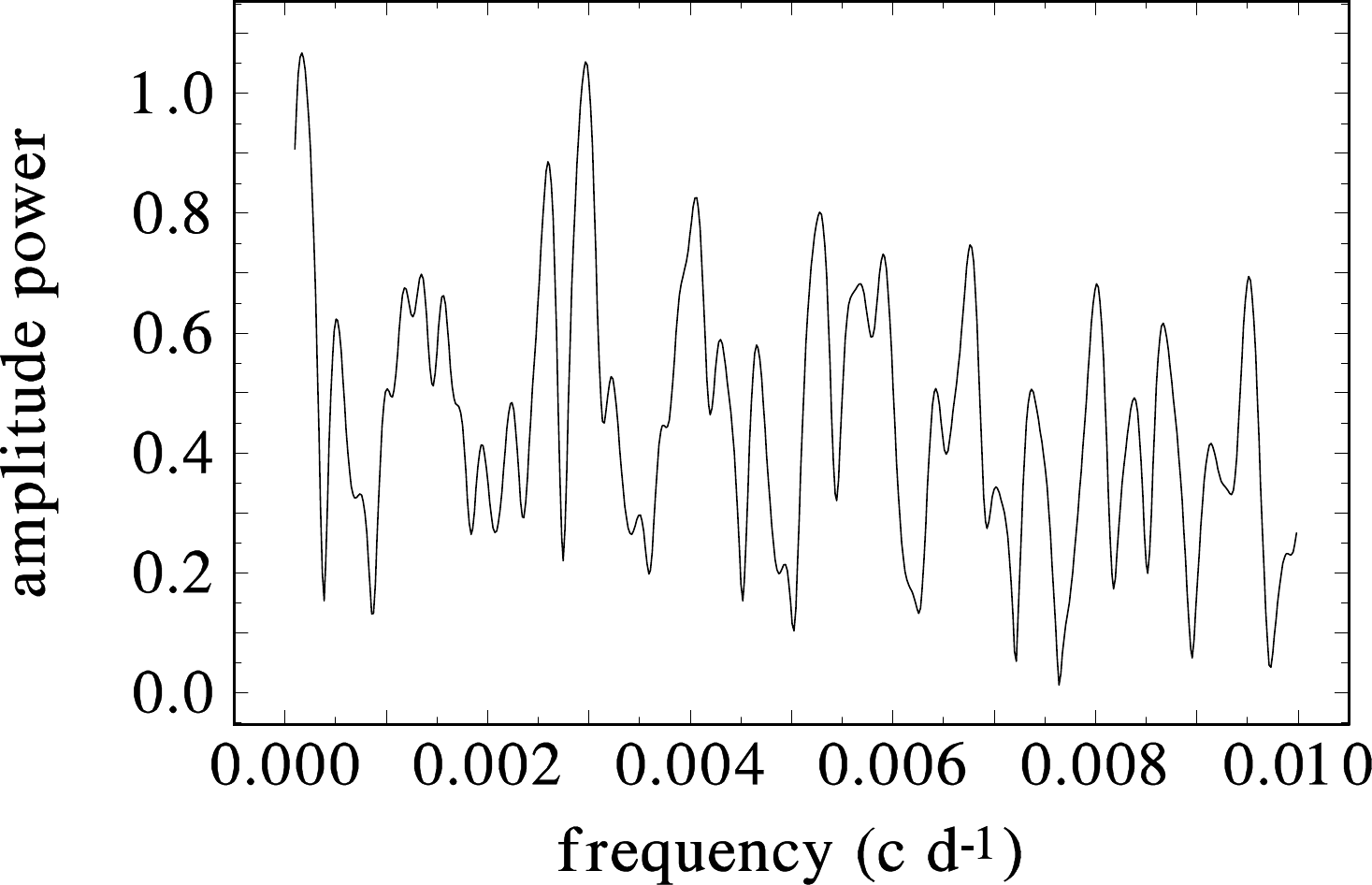}{0.4\textwidth}{Dulaney et al. (2017)}
          \fig{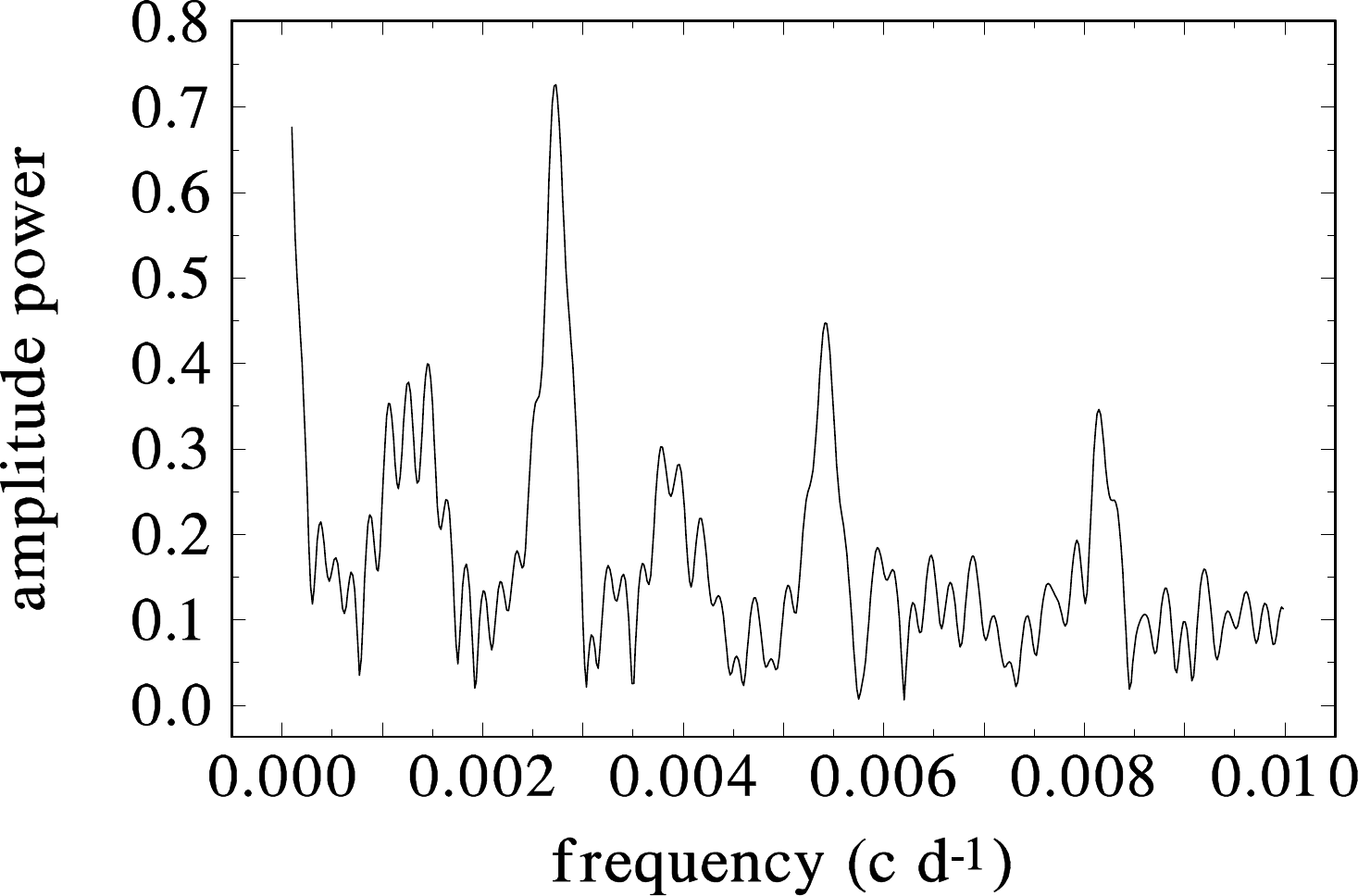}{0.4\textwidth}{spectral window}
}
\gridline{\fig{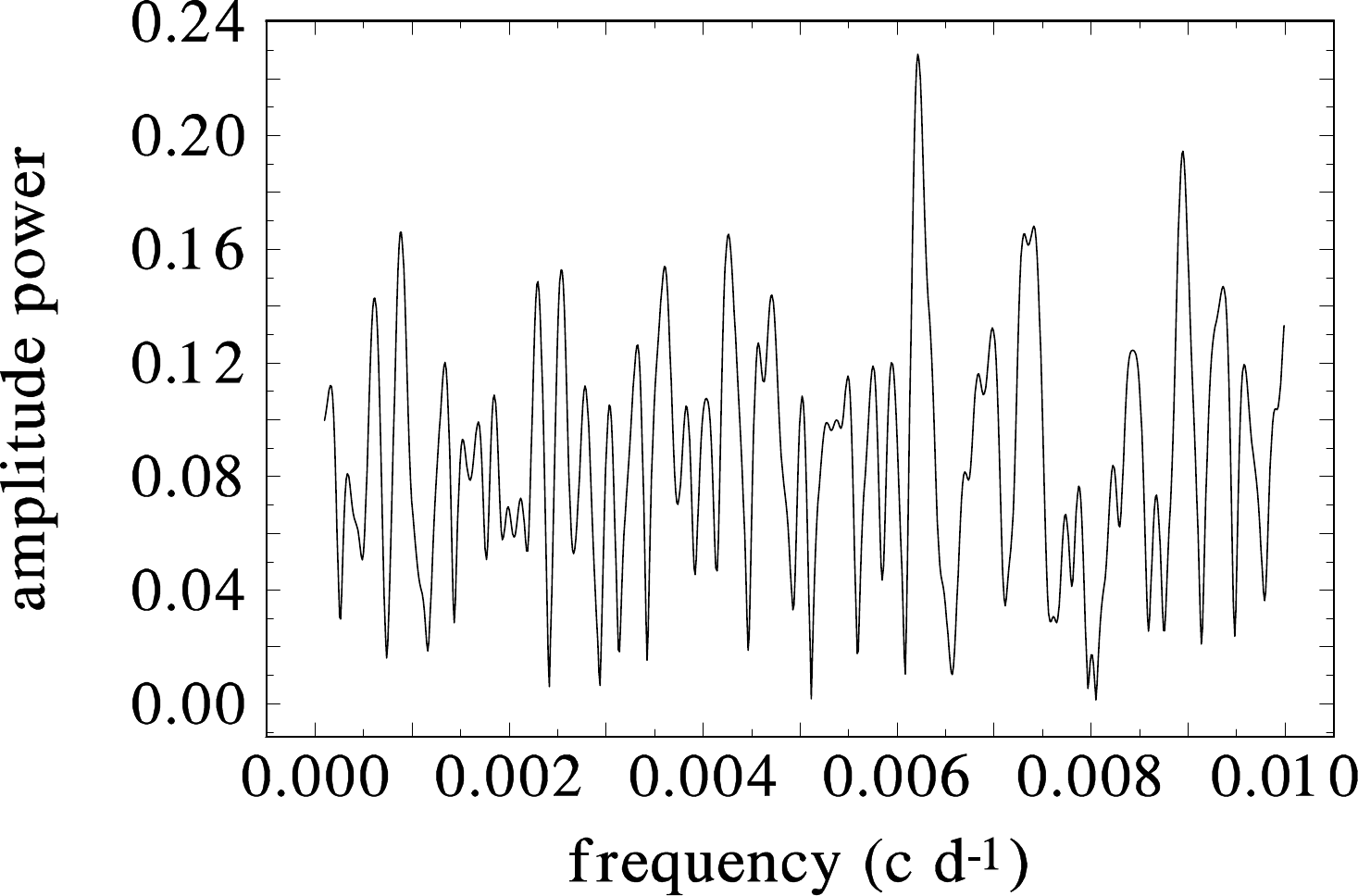}{0.4\textwidth}{Our \ha emission RVs without Reticon:\\ original spectra}
          \fig{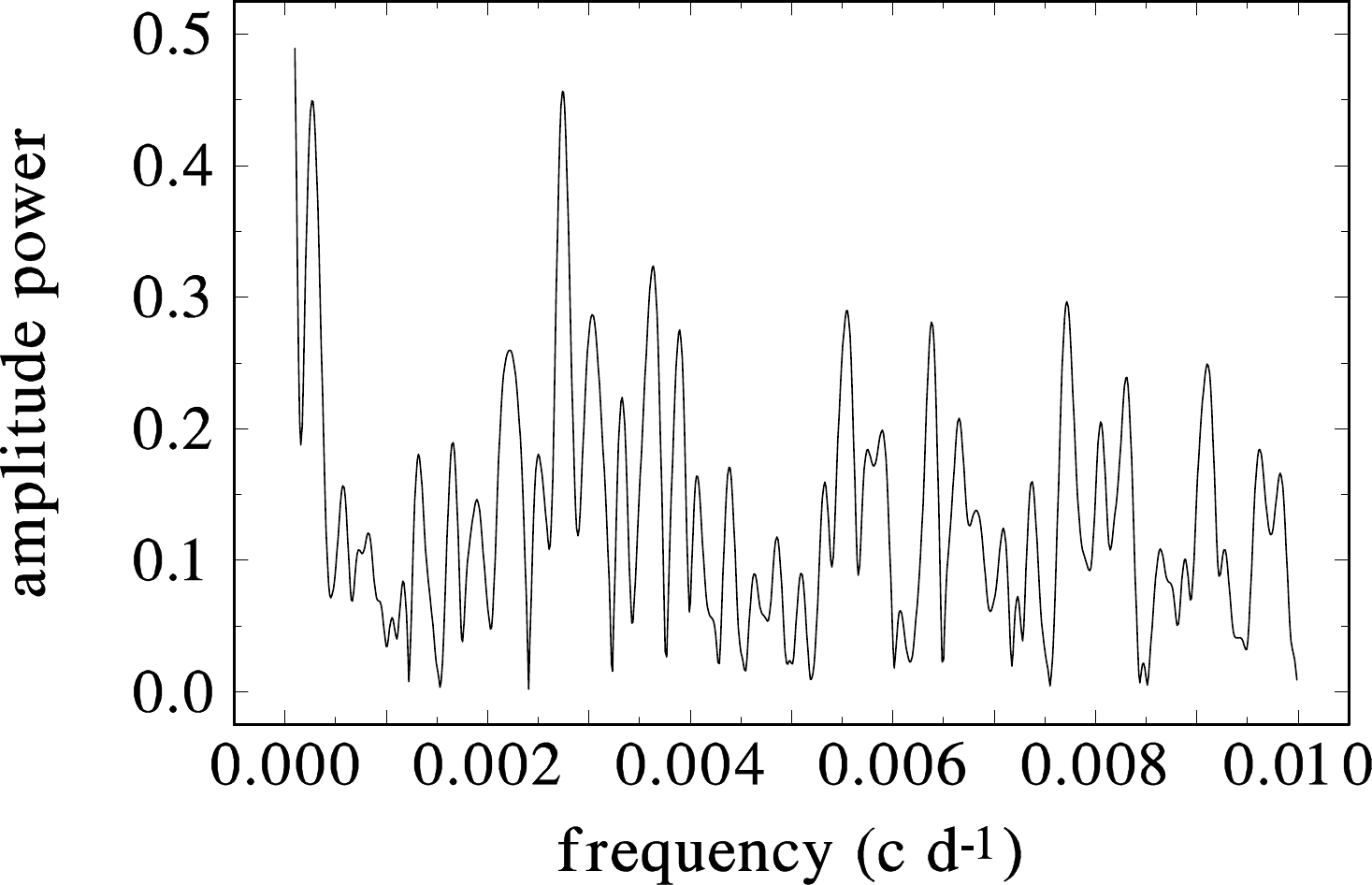}{0.4\textwidth}{spectral window}
}
\gridline{\fig{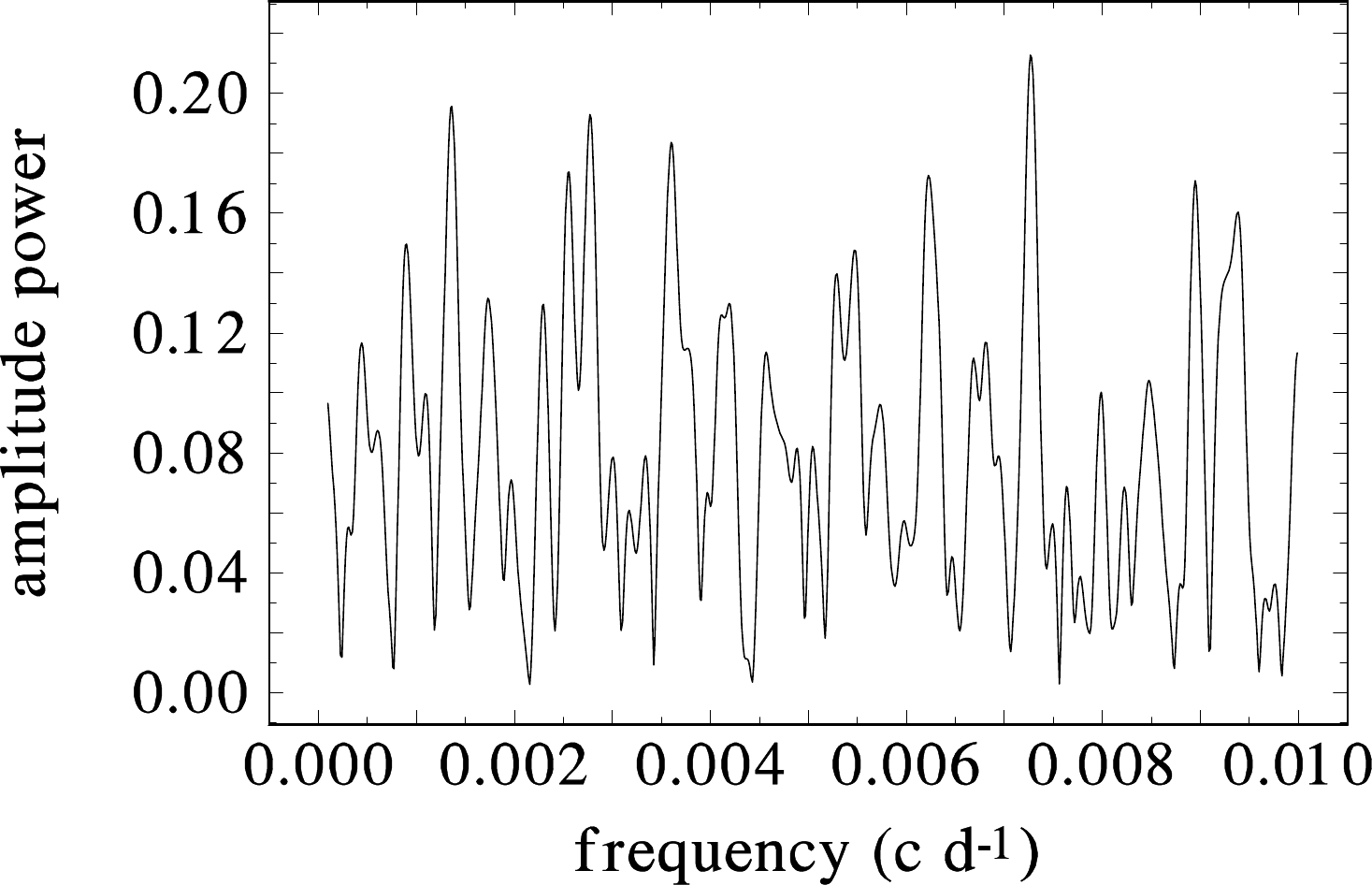}{0.4\textwidth}{Our \ha emission RVs without Reticon:\\
          spectra after telluric-line removal}
          \fig{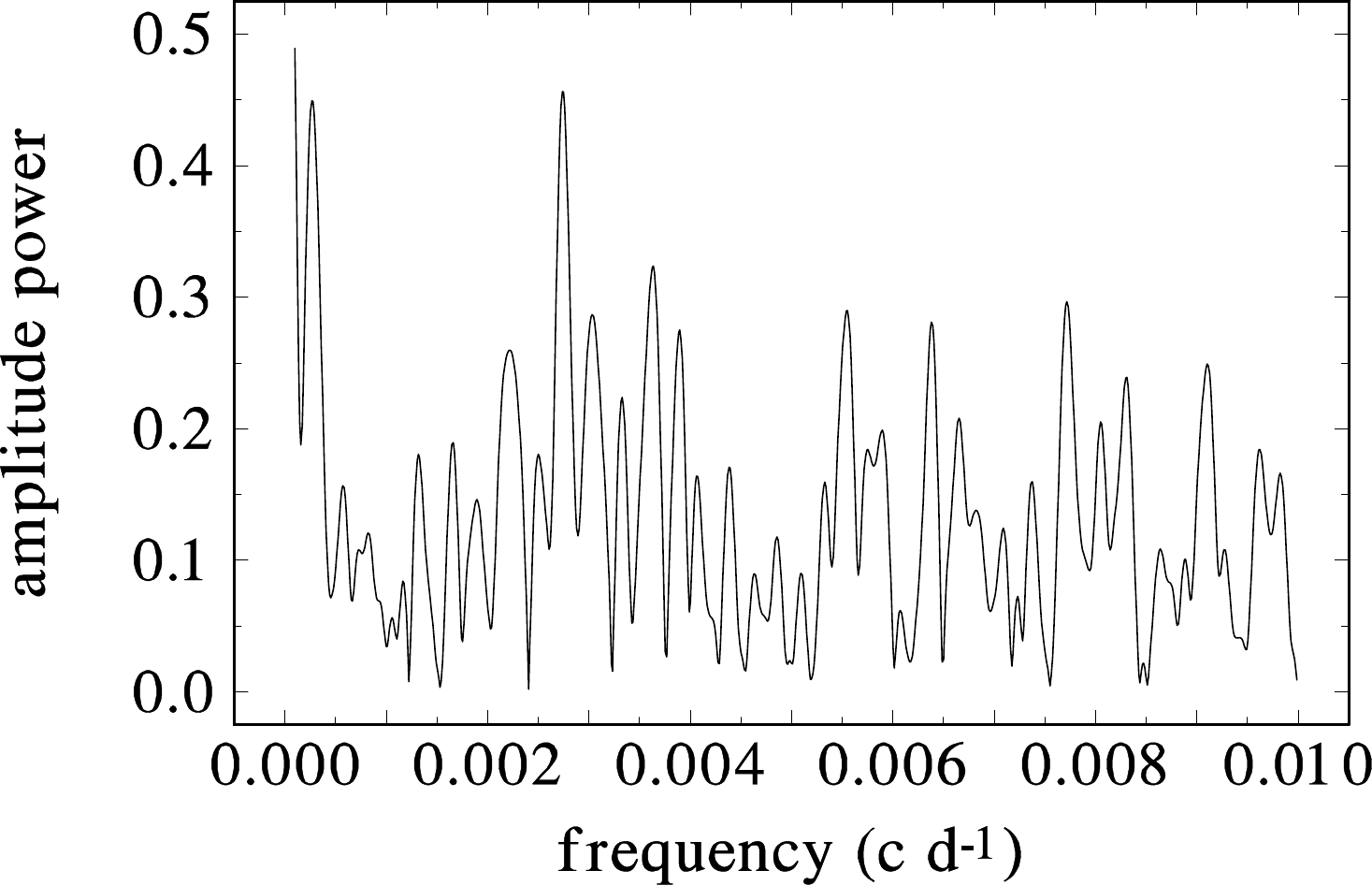}{0.4\textwidth}{spectral window}
}
\caption{Amplitude periodograms of several RV data sets based on the
\citet{deem75} discrete Fourier transform method are shown in the left
panels and the corresponding spectral windows in the right panels.\label{deem}}
\end{figure*}

\subsection{Period analyses of RVs}\label{peranal}
Using the period search based on \citet{deem75} discrete Fourier transform,
we analyzed the photographic RVs published by \citet{frost26} and
\citet{jarad89} (omitting two Yerkes RVs denoted as poor), RVs derived by
\citet{dul2017}, and our \spefo \ha emission-line RVs for periodicity
over possible binary periods from 100 to 10000~d. All correponding
periodograms and window functions are in Fig.~\ref{deem}.
We discuss these three different RV sets separately.

\subsubsection{Photographic RVs}\label{phgrv}
The message from older photographic spectra is somewhat unclear and
confusing. RVs from the seven Yerkes spectra published by \citet{frost26}
range from $-1$ to $+53$ \ks, which is too much to be explained by
measuring uncertainties only. Moreover, their robust mean is $+28\pm8$~\ks,
substantially higher than $\sim9$~\kms invariably found from the more
recent spectra. Similarly, since \citet{jarad89} correctly derived the 133~d
orbital period for another Be star, $\zeta$~Tau, with a~lower semiamplitude
than what they found for \be, it would not be easy to dismiss their result for
\bc as an~artefact of medium-dispersion photographic spectra. We note, however,
that the full range of their RVs, 62~\ks, is comparable to that of the Yerkes
RVs. One is, therefore, led to the suspicion that there were real RV variations
of \bc in the past. As a check, we measured RVs of several stronger lines
in the blue parts of the spectra at our disposal. We found that they do not
indicate any similar large RV changes above some $2-3$~\ks, all RVs clustered
near $9-10$~\ks. (We have not analyzed our blue RVs in more detail since there
is no possibility to check the small possible RV zero-point shifts in
regions without telluric lines.)

Not surprisingly, the period analysis of the combination of \citet{frost26}
and \citet{jarad89} RVs, which are substantially separated in time, resulted
in a number of one-year aliases. Formally the best periodicity is found for
a~period near 190~d.

To investigate these variations in a more quantitative way, we formally
used the program \fotel for binary orbital solutions and allowed different
systemic velocities for the two data sets. RVs denoted as good in
\citet{frost26} were given weight 0.5, and those denoted fair weight 0.3,
while the weights of all \citet{jarad89} RVs were set to 1.0.
The following elements led to a fit with lowest rms of 4.65 \kms
for \citet{frost26} RVs, and 6.91 \kms for the \citet{jarad89} RVs:

\centerline{$P=189\fd917\pm0\fd017$,}
\centerline{$T_{\rm periast.}= {\rm RJD}~31989.9\pm2.8$,}
\centerline{$e=0.54\pm0.11$, $\omega=302^\circ\pm14^\circ$,}
\centerline{$K=24.4\pm3.6$ \ks,}
\centerline{$\gamma_{\rm Frost\, et\, al.}=27.9\pm2.4$ \ks,}
\centerline{$\gamma_{\rm Jarad\, et\, al.}=11.2\pm1.5$ \ks.}

\begin{figure}
\plotone{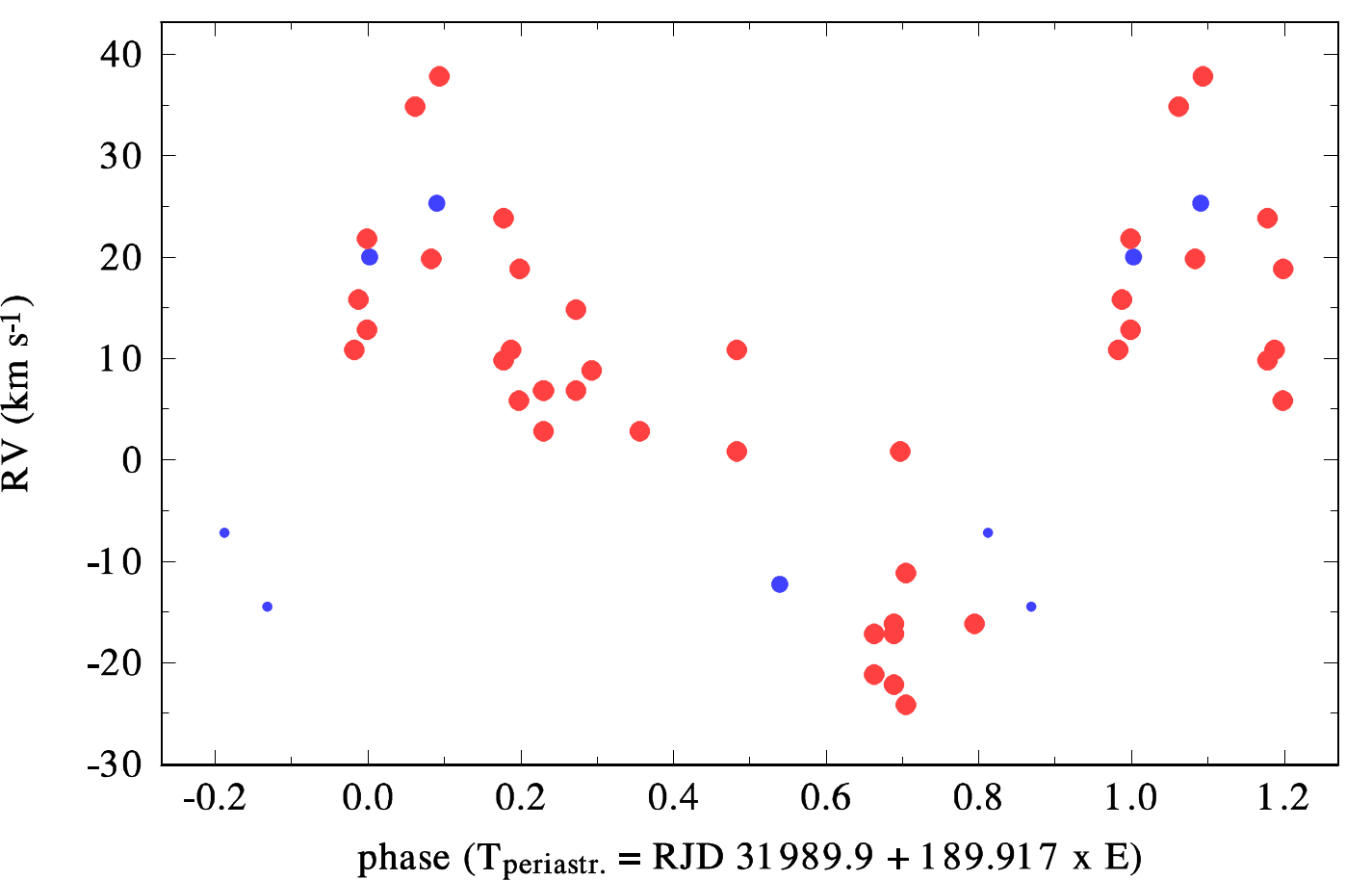}
\caption{Plot of RVs published by \citet{frost26} (blue circles)
and \citet{jarad89} (red circles) vs. phase of the 189\fd917
period. \citet{frost26} RV denoted as ``fair" are shown by
smaller symbols.}
\label{rvold}
\end{figure}

The RV curve corresponding to this fit is plotted in Fig.~\ref{rvold}.

\begin{figure}
\plotone{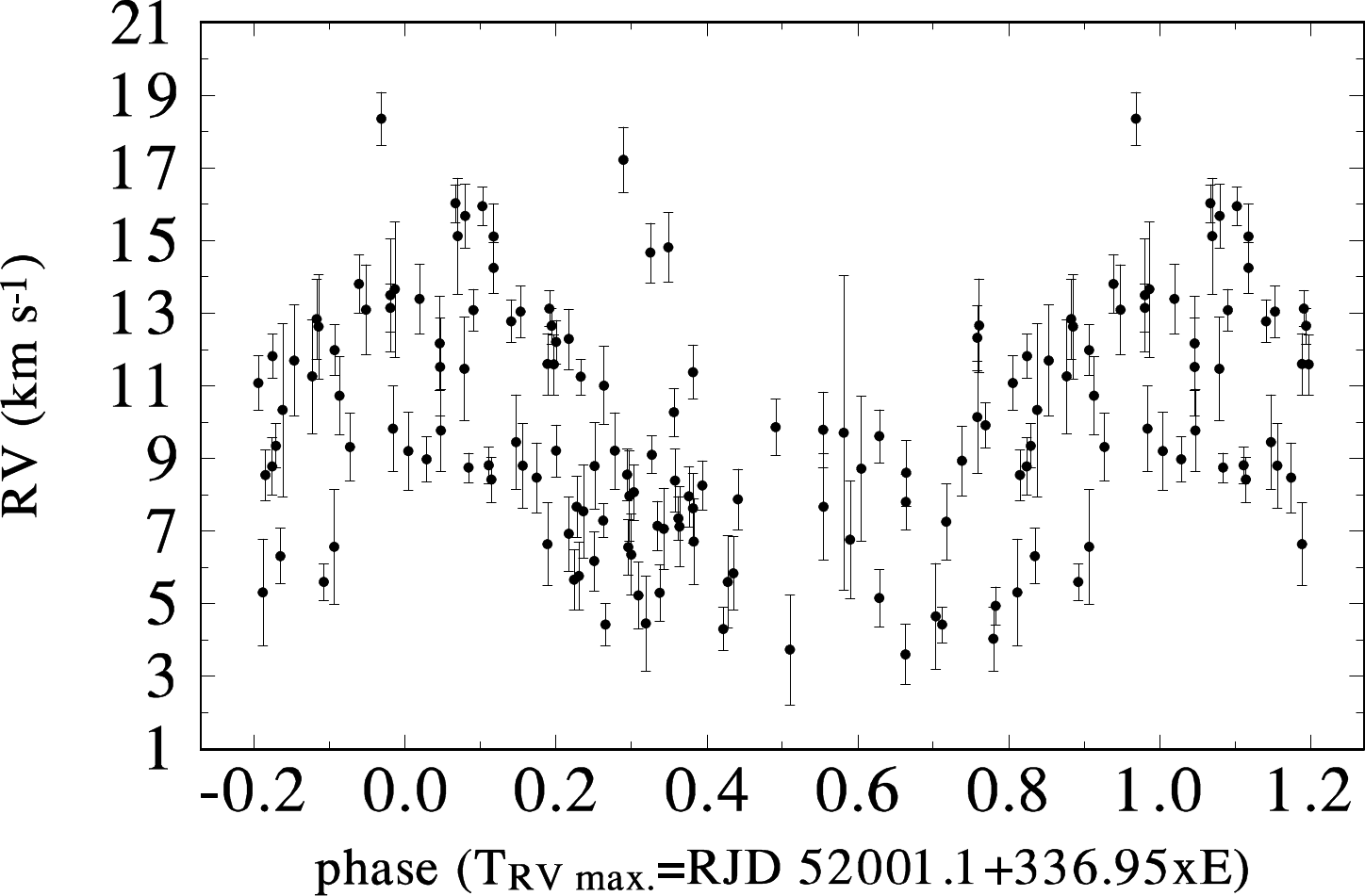}
\plotone{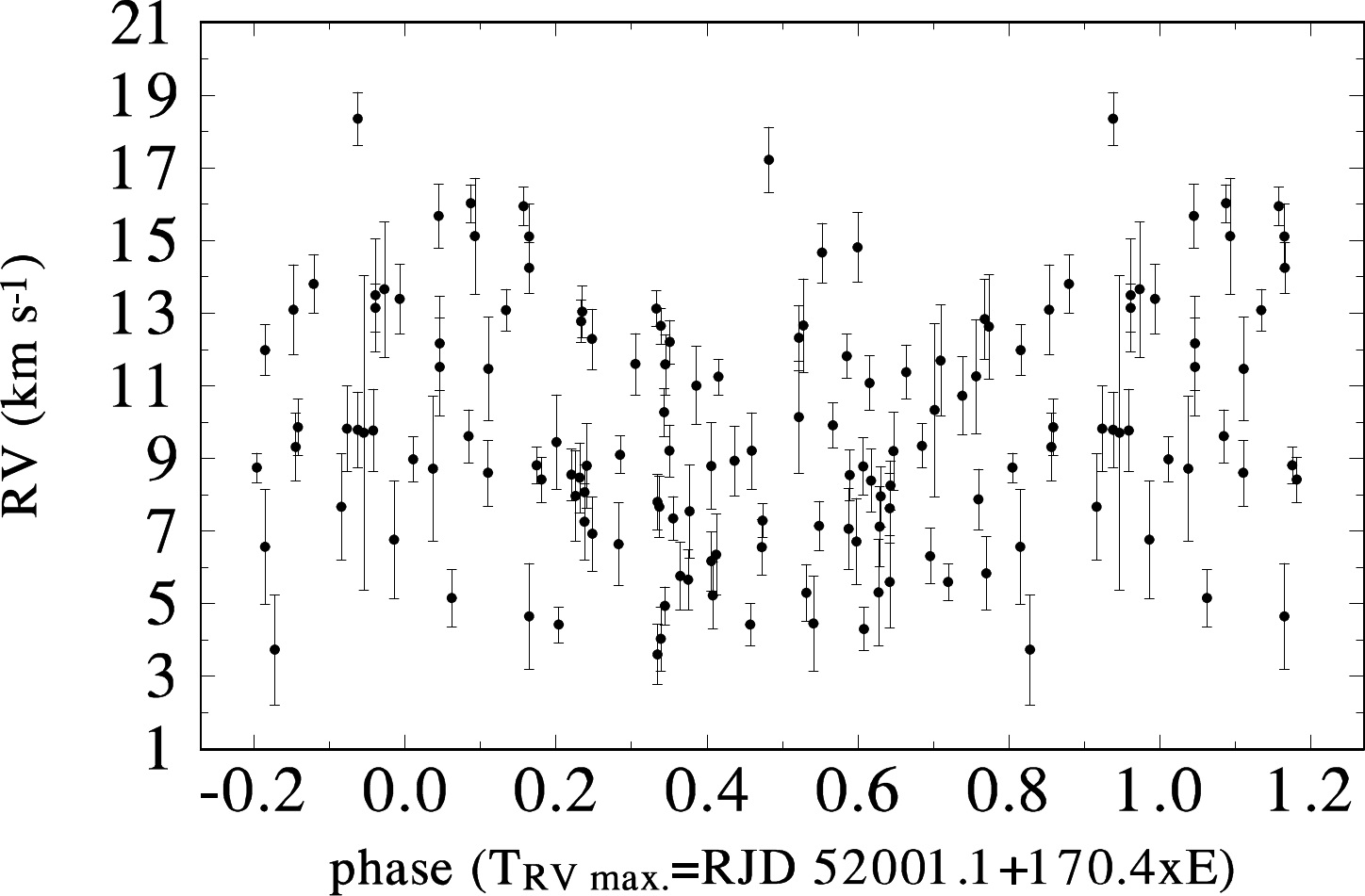}
\caption{A plot of \citet{dul2017} \ha emission RVs vs. phase of
the 336\fd95 period (top) and for their original period of 170\fd4 (bottom).
The same epoch of maximum RV, RJD~52001.2, was adopted as phase zero in
both plots.}
\label{dul337}
\end{figure}

\subsubsection{\citet{dul2017} emission-wing RVs}
The periodogram of \citet{dul2017} RVs indicate the highest peaks at
frequencies of 0.00016869~\cd, and 0.0029678~\cd. The latter corresponds
to a period of 336\fd95, i.e. roughly twice the period reported by them.
Moreover, we note that the difference of both frequencies corresponds
to the frequency of one year. The phase plot for the 336\fd95 period is
in Figure~\ref{dul337}. This phase curve looks certainly less
scattered than the plot of the same RVs vs. phase of the 170\fd4
period advocated by \citet{dul2017}, but it can simply represent
a one-year alias of the period of one year over the interval
covered by their data.

\begin{figure*}
\gridline{\fig{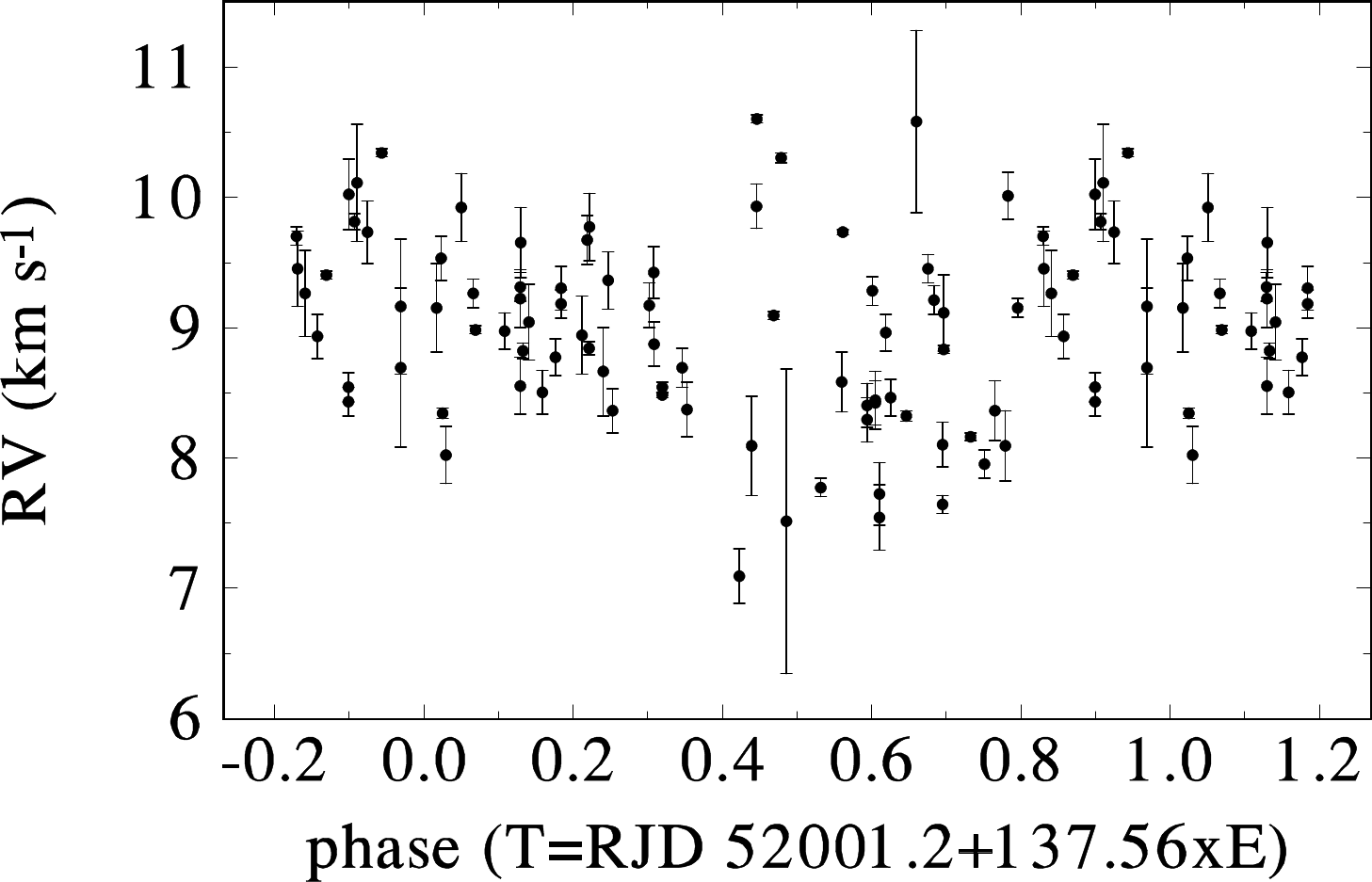}{0.46\textwidth}{original spectra}
          \fig{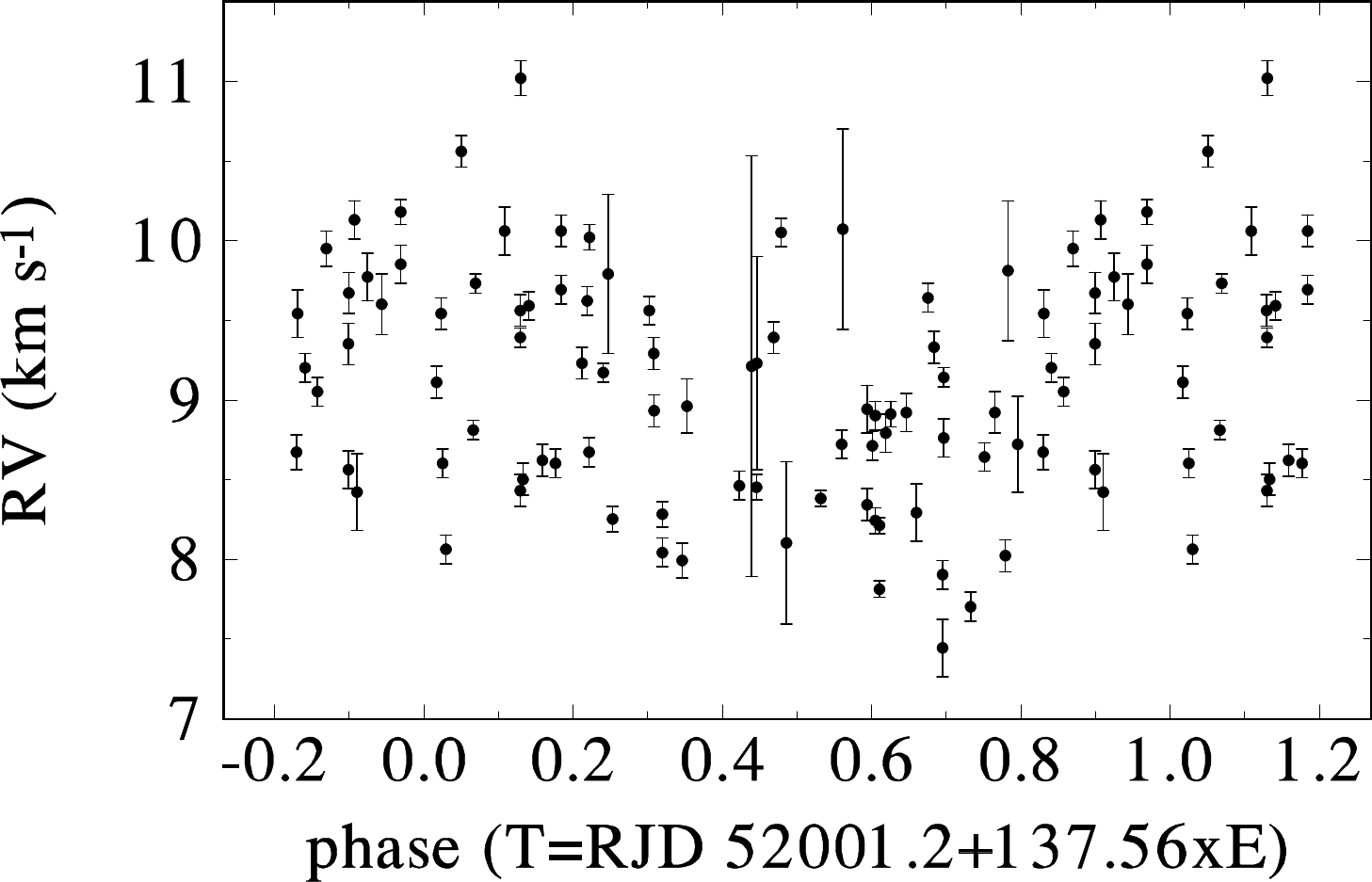}{0.46\textwidth}{telluric-corrected spectra}
}
\gridline{\fig{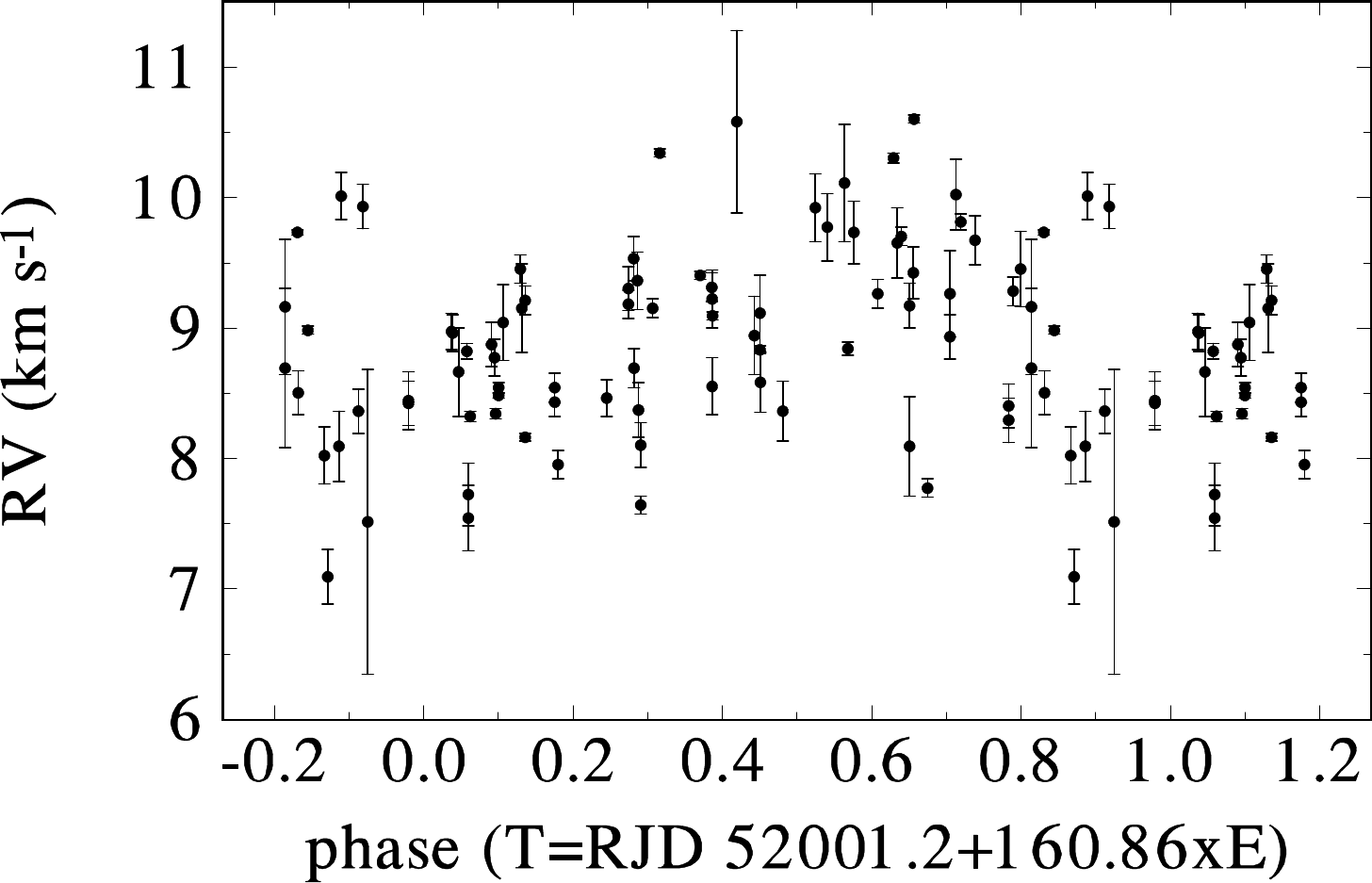}{0.46\textwidth}{original spectra}
          \fig{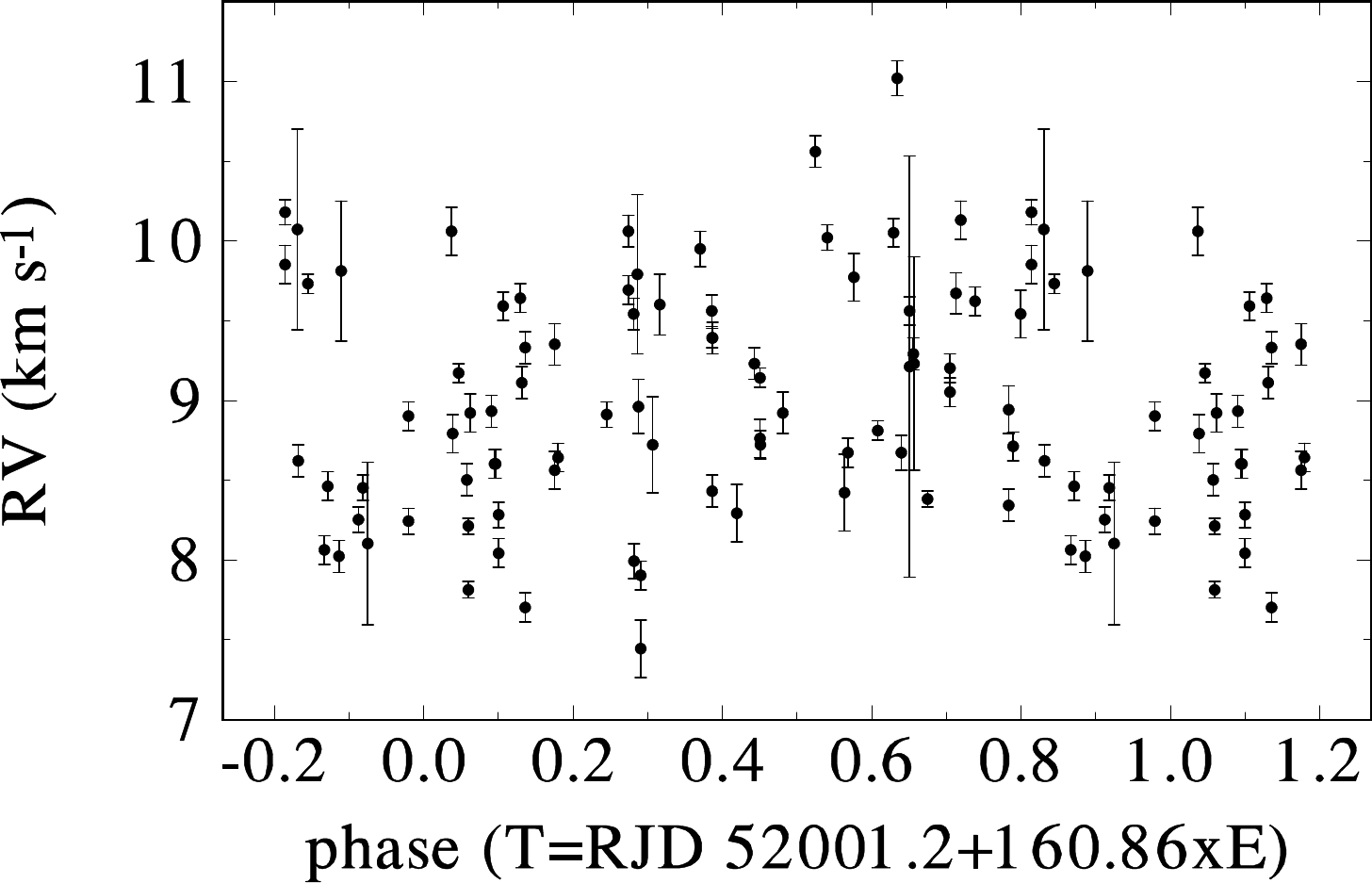}{0.46\textwidth}{telluric-corrected spectra}
}
\caption{Phase plots of our \ha emission-line RVs, both from the
original spectra and from the spectra after removal of telluric lines,
for the ``period" of 160\fd86 found as the best one from the original
spectra, and for 137\fd56 found from the spectra after the telluric-line
removal.}
\label{phasenew}
\end{figure*}

\subsubsection{Our \ha emission RVs}
We derived the periodograms for both, the RVs from the original spectra
and from the spectra after the removal of telluric lines (see Fig.~\ref{deem}).
It is seen that different best-fit frequencies of 0.0062166~\cd,
and 0.0072695~\cd were detected for the original, and telluric-corrected
spectra, respectively. These correspond to ``periods" of 160\fd86, and
137\fd56. The phase plots for both these ``periods" and both sets of RVs
are in Fig.~\ref{phasenew}. One can see that neither phase curve
is very convincing. Moreover, it is obvious that no strong peak was
detected near frequency of 0.00587~\cd, which corresponds to the period
of 170\fd4 reported by \cite{dul2017} -- cf. Fig.~\ref{deem}.
We thus re-iterate our conclusion that our RVs do not indicate RV
variability.

\section{Could the presence of telluric lines confuse
the automated RV measurements?}
RVs are often measured using various automatic routines
or pipelines. As already mentioned, there is a~large number of telluric lines
in the \ha region -- see Fig.~\ref{haspec}.
If these lines are not removed properly, they could possibly affect or even confuse
the automatic procedures that measure the RVs. In the course of the year the
positions of the telluric lines with respect of the line profile change periodically.
If the telluric lines are not removed correctly, they could easily
cause apparent periodic RV changes.

To investigate how significant the line blending of numerous telluric
lines with the observed \ha profile can be, we carried out tests on
artificially created data for the observing times of the Ritter spectra.

\begin{figure}
\plotone{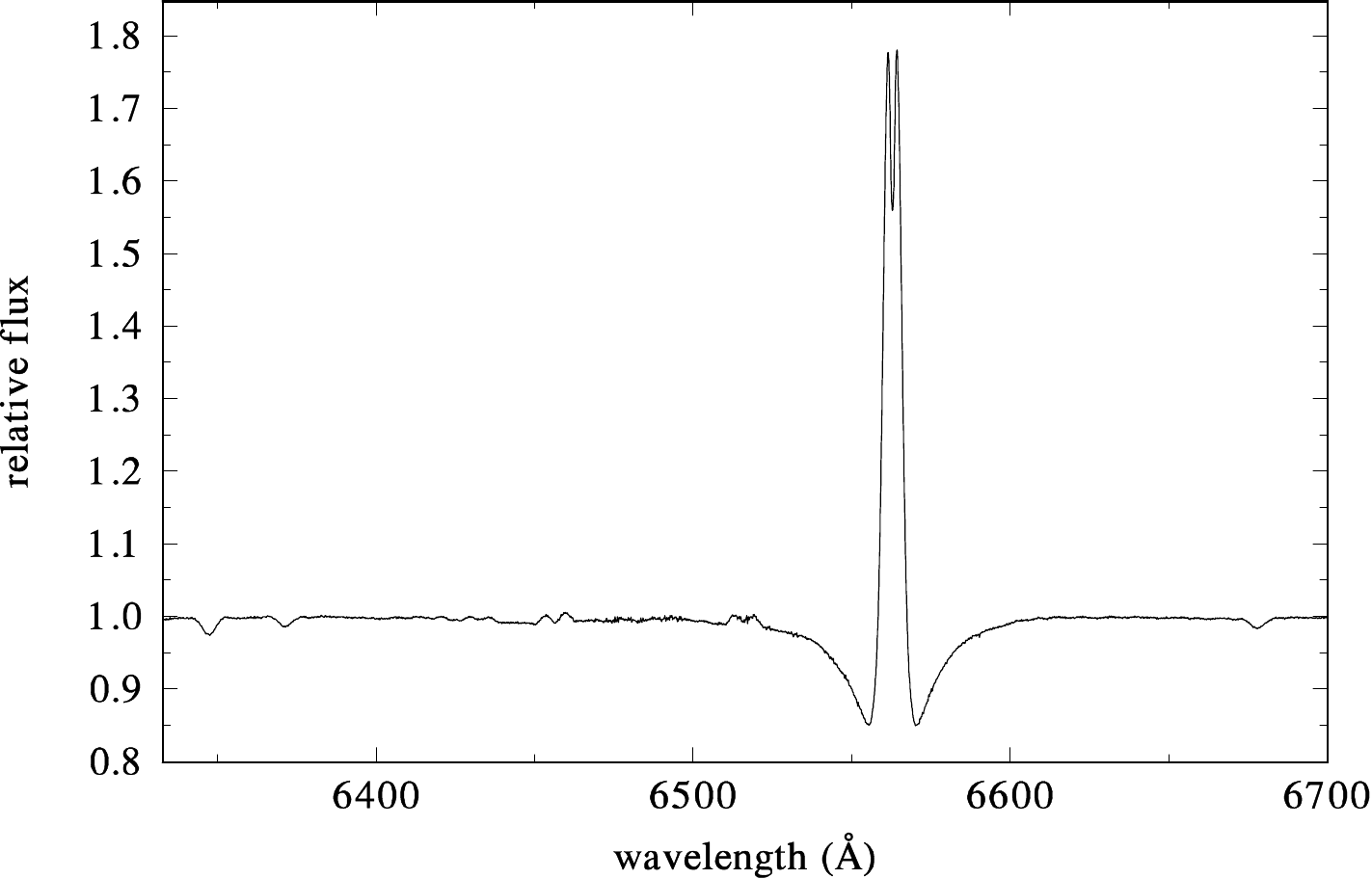}
\plotone{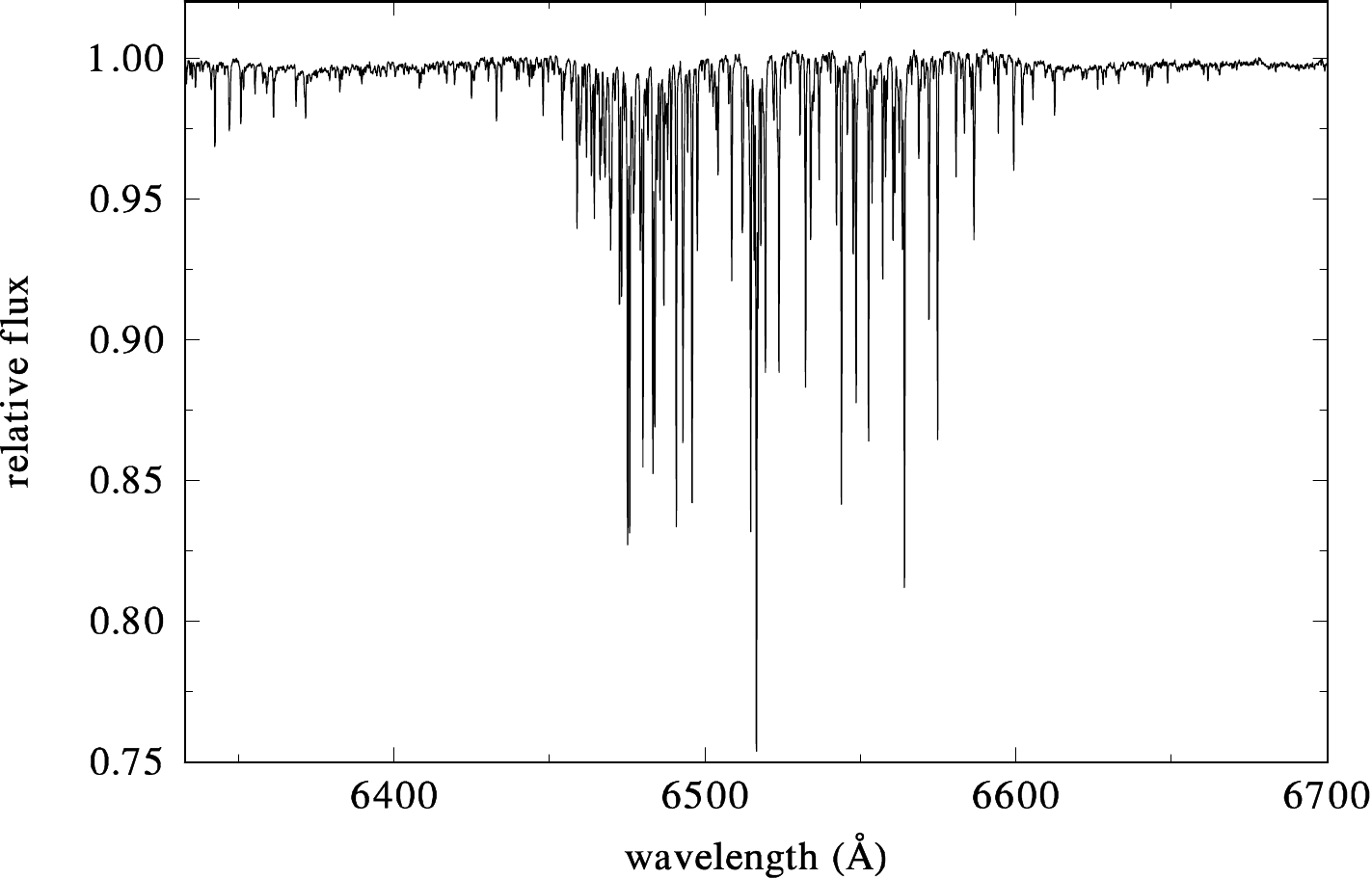}
\caption{Top: The mean disentangled DAO red spectrum of \be, with several
clearly visible spectral lines: absorption lines of the
\ion{Si}{2}{\,\,6347\&6371}~\ANG\ doublet, double emission lines of
\ion{Fe}{2}{\,\,6456} and \ion{Fe}{2}{\,\,6519\&6525}~\ANG\ blend,
\ha emission, and \ion{He}{1}{\,\,6678}~\ANG\ absorption.
Bottom: The mean disentangled spectrum of telluric lines.
}
\label{korel-dao}
\end{figure}

We realized that the spectral resolution of the DAO spectra is similar to
that of the Ritter spectra studied by \cite{dul2017}. We therefore used this
richest and longest homogeneous series, covering the red spectral region,
and applied the \korel disentangling
\citep[][and references therein]{korel3} to them. We assumed a constant
spectrum of \bc (simulating it as a very long-period orbit with virtually
zero semi-amplitude of RV changes) and we also disentangled the spectrum
of telluric lines, allowing for their time variability. The resulted
spectra are shown in Fig.~\ref{korel-dao}.

Then we multiplied the mean DAO stellar spectrum
in the neighbourhood of \ha by the mean spectrum of telluric lines, each
time shifted in RV to the respective values of the heliocentric RV corrections.
The multiplication is a better justified operation because the telluric spectral
lines do not have a continuum, thus the strength of the telluric line which
absorbs at stellar continuum will be different from the strength of the same
line absorbing in the strong stellar line (see the discussion on this topic
in \citealt{hechrud2005} and \citealt{had2006}).
We omitted the spectrum with RJD~52975.8496 with tabulated \ha emission RV
of $-1.88$~\kms in the electronic table of \citet{dul2017},
since this RJD is out of the increasing order of RJDs and the same
value repeats a few lines later with a RV of $-3.41$~\ks. Then we
imported all artificial spectra to \spefo and measured their
$V/R$ ratio. We also measured RVs on the steep wings of the \ha line
using an automatic procedure, namely a~simplified double-slit method,
commonly used in solar physics. This method utilises two artificial
slits placed in a fixed distance in wavelength on the steep wings of the
spectral line. The double slit is then moved across the line profile
to balance the transmission through both slits. The position of the double
slit then corresponds to the RV of the spectral line. The method is known
to be performing well on symmetric lines, which roughly corresponds
to our case.

 In other words, we used an~\ha profile with constant RV and $V/R$ ratio.
We found that the $V/R$ ratio of the disentangled DAO \ha profile differs
a bit from one, having $V/R=0.99792$. To test the hypothesis that the real
\bc profile has $V/R=1.0$ sharply, we multiplied all $V/R$ ratios measured
on the simulated spectra by an inverse factor of 1.00208.

 We underline that we are aware that such simulation cannot reproduce
the observed spectra well enough. Not only that our simulated spectra are
essentially noiseless in contrast to the observed ones but the strength
of the telluric lines in the real spectra certainly varies from one to
another, while our simulation used a mean telluric spectrum of constant
intensity. Yet, our results are very illuminating and clearly show that
the blending of the \ha profile with telluric lines, which are changing
their positions with time during the year lead to apparent RV and $V/R$
changes, especially when the peak intensities of the emission
line are modest, not exceeding the continuum level by two to three
times.

\begin{figure}
\plotone{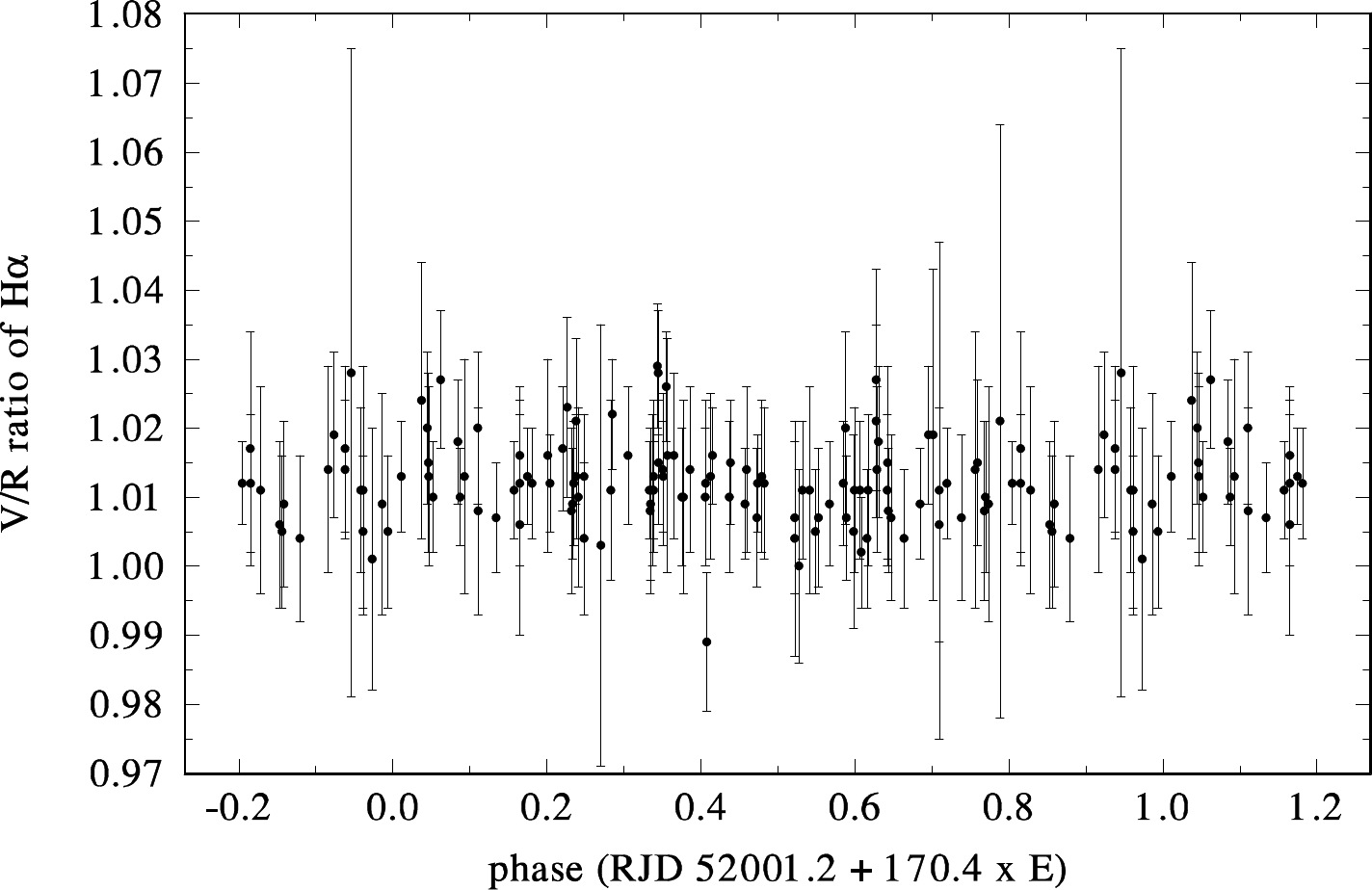}
\plotone{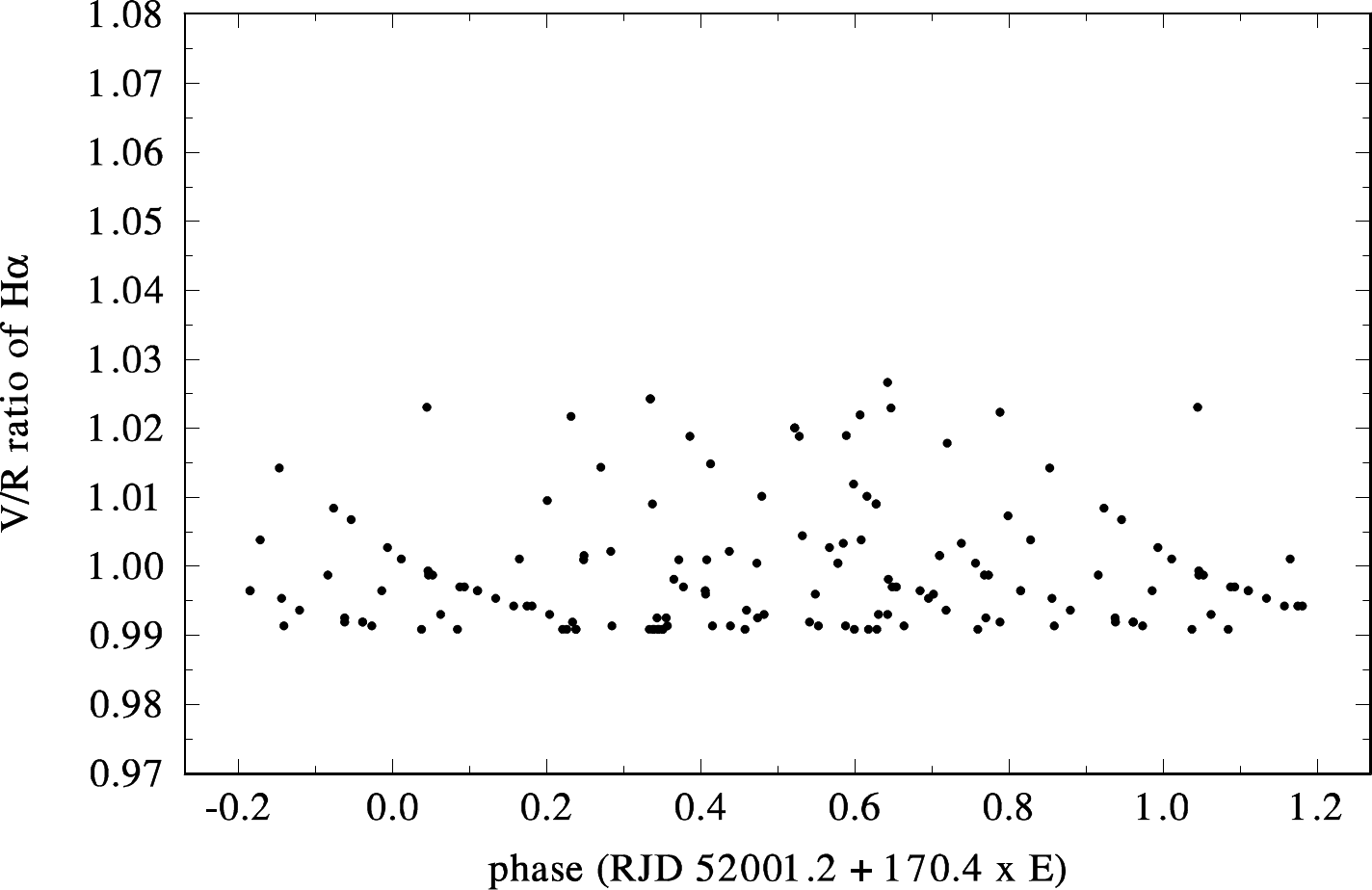}
\caption{Phase plots of the $V/R$ ratio of the double \ha emission for
the ephemeris of \citet{dul2017}
$T_{\rm max.RV}=RJD~52001.2+170\fd4\times E$.
Top: Measurements of \citet{dul2017} on the Ritter spectra;
Bottom: Simulation of Ritter spectra based on combined disentangled DAO
line profile of \ha multiplied by the spectrum of telluric lines
(see the text for details).}
\label{vrsimul}
\end{figure}

 We first discuss the results for the $V/R$ ratio.
We note that the times of observations of the Ritter spectra are such
that the vast majority of them have heliocentric RV corrections from
$-29$ to $+3$~\kms and there are only 17 spectra (out of 123), which
have heliocentric RV corrections larger than +15.
We suspect that one consequence of this fact is that almost all $V/R$
ratios measured by \citet{dul2017} are larger than one. In Fig.~\ref{vrsimul}
we compare the phase plot of the $V/R$ ratios measured by \citet{dul2017}
on real Ritter spectra with that for the artificially created spectra.
We note that our result for artificially created spectra shows
a~remarkable agreement with the real Ritter spectra as far as the range of
the $V/R$ ratios is concerned.
Since our modelling used a constant \ha profile with $V/R$ tuned to one,
this seems to constitute an indirect but quite compelling argument
that the $V/R$ changes derived by \citet{dul2017} are still partly
affected by spurious effects due to blends with the telluric lines.

\begin{figure*}
\gridline{\fig{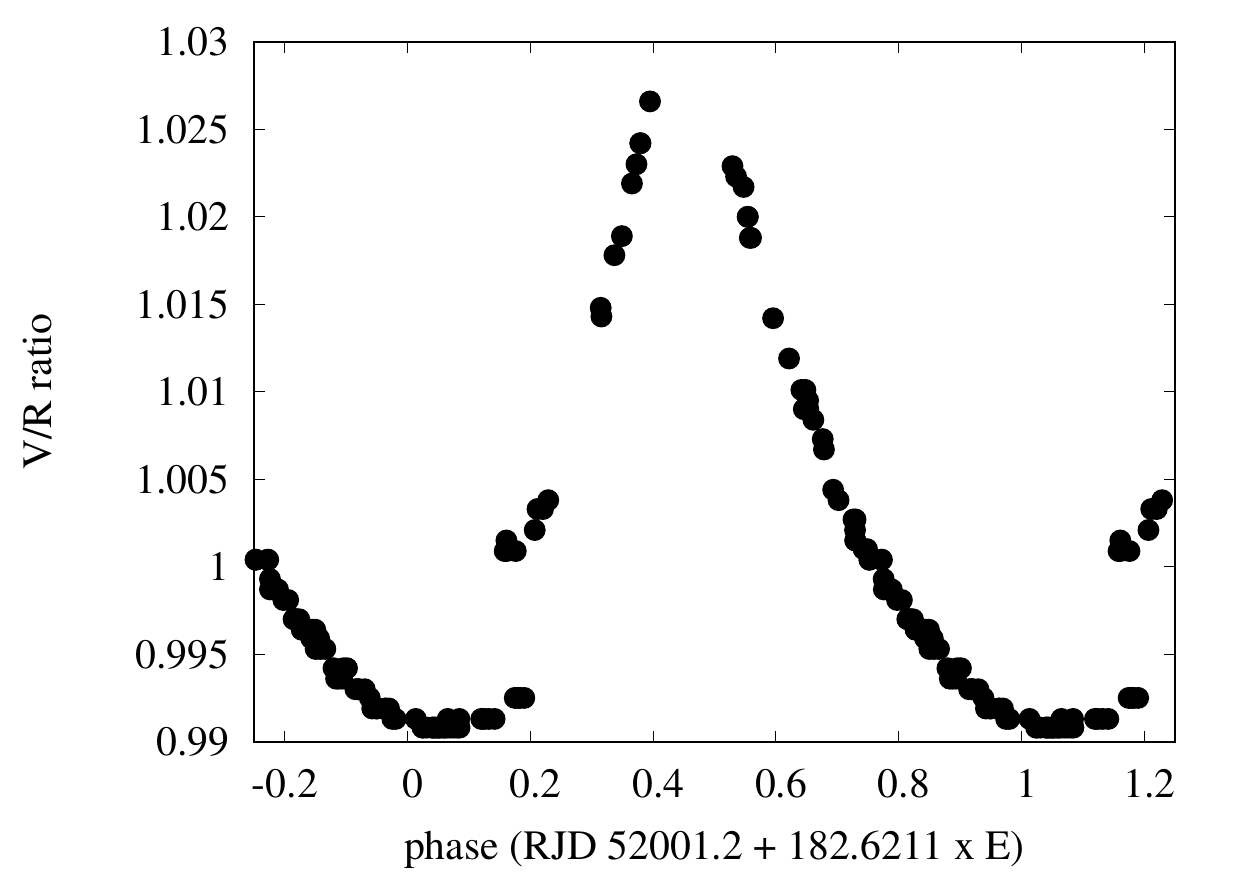}{0.46\textwidth}{(a)}
          \fig{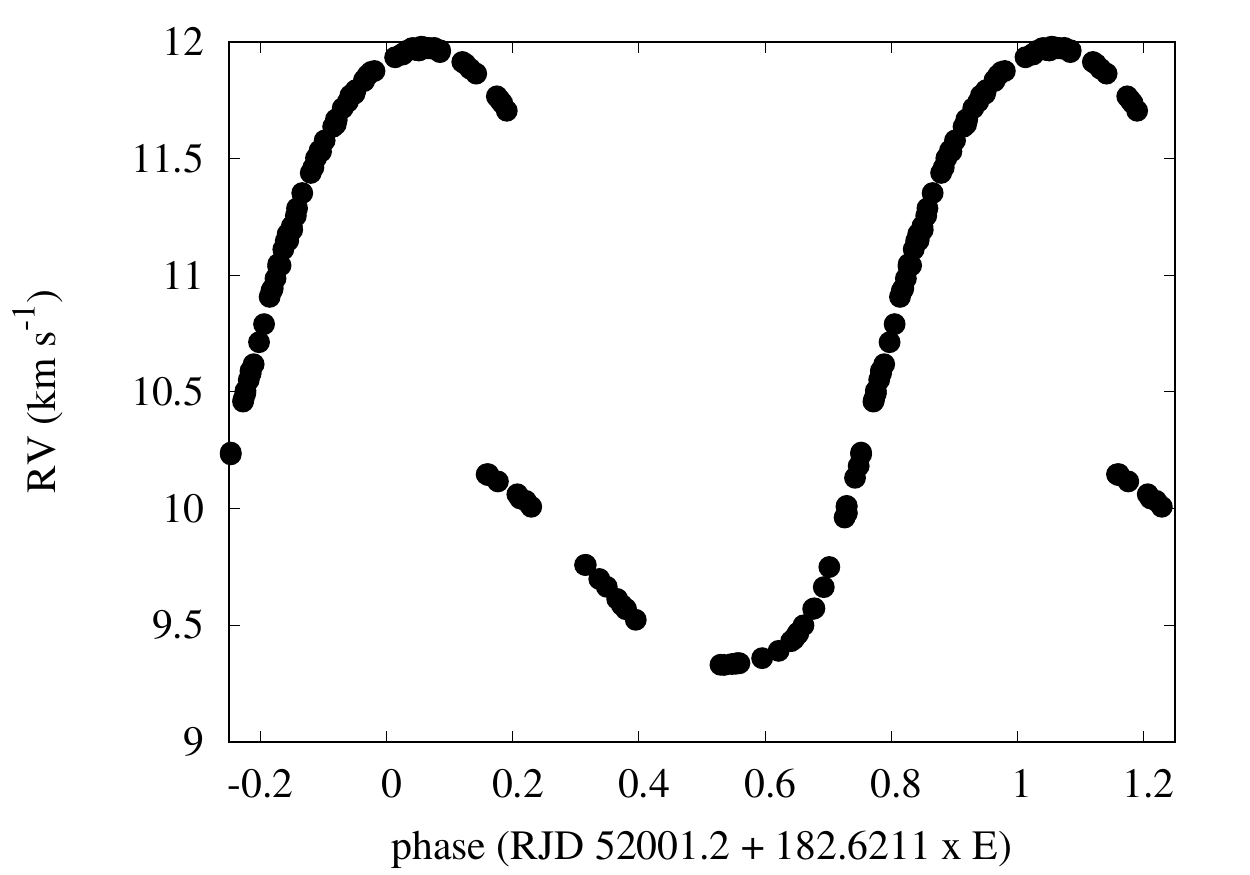}{0.46\textwidth}{(b)}
          }
\gridline{\fig{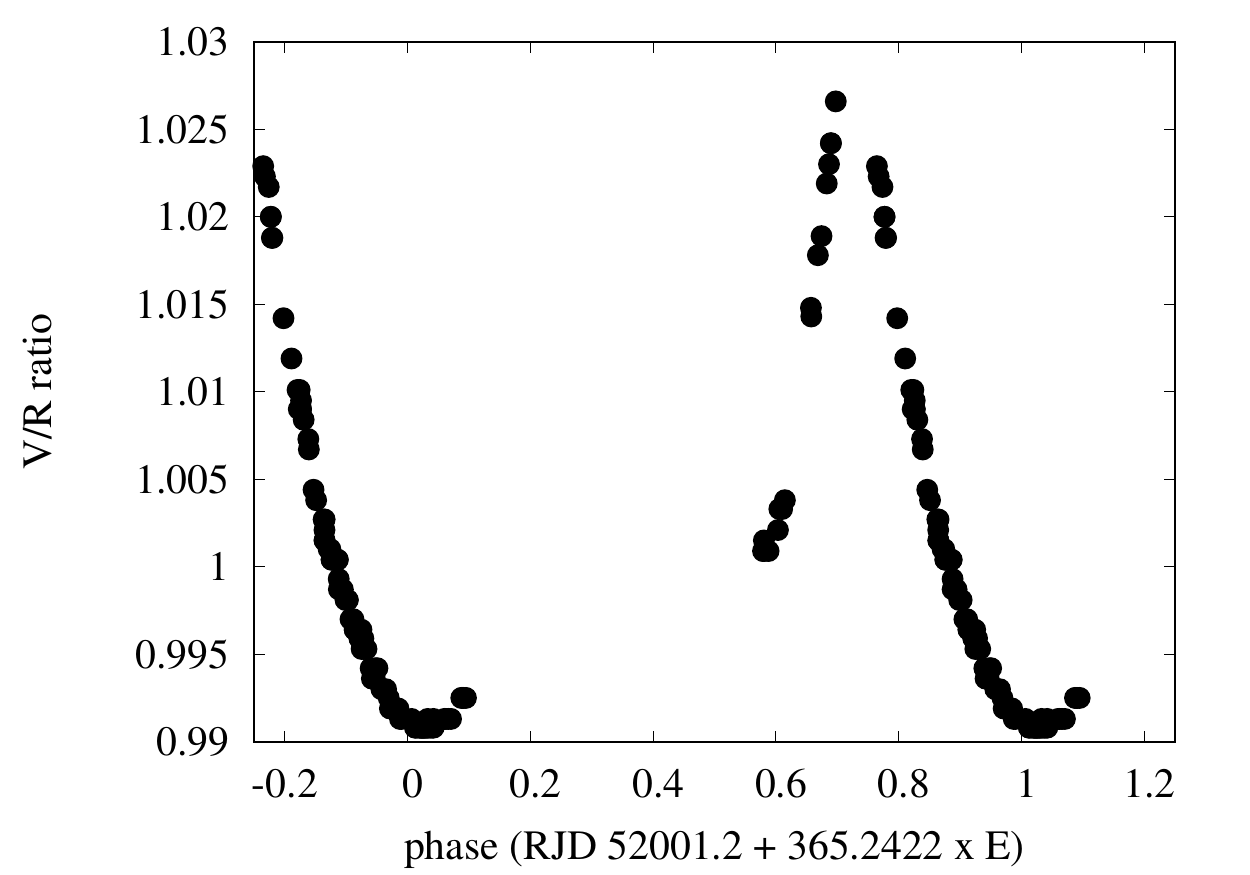}{0.46\textwidth}{(c)}
          \fig{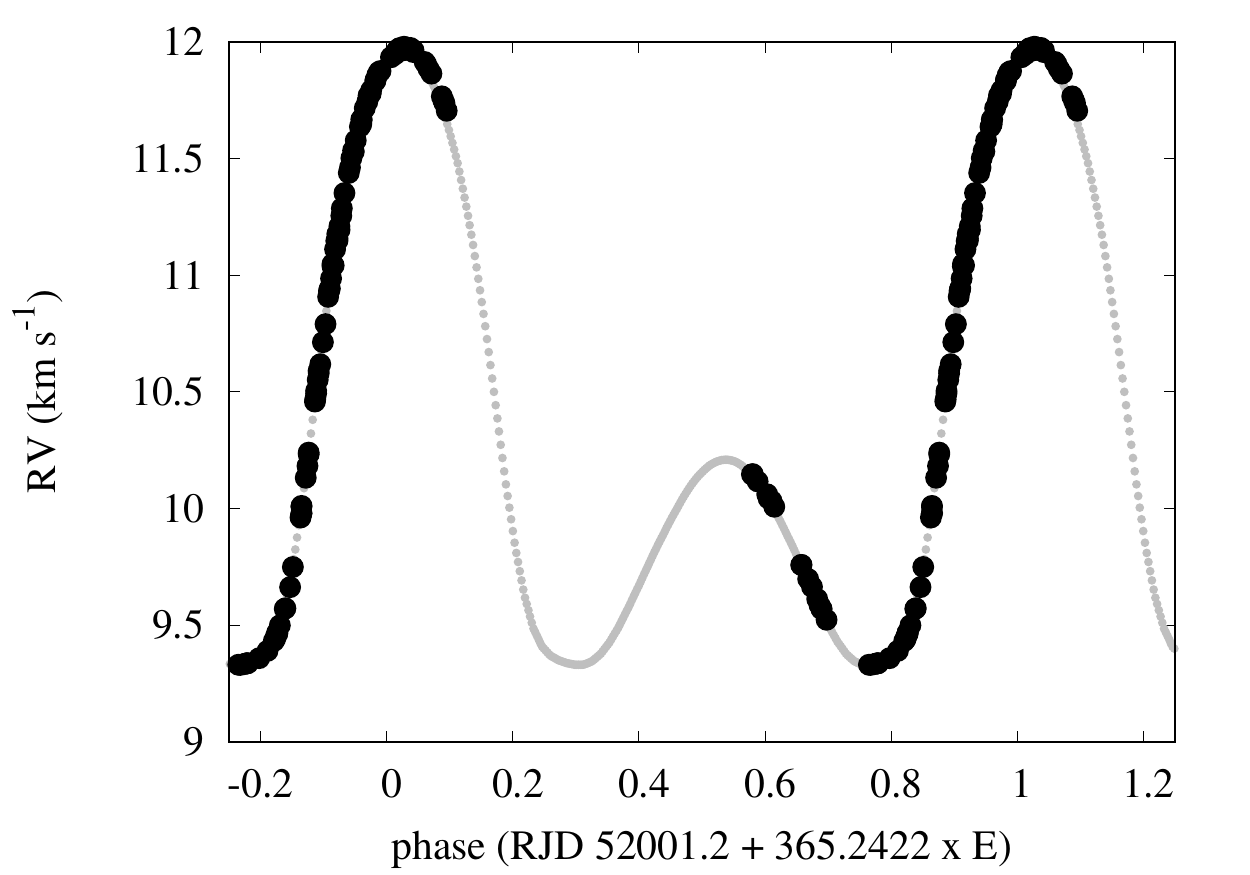}{0.46\textwidth}{(d)}
          }
\caption{Phase plots of the $V/R$ ratio (left columns) and
RV (right columns) of the \ha emission wings measured on the simulated
Ritter spectra. The phase diagrams are plotted for the periods of
half of the tropical year (upper row)
and the period of one tropical year (bottom row). In panel (d), the gray
points were calculated for simulation covering every day over one tropical
year, whereas the black points represent the dates of Ritter spectra.}
\label{simulvrrv}
\end{figure*}

We carried out a period analysis of the $V/R$ ratio to find out that the best
period is one half of the tropical year. The corresponding phase plot
is in Fig.~\ref{simulvrrv}a.

We also carried out period analyses of RVs of the \ha emission
wings measured by the automatic procedure. The best fit was also obtained
for one half of the tropical year -- see Fig.~\ref{vrsimul}b. We note that
the phase curves of RVs measured by the automatic method we used
are very smooth. We explain this by the fact that the simulated spectra are noiseless
and the slit method inevitably feels the variable blending on both line
wings. We note that the RV curve shows a strange behavior in phases from 0.1 to 0.3\,.
This corresponds to a large discontinuity at the same phase seen in phase curve of RVs
in Fig.~\ref{simulvrrv}b,
which is caused by the gap in observations in autumn, when the star was below the horizon
at night (this gap is demonstrated in Fig.~\ref{simulvrrv}d).

A related question is why does the periodic analysis prefer the period of one
half of the tropical year to the full tropical year, which should naturally
be present in the simulated spectra since it was introduced by our simulation.
As we demonstrate in Fig.~\ref{simulvrrv}d, the simulated
spectra indeed have one-tropical-year periodicity with two half-year waves
of different amplitudes (the gray points that were computed for automatic measurements
covering one tropical year). In all probability, it is a consequence of asymmetric
distribution of stronger telluric lines with respect to the laboratory
wavelength of the \ha line. As seen in Fig.~\ref{simulvrrv}d, the Ritter
spectra were almost exclusively exposed in such dates that their
positions cover preferrably the wave with a larger amplitude. A period
search seeking for a single-wave RV curve then naturally gets confused
due to this selection effect and finds a half-tropical-year
period as the best one describing the periodicity in the data.
Finally, we note that the RVs from our simulated spectra have a phase of
maximum RV identical to that found by \citet{dul2017} in their real
measurements. This indicates that the period suggested by
them might indeed be spurious due to residual telluric contamination.

  We conclude that the spurious effects due to blending of the \ha profile
of \bc with telluric lines, varying in position and strength, are inevitable.
This represents a serious warning that any periods close to
a year or half of it must be considered with caution.

\section{Period analysis of light changes with wavelets}
\begin{figure*}
\includegraphics[width=\textwidth]{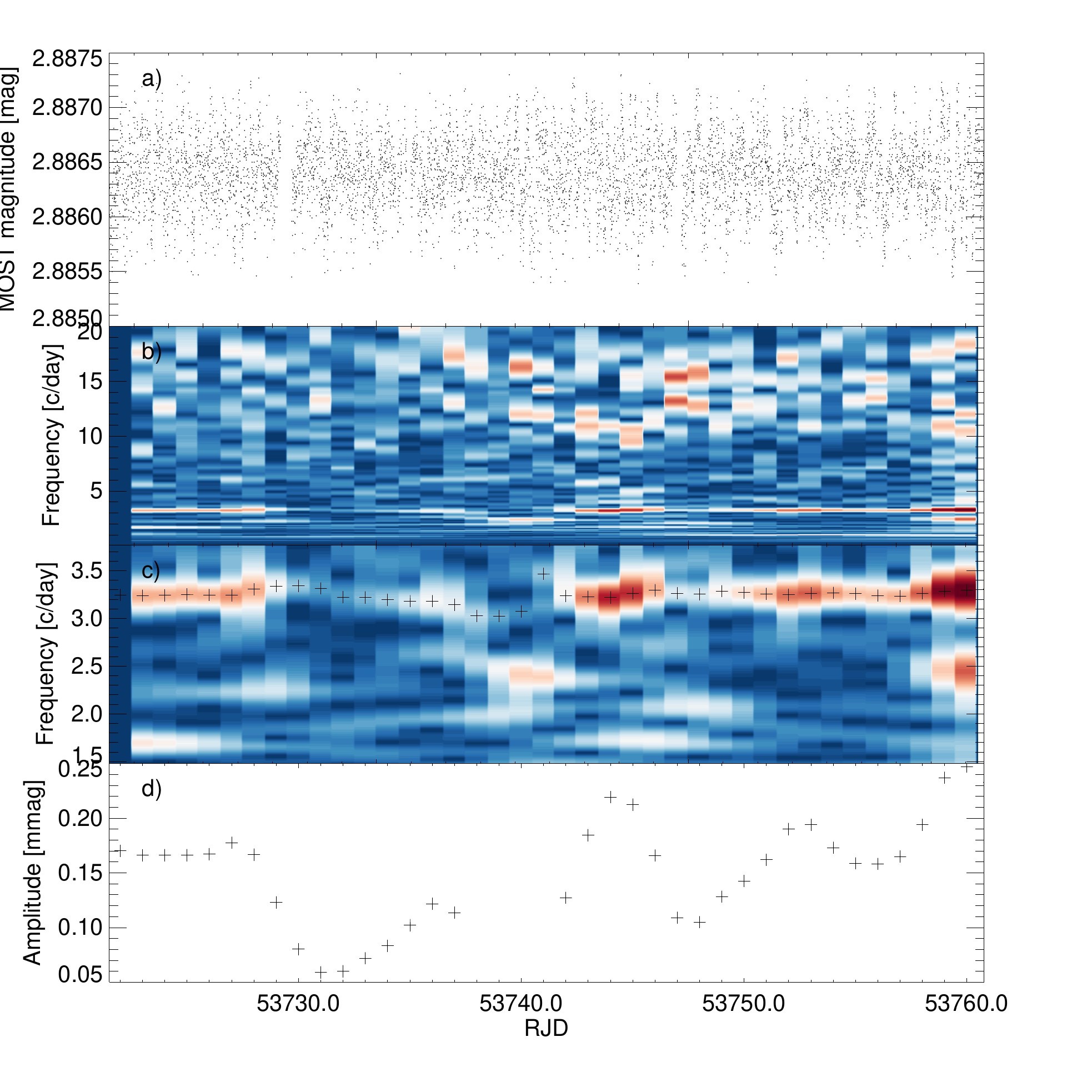}
\caption{Period analysis of MOST photometry of \bc (a) The observations, (b) the WWZ spectrum, (c) the zoom in the range of the dominant frequency with the peak identified, and (d) its amplitude over the interval of observations.}
\label{fig:wwz}
\end{figure*}

The star was almost continuously monitored by the Canadian MOST satellite for
more than 39 days with a very high cadence (an average separation between
the consecutive measurements is about 8 minutes -- see Fig.~\ref{fig:wwz}a).
Such a~data set allows to study photometric periods and their evolution.

The standard methods (for instance a Fourier analysis) were used to analyse
the same data set in the past \citep[e.g.][]{saio2007}. A~number of
short periods was identified and interpreted as evidence for non-radial
pulsations. The classical methods do not allow to learn about the evolution
of the periods over the interval of observations.

In order to study such a case we use the method that is based on the~wavelet
transform \citep[e.g.][]{1982Geop...47..203M}. For details of the method
and its trial application, see \citet{svanda2017}.
The wavelet transform can provide a~good time resolution for high-frequency
signals and a~good frequency resolution for low-frequency ones. That is
a~property well suited for real signals. The wavelet transform is thus
a~convenient method for the analysis of periodic signals with cycle lengths,
amplitude and/or phase, which evolve with time. If applied to irregular data
sets, the wavelet transform responds to irregularities rather than to actual
changes in the parameters of the signal. \citet{1996AJ....112.1709F}
introduced a~wavelet Z-transform (WWZ) that corrects the amplitude of
the detected period to the number of points available to properly describe
the given periodicity, thereby dealing with gaps and irregularities in
the dataset. The WWZ approach represents the principal method to investigate
the periodicities seen in the MOST photometry.

\begin{figure}
\includegraphics[width=0.5\textwidth]{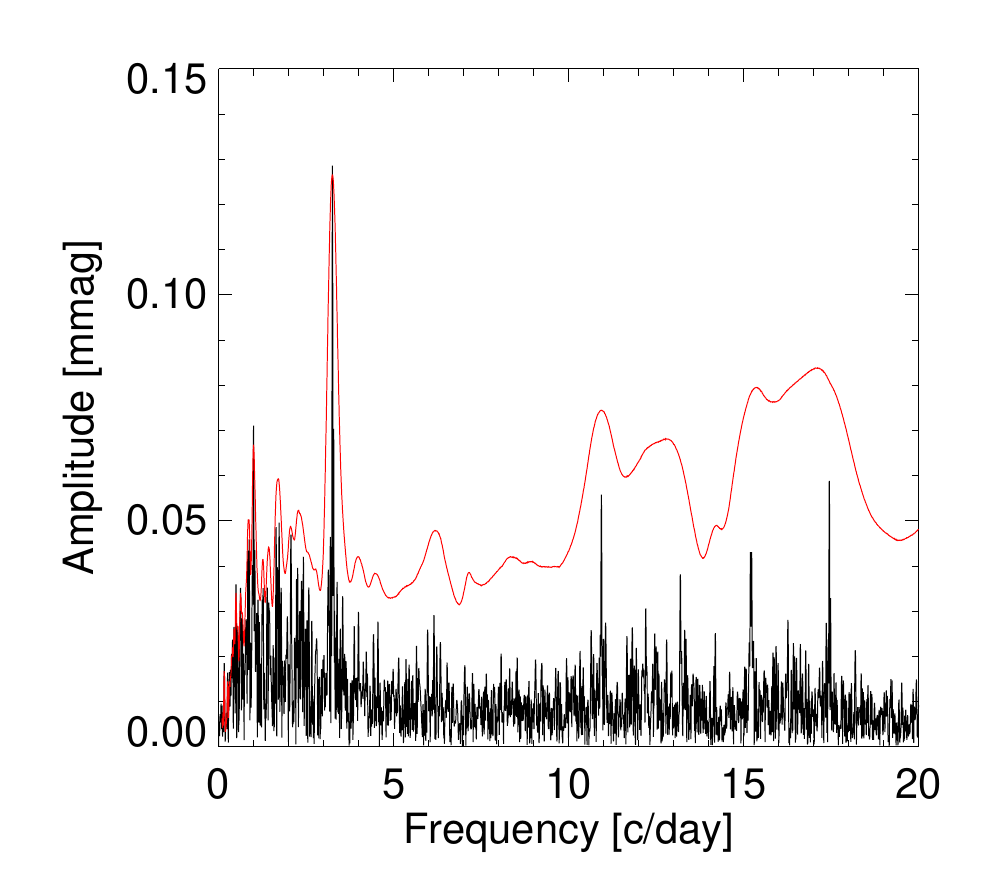}
\caption{An average periodogram obtained from WWZ analysis (as a time average of Fig.~\ref{fig:wwz}b -- the red line)
compared to the \breger periodogram (black curve). }
\label{fig:wwz_overall_spectrum}
\end{figure}

\begin{figure*}
\gridline{\fig{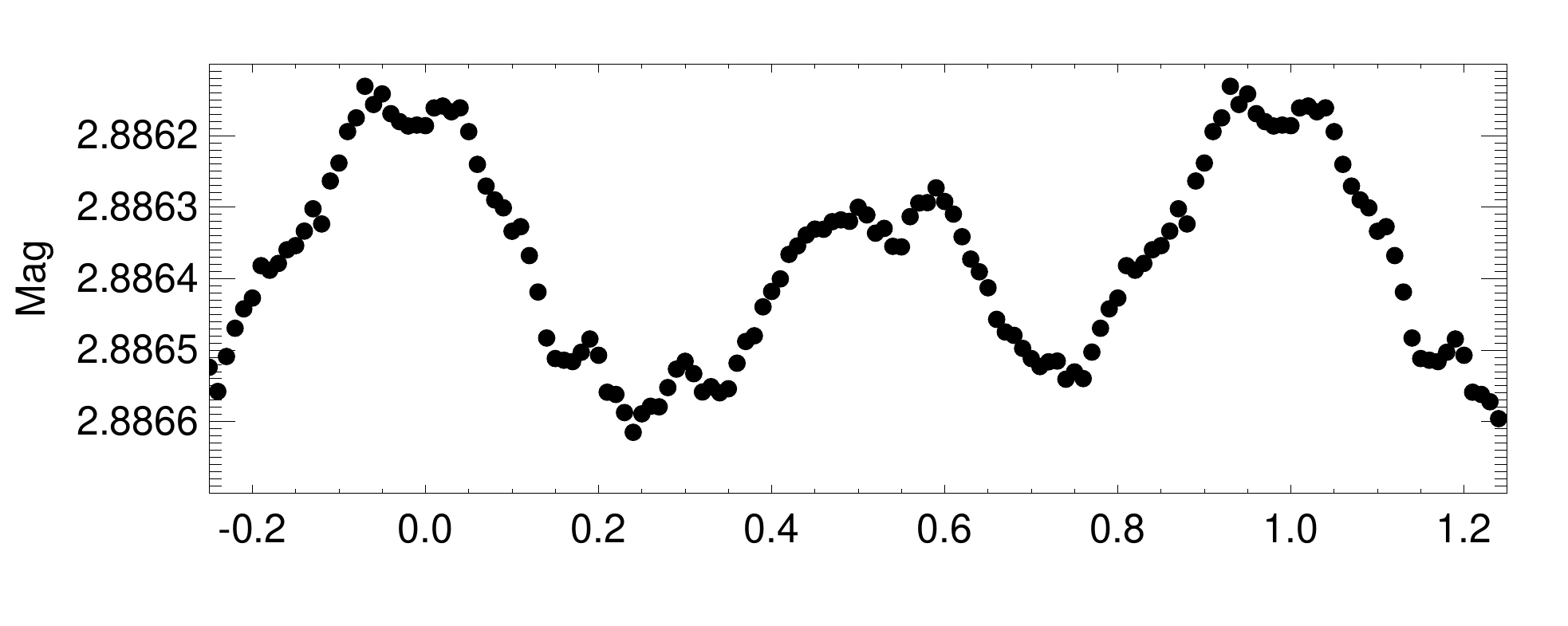}{0.46\textwidth}{RJD 53721.5 - 53729.2 \ \ \  P=0.613 d}
          \fig{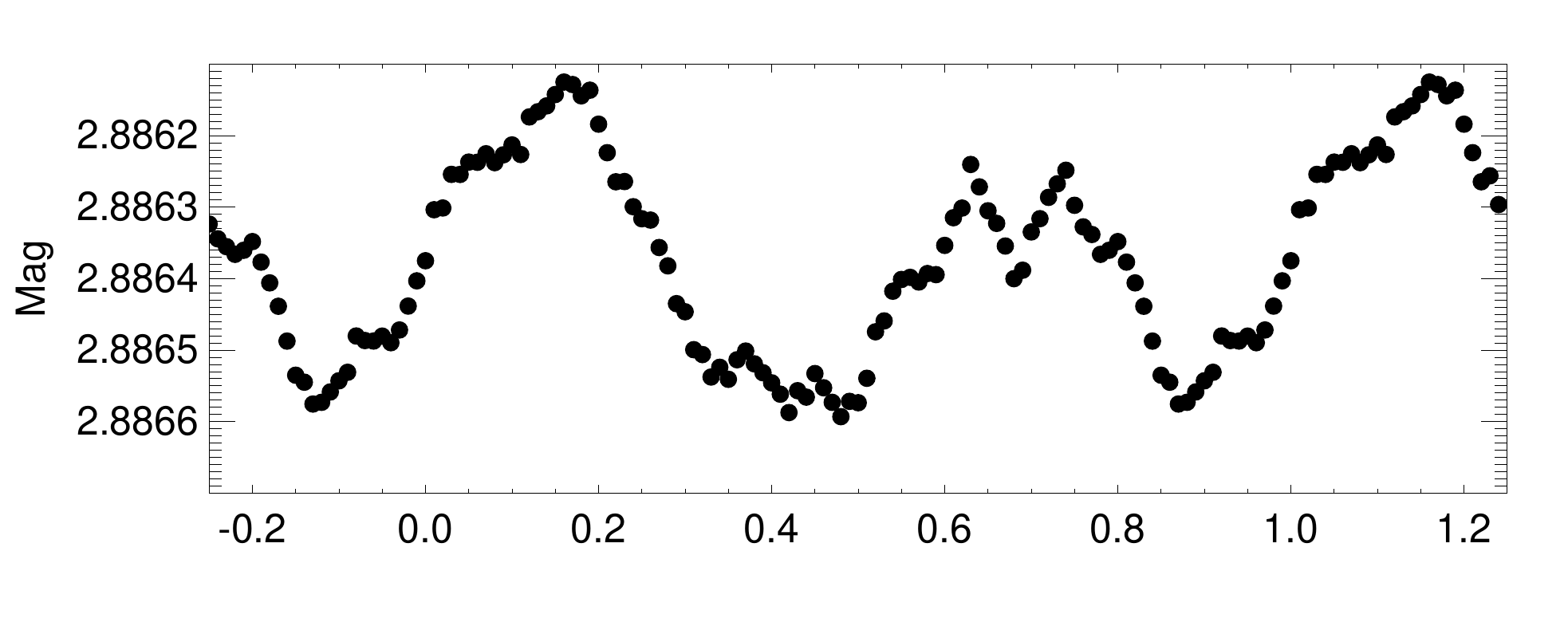}{0.46\textwidth}{RJD 53721.5 - 53729.2 \ \ \ P=0.617 d}
}
\gridline{\fig{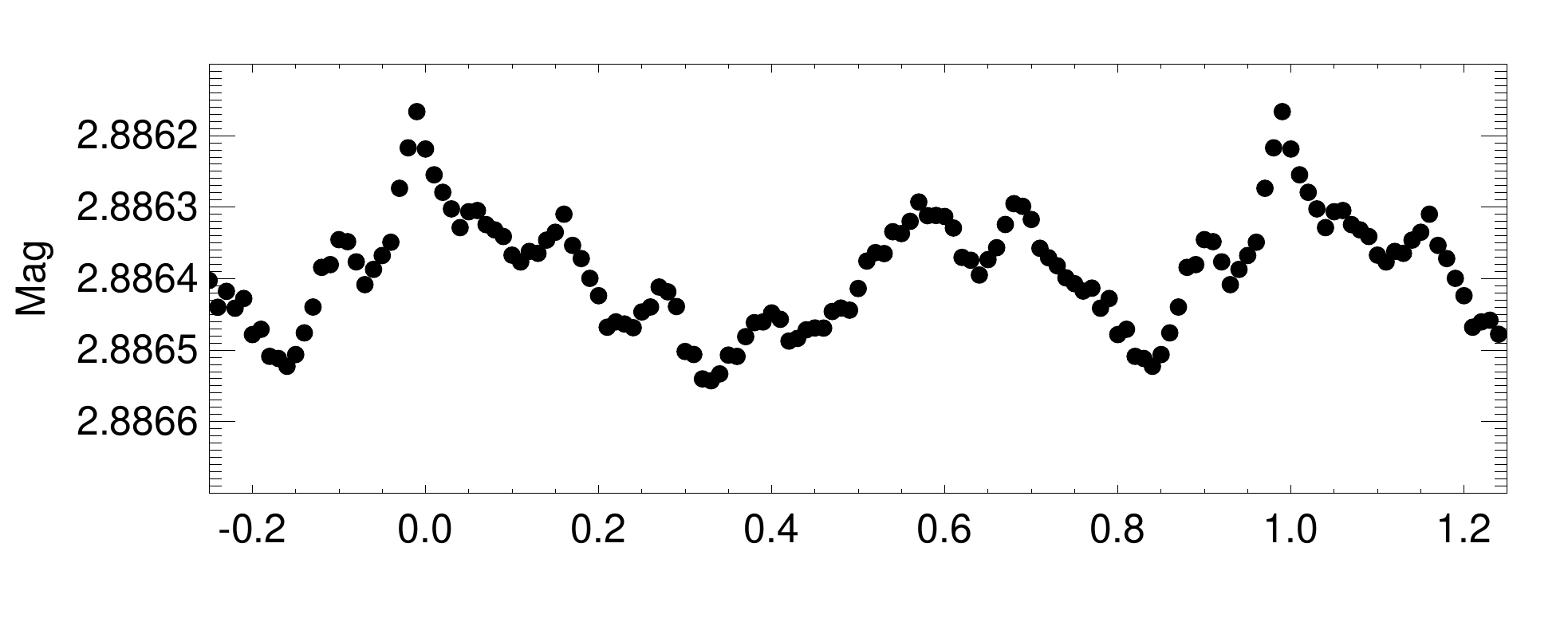}{0.46\textwidth}{RJD 53729.7 - 53737.2 \ \ \ P=0.620 d}
          \fig{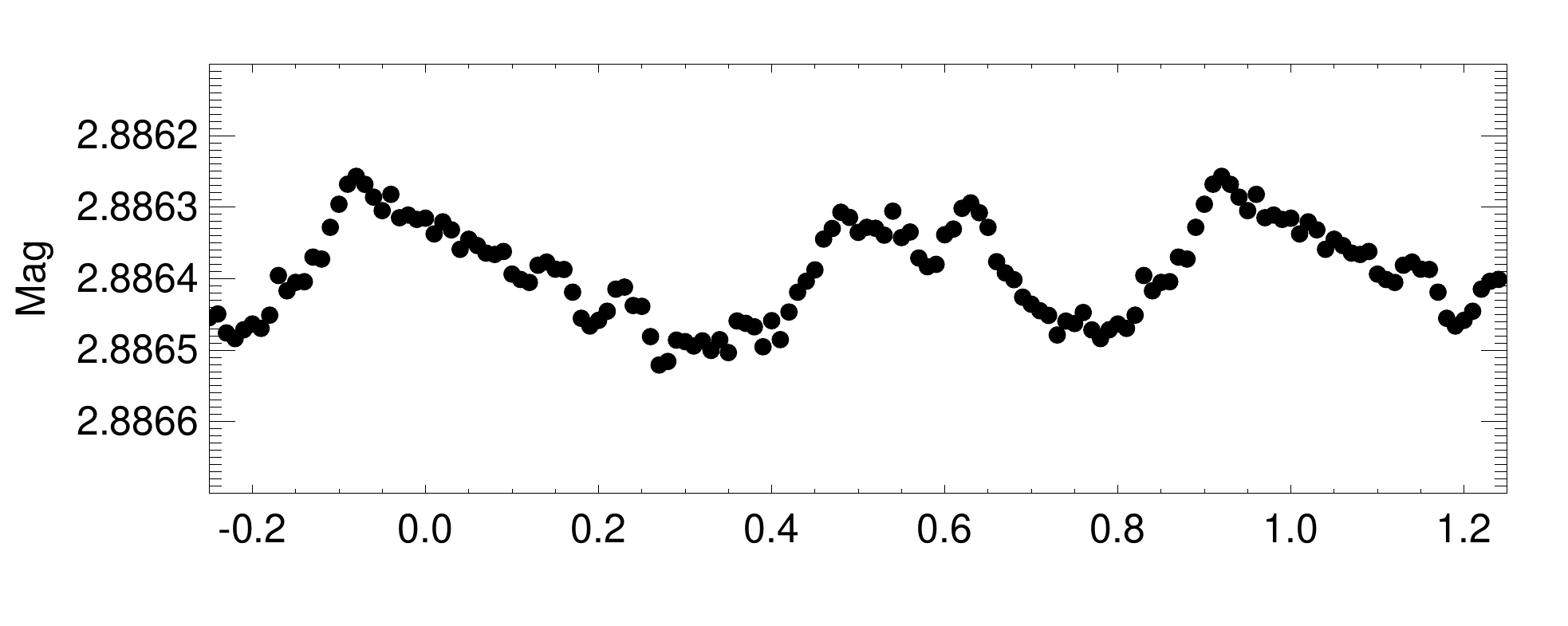}{0.46\textwidth}{RJD 53729.7 - 53737.2 \ \ \ P=0.617 d}
}
\gridline{\fig{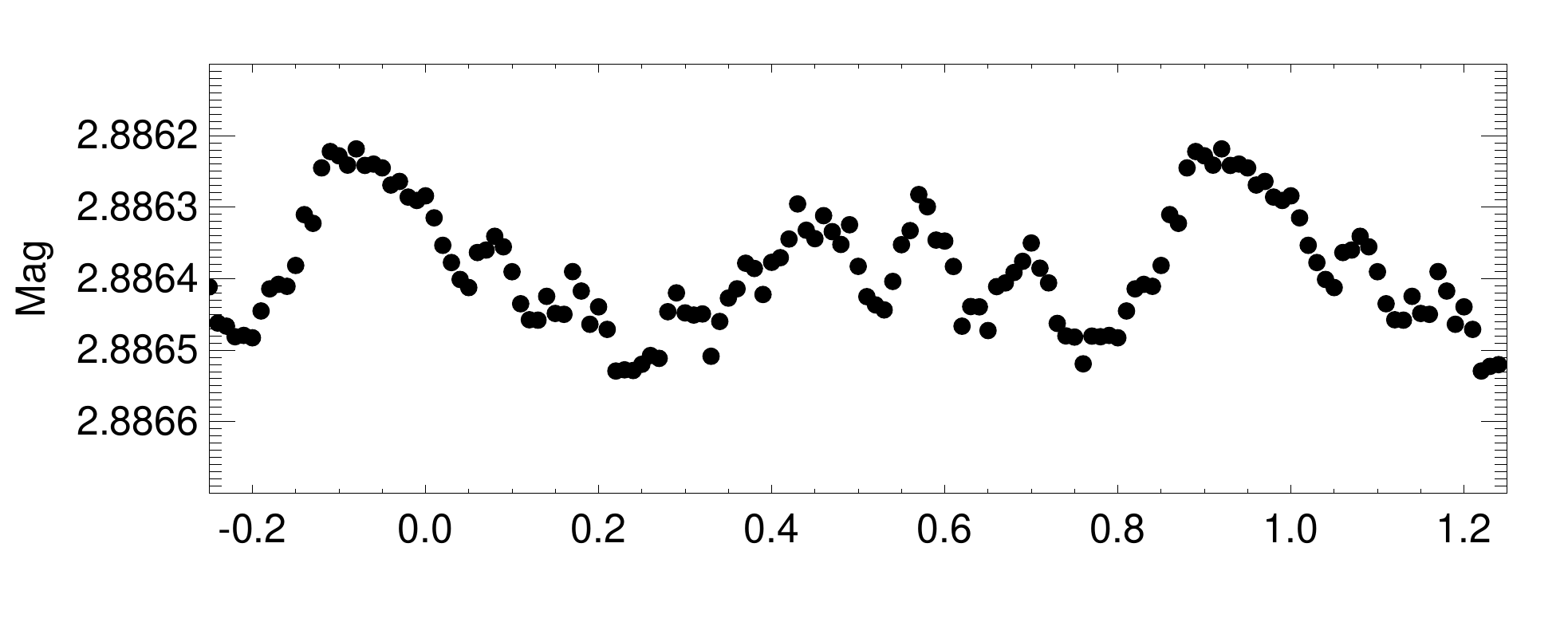}{0.46\textwidth}{RJD 53737.2 - 53745.1 \ \ \ P=0.628 d}
          \fig{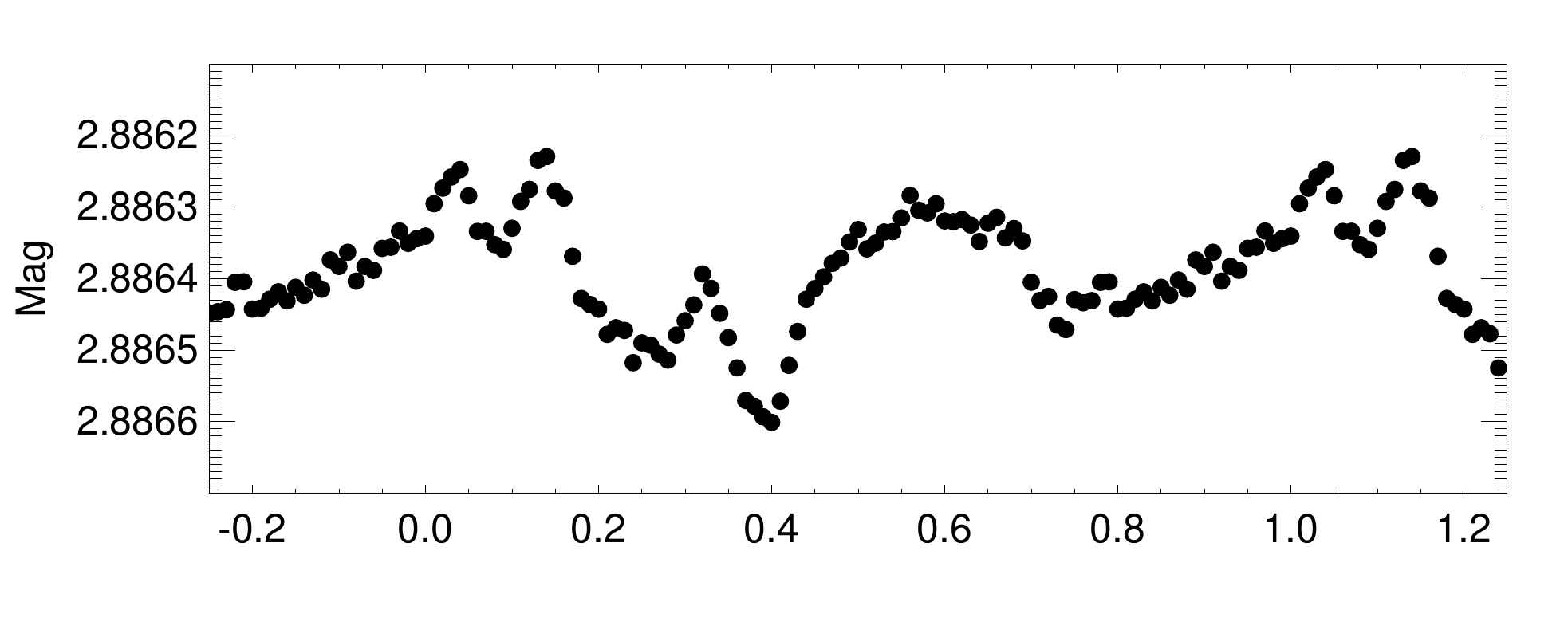}{0.46\textwidth}{RJD 53737.2 - 53745.1 \ \ \ P=0.617 d}
}
\gridline{\fig{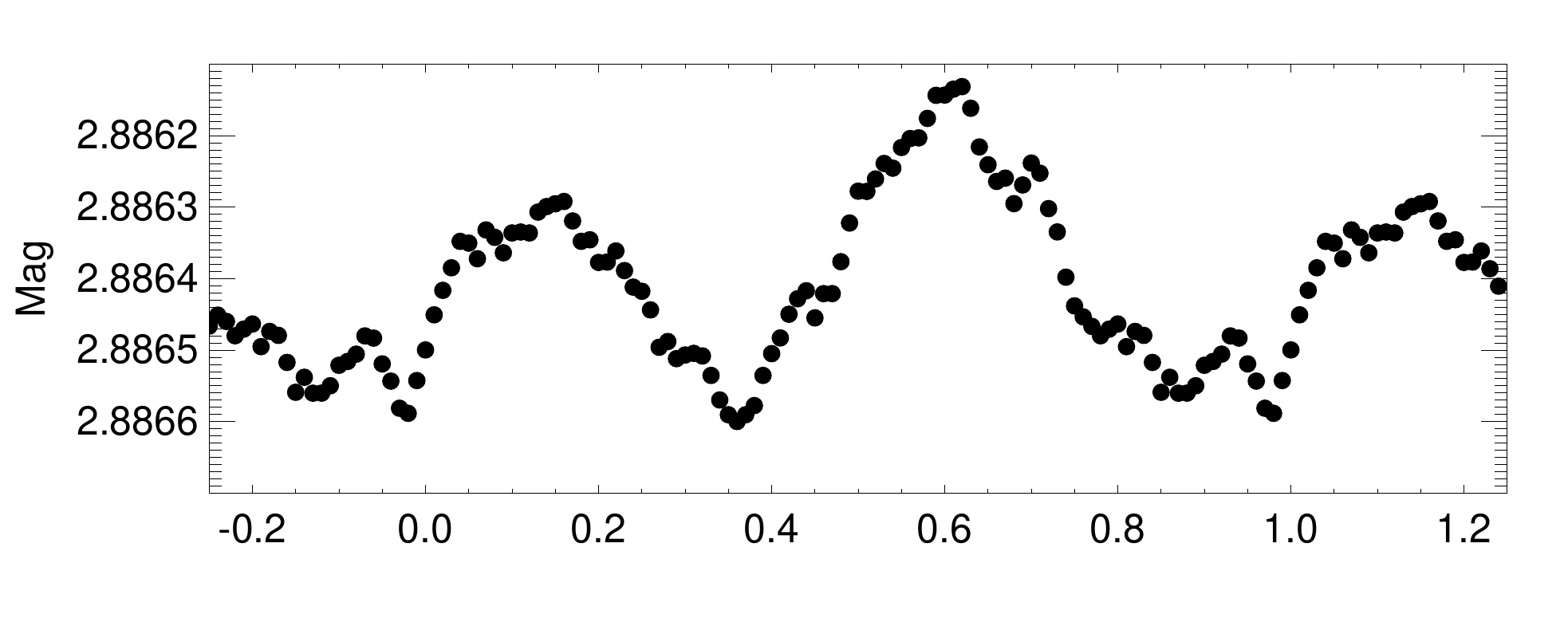}{0.46\textwidth}{RJD 53745.1 - 53752.9 \ \ \ P=0.612 d}
          \fig{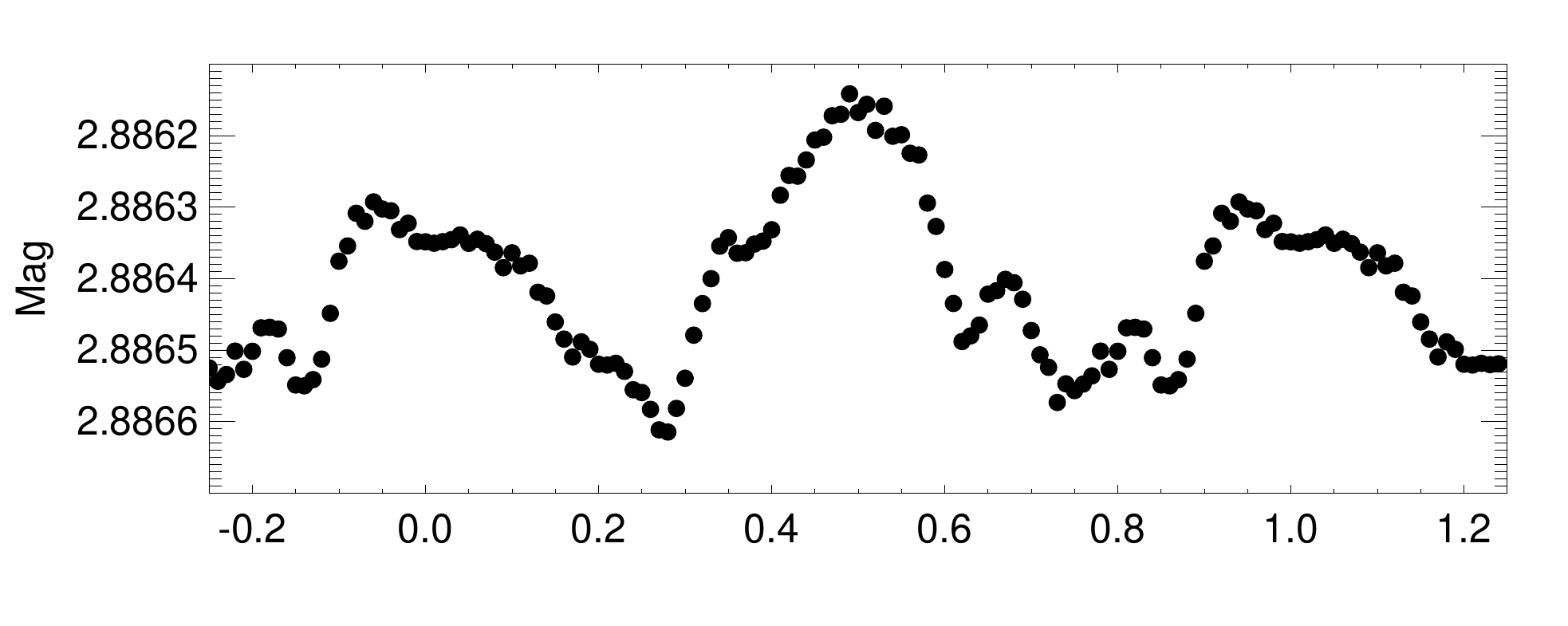}{0.46\textwidth}{RJD 53745.1 - 53752.9 \ \ \ P=0.617 d}
}
\gridline{\fig{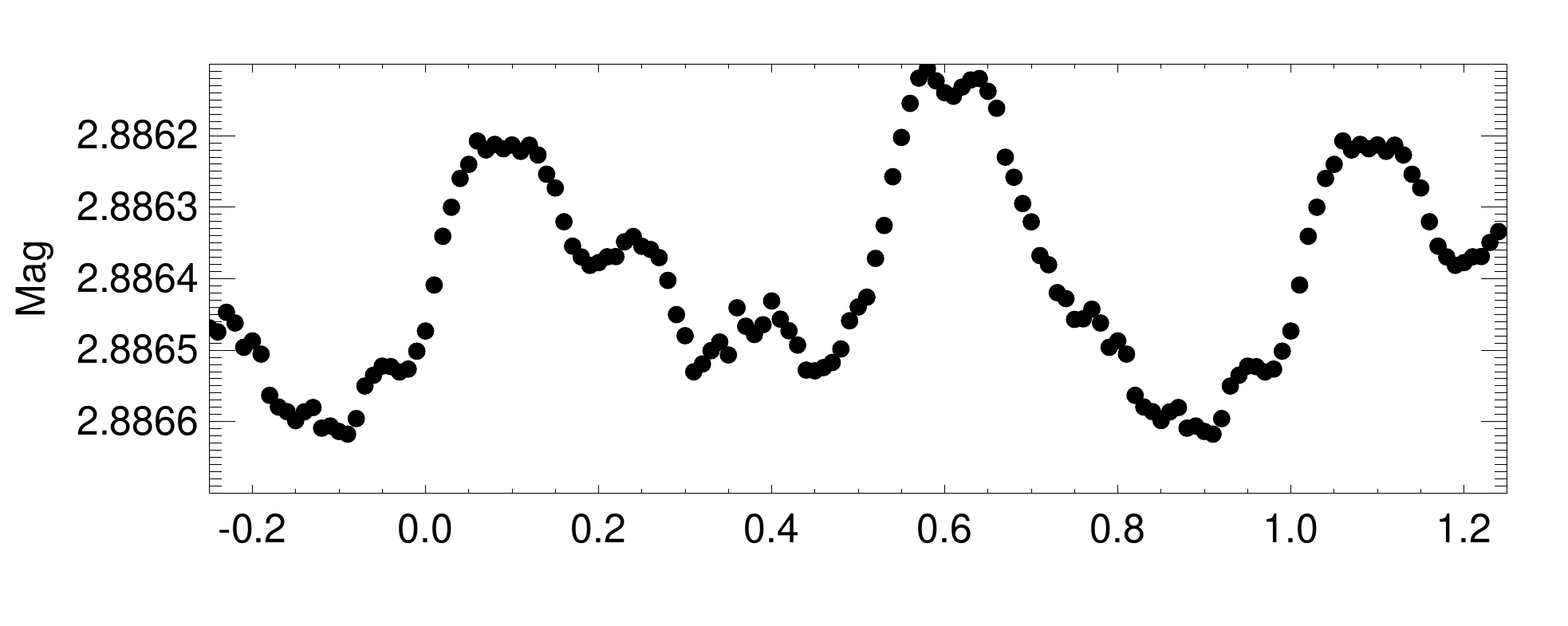}{0.46\textwidth}{RJD 53752.9 - 53760.8 \ \ \ P=0.613 d}
          \fig{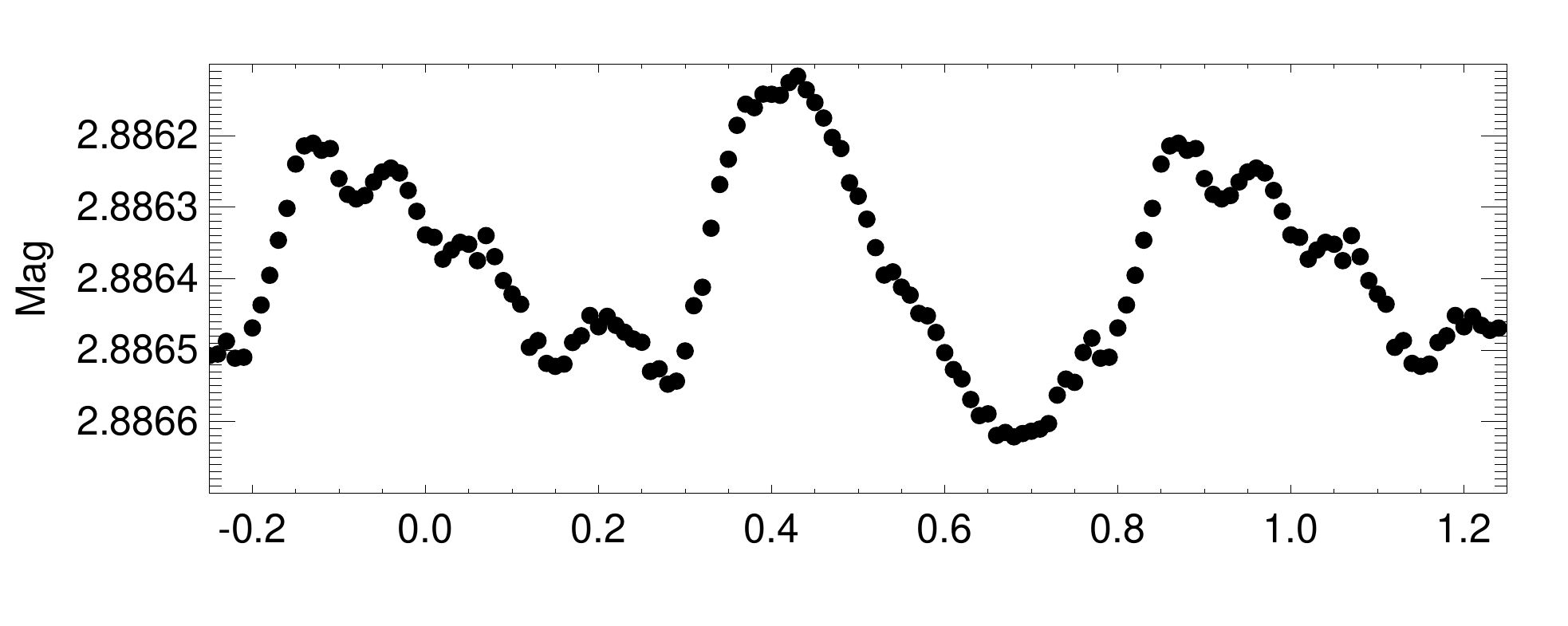}{0.46\textwidth}{RJD 53752.9 - 53760.8 \ \ \ P=0.617 d}
}
\caption{Double-wave light curves constructed from the MOST photometry
for five non-overlapping intervals. The corresponding intervals of RJDs
and local values of the period are given for each plot.
The left panels correspond to locally derived values of the period while
the right ones are for the constant mean period. On abscissa, the phase calculated
for each local value of the period, with phase zero set always at RJD 53740, is given
for each plot.}
\label{doublew}
\end{figure*}

The overall wavelet spectrum is plotted in Fig.~\ref{fig:wwz}b. It can be
seen that except for the enhanced power at around 3.25\cd{}, none of
the detected periods is stable over the whole interval of observations.
From time to time, a~strong power near 12~\cd{}, and 17~\cd{} appears and
disappears again. The averaged periodogram over the whole interval of
observations is plotted in Fig.~\ref{fig:wwz_overall_spectrum} in comparison
with a periodogram obtained with the program \breger based on the
\citet{deem75} method \citep{2004IAUS..224..786L}.
The resolution in frequency is poorer for WWZ spectrum, which is a~natural
property of this method because for time instant only a~limited
interval of observations is considered. Nevertheless, all important
frequencies below 5~\cd{} are detected by both approaches. The two differ
for higher frequencies, where the peak-width is much broader for
the WWZ spectrum than for the \breger spectrum. Yet, the \breger peaks
around 11~\cd, 13~\cd, 15~\cd{} and 17~\cd{} can be identified in
the average WWZ periodogram as well as transients. When looking at the full WWZ spectrum
in Fig.~\ref{fig:wwz}b one sees that these high-frequency periods were
detected only episodically in the second half of the observational interval
and were virtually absent in the first half.

 The strongest periodicity around 3.25~\cd{} was quite stable over the whole
interval (see Fig.~\ref{fig:wwz}c) with the cycle length of $0.307 \pm 0.004$~days
and underwent smooth changes around its mean value. Its significance (amplitude) also evolved slightly
in time (Fig.~\ref{fig:wwz}d), this periodicity was on average responsible
for the observed photometric changes with the amplitude of 0\m00014$\pm$0\m00005.
The exception is seen in an interval between RJD 53738 and 53742, where
a different cycle of length about 2.4~\cd{} with a secular evolution dominates the spectrum.
Cycles of a similar length are episodically present also on other dates. From seeing the
WWZ spectrum in Fig.~\ref{fig:wwz}c one could even track a non-monotonic secular evolution
of such cycle.

\cite{saio2007} claimed to detect two separate peaks in the spectrum with
frequencies $3.257\pm0.001$~\cd{} and $3.282\pm0.004$~\cd{} that they
interpreted as a~coupling between the rotation of the star and
the non-radial $g$-mode pulsations. They further reported weak frequency
peaks with values of $3.135\pm0.006$~\cd{} and $3.380\pm0.002$~\cd.
The detection of separate peaks is not confirmed by the WWZ method.
The multiperiodicity of the close periods, when typically the $g$-modes
create frequency groups that are also observed on Be stars \citep{Rivinius2016},
manifest themselves as beat-like structures in the wavelet spectrum at a fixed
frequency \citep[cf. also][]{Papics2017}. The wavelet spectrum of \bc{} does
not contain such features and it is qualitatively different
from the typical wavelet spectrum of pulsating Be stars
\citep[see][and their Figs. 2, 4, and 6]{Rivinius2016}.

Instead of distinct multiperiodic representation we observe a~slowly
changing value of the dominant frequency oscillating about the value
of $\sim$3.257~\cd{}. If such an~evolving signal is analysed using
the standard periodogram methods, the method will find a~solution
in a sum of periodic signals with arbitrary periods. Even though
a representation of a times series with a set of periodic signals may be
sufficient for some applications, such a representation does not
reveal possible real physical changes. We do not claim that \bc{} is not
a pulsating star, we only demonstrate that there is no clear evidence of pulsations
in the MOST observations.

  There has been a long debate about the nature of the low-amplitude
(multi?)-periodic light variations observed for many Be stars. Since
\citet{hec84} called attention to the fact that these variations often
occur with a double-wave curve and this was confirmed on a larger sample
of stars by \citet{balona86}, small-amplitude periodic light changes
were found for a number of Be stars.
While P.~Harmanec and L.A.~Balona, among others, advocated the rotational
modulation of the stars in question or their circumstellar envelopes,
investigators like C.T.~Bolton, J.R.~Percy, D.~Baade or G.A.H.~Walker with
their collaborators advocated interpretation in terms of non-radial pulsations.
Readers are referred to, e.g. \citet{baba94} and many other papers by
the above authors or to a~more recent repetition
of the debate \citep[e.g.][]{Balona2013, Rivinius2013, rivi2013}.

 \citet{hec98} analyzed extended series of
observations of $\omega$~CMa, interpreted as non-radial pulsator by Baade,
and showed that the $O-C$ analysis of its RV changes with the 1\fd37 period
indicate mild secular cyclic variations of this period, possibly related to the strength
of the \ha emission. He also showed that a Fourier analysis of all RVs can misinterpret
a single periodicity with mildly variable period as multiperiodicity of several
close periods. It was also found that the shape of the periodic light variations
can change gradually, even within one season, from double-wave to single-wave
shape \citep[cf., e.g.][]{luis89, luis92, boulder}.
This also complicates the period analysis based on Fourier analyses.

In the case of \bc, the dominant period changing in time also represents the real
changes of the light curve. To demonstrate this effect, we split the whole 39-day
interval into 5 intervals of equal length. For each interval we determined the local value
of the dominant period around 3.257~\cd{} and construct the light curve as a function of
phase. We soon found that phase plots of the data with a half of this frequency
define a double-wave light variation, with unequal minima and maxima.
For a better clarity, the light curves were binned with a
binsize of $0.025$ in phase, the sampling in phase has a step of $0.01$.

The light curves from individual data subsets plotted in the left panels
of Fig.~\ref{doublew} show that the shape of the light curve varies indeed
from one subset to another. The spread of the points around
the mean curve is larger for the second and third subsets since,
as it is evident from Fig.~\ref{fig:wwz}c,
in those time intervals the value of the dominant period is changing rapidly.
On the other hand, in the first, fourth and fifth time subset the period was
quite stable, which led to a ``cleaner'' appearance of the light curve.

The plots also demonstrate that the choice of twice longer period then the dominant
local period detected with the wavelet analysis was a correct one.
The depths of the two maxima and minima are clearly different.
Furthermore the phase shift of the curves also varies from one subset to
another. As mentioned above, such a behaviour is characteristic for rapid light
changes of other Be stars, too -- see, for instance, the rapidly changing light curve
of the archetype rapidly variable B6e star $o$~And \citep{boulder,sareyan98,sareyan2002}
and for a number of other Be stars \citep{luis89,luis92}. That all is consistent
with the rotational modulation and physical changes taking place and inconsistent with
the multiperiodic oscillations.

On the other hand the variations of the peak frequency in the five intervals is small, thus it is almost
impossible to convincingly show the effects of the period determination on the phase light curves.
To show this weak effect, we plotted the same data for a constant period
of 0\fd617 the right panels of Fig.~\ref{doublew}. The differences between
the corresponding panels of Fig.~\ref{doublew}
are small and probably beyond statistical significance.

 In passing, we admit that in view of the very low amplitude of
the light changes, our interpretation is only tentative.
The majority of investigators currently favours the hypothesis of
non-radial pulsations to explain the rapid light- and also line-profile
changes of Be stars.

\section{Probable physical properties}
 There is actually a rather large range of uncertainty of the basic
physical properties of \be.

We used the \pyt program of J.A.~Nemravov\'a
\citep[see][for the most detailed description]{ksitau}
that interpolates in a~precalculated
grid of synthetic stellar spectra sampled in the effective temperature,
gravitational acceleration, and metallicity.
The synthetic spectra are compared either to individual observed spectra
or to the disentangled ones for one, two or more components of
a~multiple system (normalised to the joint continuum
of all bodies) and the initial parameters are optimised by minimisation
of $\chi^2$ until the best match is achieved.  In our case, we
used the blue parts of the high-S/N spectra in three regions:
$4000 - 4190$~\ANG, $4250 - 4420$~\ANG, and $4440 - 4600$~\ANG.
We fitted the individual spectra and as another trial, we also
disentangled the spectra with \korel and fitted the disentangled
spectrum. The result is summarised in Table~\ref{pyt}. The difference
between the two fits probably characterises the real uncertainty of
the values obtained.

For comparison, \citet{fremat2005} derived \tef = $11772\pm344$~K,
\lgg = $3.811\pm0.062$, and \vsin $230\pm14$~\ks, while \citet{arne54}
obtained \vsin$ = 260$~\kms. Reproducing the optical and UV spectral
energy distribution, \citet{wheel2012} concluded that the disk is in Keplerian
rotation and seen under an inclination of about $45^\circ-50^\circ$.
They adopted \tef = 12000~K and \lgg = 4.0 as typical for a B8e star
and estimated the equatorial radius of the star of 4.68~\rs, $A_{\rm V}=0$\m05,
and polar temperature of 15000~K.

\begin{deluxetable}{lrrrrrrrlrrr}
\tablecaption{Estimates of the radiative paramaters of \bc from the blue
spectra at our disposal with the program \pyte. Results of the fit for
individual spectra, and for the \korel disentangled spectrum are tabulated.
The solar composition was assumed.
\label{pyt}}
\tablewidth{0pt}
\tablehead{
\colhead{Quantity}&\colhead{Individual spectra}&
\colhead{\korele}}
\startdata
\tef (K)            & 12560   & 13010 \\
\lgg [cgs]          & 4.20    & 4.35 \\
\vsin (km\,s$^{-1}$)& 260     & 266 \\
\enddata
\end{deluxetable}

There is no Gaia DR2 release parallax of \bc since
the star is too bright. The most accurate parallax is from the
Hipparcos revised reduction \citep{leeuw2007}:
$p=0\farcs02017\pm0\farcs00020$. Combined with the calibrated standard
\ubv\ photometry from Hvar ($V=2$\m890, \bv$=-0$\m101, \ub$=-0$\m268)
and $A_{\rm V}=0$\m05 \citep{wheel2012}, one obtains $V_0=2$\m84,
and M$_{\rm V}=-0$\m64. Using the two extreme values of $\log$\tef,
4.071 \citep{fremat2005} and 4.114 (from the \pyt fit of \korel
disentangled blue spectrum) and bolometric corrections
from \citet{flower96}, one obtains the mean stellar radius of
3.83~\rs, or 3.52~\rs, respectively. This is in a broad agreement
with modelling results. \citet{kl2015} obtained the equatorial and
polar radii of 4.17~\rs, and 2.8~\rs, respectively, while
\citet{wheel2012} got 4.68~\rs, and 3.55~\rs. Their results imply
the mean radii of 3.42~\rs, and 4.07~\rs. For completeness,
mean radii of normal stars for the two extreme values of \tef would be
2.45, and 2.65~\rs\ \citep{mr88}.

Our \vsin of $\sim260$~\kms combined with the published values of equatorial
radius of 4.68~\rs\ and 4.20~\rs\ and assuming that the 0\fd617 period
is the stellar rotational period would lead to the inclination of the
rotational axis of $42^\circ\!\!.7$, and $49^\circ\!\!.0$, respectively.

 Assuming that the circumstellar disk lies in the stellar equatorial plane
\citet{kl2015} obtained an inclination of 43$^\circ$ and \citet{wheel2012}
estimated the range of inclinations $45^\circ-50^\circ$.  Our identification
of the 0\fd617 period with the stellar rotational period is therefore
quite consistent with these pieces of information.

\section{Conclusions}
Our detailed analyses presented in previous sections demonstrated that
the spectroscopic binary nature of \bc with an orbital period of
170\fd4 could not be confirmed. The star can still be a binary
(as conjectured from the truncation of
its circumstellar disk by \citealt{kl2015}) but there is currently
no direct proof of it, based on periodic RV changes.

 The discussion of RVs from photographic spectra in sect.~\ref{phgrv}
indicates that there could be time intervals in the past when the RV was
varying within a rather large range of values, perhaps due to
transient asymmetries in the circumstellar disk. We remind that
$V/R=1.17$ was found on an~\ha profile from March 1953 published
by \citet{under53}.

 From the wavelet analysis of the precise MOST satellite photometry
we conclude that there is only one stable (maybe slightly variable) period of
low-amplitude light changes, 0\fd617. We tentatively identify it
with the star's rotational period.

\begin{acknowledgements}
We acknowledge the use of archival spectra from the Haute Provence, Canada
France Hawaii telescope, and ESO public archives.
The reduced MOST observations of \bc were kindly provided to us by
Christopher Cameron. Some of the Ond\v{r}ejov spectra were secured by
M.~Dov\v{c}iak, M.~Kraus, R.~K\v{r}\'\i\v{c}ek, P.~Nem\'eth, J.A.~Nemravov\'a,
and M.~Wolf. We thank Jana~A.~Nemravov\'a (who left the astronomical research
in the meantime) for the permission to use her program \pyte.
We also acknowledge the use of the public version of the program \korel
written by Petr Hadrava. The tough but largely justified criticism of the
first version of this paper by an anonymous referee helped us to improve
the text and argumentation and is gratefully acknowledged.
In the initial phases of this project, P.~Harmanec and M.~\v{S}vanda were
supported from the grant GA15-02112S of the Czech Science Foundation.
In the final stage of this study, P.~Harmanec was supported from the grant
GA19-01995S of the Czech Science Foundation.
Research of D.~Kor\v{c}\'akov\'a is supported
from the grant GA17-00871S of the Czech Science Foundation.
H.~Bo\v{z}i\'c acknowledges financial support from Croatian Science Foundation under
the project 6212 ``Solar and Stellar Variability". Research of L.~Vanzi is
supported from CONICYT through project Fondecyt n. 1171364.
The use of the NASA/ADS bibliographical service, and
the Simbad database are greatly acknowledged.
\end{acknowledgements}





\appendix

\section{Appendix Tables with individual RVs and spectrophotometric
quantities}

We provide here individual RVs with RJDs derived by us for the Yerkes
photographic spectra in Tab.~\ref{oldrv} and also individual
RVs and spectrophotometric quatities for our \ha spectra - see
Tab.~\ref{ourrv}.  We tabulate the peak intensities $I_V$ and $I_R$
measured in the units of the local continuum level, their average and
ratio, and the central intensity of the \ha absorption core $I_{\rm c}$.
Individual spectrographs are identified in the last column ``Spg. No."
by numbers they have in Tab.~\ref{jousp}.

\begin{deluxetable}{rrrrrrrlrrr}
\tablecaption{RVs (in \ks) from Yerkes photographic spectra
published by \citet{frost26}\label{oldrv}}
\tablewidth{0pt}
\tablehead{
\colhead{RJD}&\colhead{RV}
}
\startdata
16796.9717 &  47.90 \\
18670.7214 &  13.40 \\
19022.7829 &  -1.30 \\
19029.7902 &  44.20 \\
19039.8355 &  20.70 \\
19092.6680 &  53.20 \\
19367.8188 &  15.60 \\
\enddata
\end{deluxetable}

\startlongtable
\begin{deluxetable*}{crrrcccccrrrr}
\tablecaption{Individual RV (in \ks) and line-strength measurements
in the \ha spectra at our disposal.
\label{ourrv}}
\tablewidth{0pt}
\tablehead{
\colhead{RJD}&\colhead{RV$_{\rm em.}$}&\colhead{RV$_{\rm em.}^{\rm clean}$}&
\colhead{dRV$_{\rm zero}$}&\colhead{rms$_{\,\rm telur.l.}$}&\colhead{$(I_V+I_R)/2$}&
\colhead{Spg. No.}
}
\startdata
49745.8724& 8.44$\pm$0.22& 8.90$\pm$0.09&-0.36&0.16&$1.7095\pm0.0095$& 1\\
49745.8813& 8.42$\pm$0.17& 8.24$\pm$0.08& 0.15&0.12&$1.7065\pm0.0035$& 1\\
51567.4879& 7.15$\pm$0.40& 5.64$\pm$0.16&-0.56&0.75&$1.6030\pm0.0010$& 2\\
51589.4298&12.43$\pm$0.03&13.01$\pm$0.09& 2.02&0.28&$1.6795\pm0.0015$& 2\\
51641.2949&12.76$\pm$0.03&14.10$\pm$0.51& 3.79&0.95&$1.6810\pm0.0130$& 2\\
51655.3022&12.21$\pm$0.15&12.09$\pm$0.24& 1.87&0.20&$1.6785\pm0.0085$& 2\\
51661.3168&11.79$\pm$0.22&11.77$\pm$0.09& 2.67&0.19&$1.6885\pm0.0085$& 2\\
52229.5462&10.58$\pm$0.70& 8.29$\pm$0.18&-0.07&0.27&$1.6735\pm0.0065$& 3\\
52234.5702& 8.83$\pm$0.03& 9.14$\pm$0.06&-0.95&0.46&$1.6825\pm0.0185$& 3\\
52234.5821& 9.11$\pm$0.29& 8.76$\pm$0.12&-0.93&0.36&$1.6895\pm0.0165$& 3\\
52362.3431& 8.46$\pm$0.14& 8.91$\pm$0.08&-0.01&0.07&$1.7775\pm0.0035$& 4\\
52706.7670& 9.31$\pm$0.11& 9.56$\pm$0.10& 1.28&0.19&$1.7910\pm0.0090$& 1\\
52706.7811& 8.55$\pm$0.22& 8.43$\pm$0.10& 1.42&0.17&$1.7860\pm0.0020$& 1\\
52706.7964& 9.22$\pm$0.22& 9.39$\pm$0.06& 1.12&0.25&$1.7935\pm0.0025$& 1\\
52749.3933& 8.09$\pm$0.38& 9.21$\pm$1.32&-1.00&0.30&$1.7645\pm0.0075$& 3\\
52750.3686&10.60$\pm$0.03& 9.23$\pm$0.67&-0.70&0.30&$1.7700\pm0.0010$& 3\\
52770.6970& 8.40$\pm$0.17& 8.94$\pm$0.15& 4.21&0.11&$1.7420\pm0.0220$& 1\\
52770.7018& 8.29$\pm$0.17& 8.34$\pm$0.10& 4.53&0.12&$1.7870\pm0.0040$& 1\\
52771.6852& 9.28$\pm$0.11& 8.71$\pm$0.09& 2.76&0.11&$1.7775\pm0.0035$& 1\\
52983.4974& 9.04$\pm$0.29& 9.59$\pm$0.09&-0.43&0.31&$1.7585\pm0.0205$& 3\\
53011.7141& 8.69$\pm$0.15& 7.99$\pm$0.11&-0.95&0.09&$1.7880\pm0.0130$& 5\\
53012.6330& 8.37$\pm$0.21& 8.96$\pm$0.17&-0.24&0.05&$1.8000\pm0.0070$& 5\\
53028.6014& 9.09$\pm$0.03& 9.39$\pm$0.10&-0.47&0.47&$1.7715\pm0.0085$& 3\\
53079.7813& 9.26$\pm$0.33& 9.20$\pm$0.09& 0.87&0.12&$1.7885\pm0.0115$& 1\\
53097.3956& 9.16$\pm$0.52& 9.85$\pm$0.12& 0.55&0.35&$1.7720\pm0.0060$& 3\\
53097.3993& 8.69$\pm$0.61&10.18$\pm$0.08& 0.38&0.41&$1.7780\pm0.0020$& 3\\
53105.7900& 8.02$\pm$0.22& 8.06$\pm$0.09& 1.58&0.14&$1.7830\pm0.0120$& 1\\
53134.7252& 8.66$\pm$0.34& 9.17$\pm$0.06& 0.25&0.19&$1.7975\pm0.0025$& 1\\
53261.0526& 8.50$\pm$0.17& 8.62$\pm$0.10& 1.24&0.14&$1.7395\pm0.0025$& 1\\
53274.0536& 8.36$\pm$0.17& 8.25$\pm$0.08& 1.48&0.13&$1.7590\pm0.0030$& 1\\
53334.8232& 8.10$\pm$0.17& 7.90$\pm$0.09& 0.75&0.06&$1.8925\pm0.0055$& 6\\
53334.8527& 7.64$\pm$0.07& 7.44$\pm$0.18& 1.17&0.04&$1.8250\pm0.0030$& 6\\
53385.9364& 9.26$\pm$0.11& 8.81$\pm$0.06& 0.45&0.14&$1.7720\pm0.0100$& 1\\
53406.8866& 9.67$\pm$0.19& 9.62$\pm$0.09& 0.50&0.19&$1.7765\pm0.0015$& 1\\
53469.7531& 9.45$\pm$0.11& 9.64$\pm$0.09& 0.57&0.14&$1.7720\pm0.0060$& 1\\
53470.7984& 9.21$\pm$0.11& 9.33$\pm$0.10& 0.40&0.13&$1.7890\pm0.0030$& 1\\
53638.0548& 8.43$\pm$0.11& 9.35$\pm$0.13& 0.03&0.16&$1.7440\pm0.0010$& 1\\
53638.0566& 8.54$\pm$0.11& 8.56$\pm$0.12&-0.06&0.15&$1.7460\pm0.0040$& 1\\
53655.0693& 9.53$\pm$0.17& 9.54$\pm$0.10& 0.12&0.14&$1.7535\pm0.0085$& 1\\
53681.0415& 8.94$\pm$0.30& 9.23$\pm$0.10& 0.19&0.18&$1.7735\pm0.0005$& 1\\
53791.8117& 9.15$\pm$0.34& 9.11$\pm$0.10&-0.41&0.17&$1.7805\pm0.0065$& 1\\
53814.7848& 9.18$\pm$0.11&10.06$\pm$0.10&-0.78&0.14&$1.7675\pm0.0035$& 1\\
53814.7867& 9.30$\pm$0.17& 9.69$\pm$0.09&-0.96&0.19&$1.7590\pm0.0030$& 1\\
54004.0644& 8.58$\pm$0.23& 8.72$\pm$0.09&-0.29&0.11&$1.7545\pm0.0055$& 1\\
54044.9503& 8.93$\pm$0.17& 9.05$\pm$0.09& 0.69&0.12&$1.7555\pm0.0035$& 1\\
54107.0056& 8.87$\pm$0.17& 8.93$\pm$0.10&-0.61&0.16&$1.7680\pm0.0120$& 1\\
54108.5620& 8.48$\pm$0.02& 8.28$\pm$0.08&-1.04&0.43&$1.7495\pm0.0225$& 3\\
54108.5657& 8.54$\pm$0.04& 8.04$\pm$0.09&-0.38&0.41&$1.7470\pm0.0230$& 3\\
54169.8236& 8.36$\pm$0.23& 8.92$\pm$0.13&-0.42&0.18&$1.7865\pm0.0015$& 1\\
54401.0389& 9.93$\pm$0.17& 8.45$\pm$0.08& 0.32&0.42&$1.7420\pm0.0040$& 1\\
54423.7029& 7.72$\pm$0.24& 7.81$\pm$0.05&-0.09&0.06&$1.8055\pm0.0035$& 7\\
54423.7047& 7.54$\pm$0.25& 8.21$\pm$0.05&-0.06&0.07&$1.8090\pm0.0030$& 7\\
54443.0654& 7.95$\pm$0.11& 8.64$\pm$0.09&-0.12&0.22&$1.7645\pm0.0105$& 1\\
54518.8122& 9.17$\pm$0.17& 9.56$\pm$0.09&-0.05&0.19&$1.7770\pm0.0050$& 1\\
54519.6421& 9.42$\pm$0.20& 9.29$\pm$0.10& 0.16&0.19&$1.7825\pm0.0025$& 1\\
55320.4961& 9.65$\pm$0.27&11.02$\pm$0.11& 3.99&0.12&$1.8030\pm0.0020$& 8\\
55470.6614& 8.84$\pm$0.05& 8.67$\pm$0.09& 0.18&0.46&$1.7210\pm0.0030$& 3\\
55829.6123& 9.45$\pm$0.29& 9.54$\pm$0.15& 0.08&0.45&$1.6800\pm0.0130$& 3\\
55867.8024& 8.97$\pm$0.14&10.06$\pm$0.15&-3.00&0.09&$1.8090\pm0.0090$& 8\\
55871.1482& 8.82$\pm$0.06& 8.50$\pm$0.10& 0.03&0.06&$1.7655\pm0.0025$& 9\\
55877.1439& 8.77$\pm$0.14& 8.60$\pm$0.09& 0.07&0.05&$1.7625\pm0.0025$& 9\\
55976.5869&10.02$\pm$0.27& 9.67$\pm$0.13&-3.46&0.05&$1.7365\pm0.0035$& 8\\
55977.6398& 9.81$\pm$0.06&10.13$\pm$0.12&-5.20&0.06&$1.7235\pm0.0035$& 8\\
56323.7302& 7.09$\pm$0.21& 8.46$\pm$0.09&-4.12&0.07&$1.7795\pm0.0125$& 8\\
56332.3951& 7.51$\pm$1.17& 8.10$\pm$0.51& 0.01&0.58&$1.7395\pm0.0065$& 3\\
56350.6653& 8.96$\pm$0.14& 8.79$\pm$0.12& 0.35&0.15&$1.8260\pm0.0160$&10\\
56354.4928& 8.32$\pm$0.04& 8.92$\pm$0.12&-0.52&0.38&$1.7515\pm0.0025$& 3\\
56366.3279& 8.16$\pm$0.03& 7.70$\pm$0.09& 1.24&0.40&$1.7480\pm0.0020$& 3\\
56395.3447&10.34$\pm$0.03& 9.60$\pm$0.19&-0.71&0.27&$1.7500\pm0.0020$& 3\\
56606.5390&10.30$\pm$0.04&10.05$\pm$0.09& 0.70&0.53&$1.7365\pm0.0055$& 3\\
56613.8631& 7.77$\pm$0.07& 8.38$\pm$0.05&-1.53&0.05&$1.8375\pm0.0035$& 8\\
56647.8767& 8.09$\pm$0.27& 8.02$\pm$0.10&-0.32&0.09&$1.8175\pm0.0125$& 8\\
56648.3346&10.01$\pm$0.18& 9.81$\pm$0.44& 1.04&0.32&$1.7085\pm0.0365$& 3\\
56681.6714& 8.34$\pm$0.04& 8.60$\pm$0.09&-0.07&0.06&$1.8420\pm0.0120$& 8\\
56712.2316& 9.36$\pm$0.22& 9.79$\pm$0.50&-0.30&0.37&$1.7550\pm0.0100$& 3\\
56929.9217& 9.70$\pm$0.07& 8.67$\pm$0.11& 0.54&0.12&$1.7020\pm0.0050$& 8\\
56962.8833& 8.98$\pm$0.03& 9.73$\pm$0.06& 1.02&0.08&$1.8095\pm0.0105$& 8\\
57078.5964&10.11$\pm$0.45& 8.42$\pm$0.24& 0.99&0.36&$1.8755\pm0.0335$&10\\
57080.5584& 9.73$\pm$0.24& 9.77$\pm$0.15& 1.08&0.30&$1.8290\pm0.0030$&10\\
57396.5672& 9.77$\pm$0.26&10.02$\pm$0.08& 0.28&0.42&$1.7460\pm0.0150$& 3\\
57443.2592& 9.73$\pm$0.02&10.07$\pm$0.63& 1.18&0.53&$1.7500\pm0.0080$& 3\\
58163.2564& 9.15$\pm$0.07& 8.72$\pm$0.30&-0.97&0.53&$1.7770\pm0.0040$& 3\\
58173.4674& 9.40$\pm$0.03& 9.95$\pm$0.11&-2.01&0.61&$1.7620\pm0.0040$& 3\\
58198.3383& 9.92$\pm$0.26&10.56$\pm$0.10&-0.19&0.44&$1.7635\pm0.0055$& 3\\
\enddata
\end{deluxetable*}

\bibliography{betacmi}
\end{document}